\documentclass[useAMS,usenatbib]{mn2e}
\pdfoutput=1

\newcommand\Msol{$M_{\odot}$}
\newcommand\Lsol{$L_{\odot}$}


\usepackage{times}
\usepackage[pdftex]{graphicx}
\usepackage{aas_macros}
\usepackage{amssymb}
\usepackage{natbib}
\usepackage{multirow}
\usepackage{xspace}
\usepackage{pdflscape}
\usepackage{multicol}
\usepackage{enumitem}
\usepackage{color}
\usepackage{setspace}
\usepackage{amsmath}
\usepackage{rotating}
\usepackage{ marvosym }

\newcommand\sersic{S\'{e}rsic }
\newcommand\sersics{S\'{e}rsics }
\newcommand\Sersic{S\'{e}rsic}

\newcommand\second{^{\prime\prime}}
\newcommand\minute{^{\prime}}

\title[Beyond \sersic + exponential in Coma]
{Beyond \sersic + exponential disc morphologies in the Coma Cluster}
\author[J.T.C.G. Head et al.]
{
Jacob T.C.G. Head$^1$\thanks{Email: j.t.c.head@durham.ac.uk}, John R. Lucey$^1$, Michael J. Hudson$^{2,3}$\\
$^{1}$Department of Physics, University of Durham, South Road, Durham, DH1 3LE, UK\\
$^{2}$Department of Physics and Astronomy, University of Waterloo, Waterloo, Ontario, N2L 3G1, Canada\\
$^{3}$Perimeter Institute for Theoretical Physics, 31 Caroline St. N., Waterloo, ON, N2L 2Y5, Canada\\
}
\begin{document}

\date{Submitted: 03-11-2014 Accepted: 21-07-2015}
\pagerange{\pageref{firstpage}--\pageref{lastpage}} \pubyear{2015}
\maketitle{}
\label{firstpage}
%%%%%%%%%%%%%%%%%%%%%%%%%%%%%

\begin{abstract}
Galaxies are not limited to simple spheroid or bulge + disc morphologies. We explore the diversity of internal galaxy structures in the Coma cluster across a wide range of luminosities ($-17$\,$>$\,$M_g$\,$>$\,$-22$) and cluster-centric radii ($0$\,$<$\,$r_{\rm{cluster}}$\,$<$\,1.3 $r_{200}$) through analysis of deep Canada-France-Hawaii Telescope $i$ band imaging. We present 2D multi-component decomposition via {\footnotesize{\tt GALFIT}}, encompassing a wide range of candidate model morphologies with up to three photometric components. Particular focus is placed on early-type galaxies with outer discs (i.e. S0s), and deviations from simple (`unbroken') exponential discs. Rigorous filtering ensures that each model component provides a statistically significant improvement to the goodness-of-fit. The majority of Coma cluster members in our sample (478 of 631) are reliably fitted by symmetric structural models. Of these, 134 ($28\%$) are single \sersic objects, 143 ($30\%$) are well-described by 2 component structures, while 201 ($42\%$) require more complex models. Multi-component \sersic galaxies resemble compact psuedobulges ($n\sim$\,2, $R_e \sim$\, 4 kpc) surrounded by extended Gaussian-like outer structures ($R_e > 10$ kpc). 11\% of galaxies ($N=52$) feature a break in their outer profiles, indicating `truncated' or `anti-truncated' discs. Beyond the break radius, truncated galaxies are structurally consistent with exponential discs, disfavouring physical truncation as their formation mechanism. Bulge luminosity in anti-truncated galaxies correlates strongly with galaxy luminosity, indicating a bulge-enhancing origin for these systems. Both types of broken disc are found overwhelmingly ($>70\%$) in `barred' galaxies, despite a low measured bar fraction for Coma ($20\pm2\%$). Thus, galaxy bars play an important role in formation of broken disc structures. No strong variation in galaxy structure is detected with projected cluster-centric radius.

\end{abstract}


\section{Introduction}\label{intro}
Lenticular (S0) galaxies occupy the crux of the Hubble sequence, representing the morphological intermediate between disc-dominated spiral galaxies and spheroidal ellipticals. However, it remains unclear whether S0s are {\it evolutionary} intermediates between star-forming late-type galaxies and mainly passive early-type galaxies (ETGs). This evolutionary link has been extensively investigated with emphasis on the transformation of spirals into S0s via quenching their star formation (see \citealp{Barr2007,Aragon2008,Barway2009}).

Classically, S0s comprise a spheroid-shaped bulge component and a smooth disc with little or no interstellar dust or star formation. These bulge and disc structures are well described by \sersic ($r^{\frac{1}{n}}$; \citealp{Sersic}) and exponential profiles respectively. Conversely, giant elliptical galaxies are traditionally viewed as smooth, single spheroid systems well-described by a de Vauccouleur's profile (\sersic $n=4$; \citealp{dVauc}). The morphological distinction between these two classes can be unreliable depending on disc strength, galaxy inclination, or observation depth \citep{Kent1985,Rix1990,Jorgensen1994,vdBergh2009b}.

\cite{vdBergh1976} introduced the idea that the S0 morphology encompasses multiple distinct classes of galaxy (S0a-c; analogous to the spiral Sa-c types), differing in luminosity and evolutionary pathway (see also \citealp{vdBergh1990,vdBergh2009a}). This concept was supported by kinematic studies of S0s (e.g. \citealp{Dressler1983}), which demonstrated equivalence of the rotational properties of disks in S0s and spiral galaxies. More recently, this idea has been developed further by the ATLAS$^{\rm 3D}$ group (e.g. \citealp{Emsellem2011,Cappellari2011}, see also \citealp{Kormendy2012}). In this paradigm, most ETGs form a continuous sequence of rotating, quiescent galaxies with specific angular momentum increasing with Hubble $T$ stage. Hence, with sufficient signal-to-noise (S/N), discs should be detectable in many galaxies classically typed as elliptical. 

With increasing local environment density, the morphological fraction of galaxies becomes increasingly dominated by ETGs (particularly S0s). Conversely, spiral galaxies are rare in the dense cluster environment. This morphology-density relation \citep{Dressler1980} implies that the cluster environment plays an important role in the evolution of S0s from spirals (or spiral-like progenitors). The mechanisms potentially responsible for this evolution (see review in \citealp{Boselli2006}) can be broadly categorised as disc-fading (e.g. gas-stripping) or bulge-enhancing (e.g. tidal interactions/mergers). While the latter category is traditionally thought of as disc-disruptive, it has been demonstrated that S0 morphologies can survive merger-based quenching \citep{EMoral2013,Querejeta2014}.

The well-studied Coma cluster (Abell 1656) possesses one of the richest ETG populations in the local universe. As such, Coma is an excellent laboratory for studying the morphologies (e.g. \citealp{Wolf1902,Shapley1934,Dressler1980,vDokkum2014b}) and characteristics of ETGs (e.g. \citealp{Lucey1991,BowLucEll,Jorgensen1999,Hudson2010,Lansbury2014,Weinzirl2013}). In addition, Coma encompasses a wide range of environment conditions ($\sim$\,$100\times$ difference in galaxy density between the core and the virial radius), allowing in-depth investigation of radial trends of environment-mediated processes \citep{Gavazzi1989,Guzman1992,Carter2008,Gavazzi2010,Smith2012,Cappellari2013,Rawle2013}.

In \citeauthor{Head2014} (\citeyear{Head2014}; hereafter `Paper I'), we presented bulge - disc decompositions of $\sim$\,600 Coma cluster galaxies, demonstrating that $\sim$\,$\frac{1}{3}$ of Coma ETGs are well-described by an {\it archetypal} S0 (central \sersic bulge + outer exponential disc) model morphology. Focusing exclusively on these archetypal galaxies, we found that bulges of S0 galaxies resemble pseudobulges ($n\sim$\,2, $R_e \sim$\,1 kpc), while their discs were measured to be intrinsically smaller, or brighter than equivalent structures in star-forming spirals. A bulge $-$ disc colour separation of $\sim$\,0.1 mag was measured in $g-i$ ($\sim$\,$0.2$ mag in $u-g$), indicating either a $\sim$\,2-$3\times$ age difference, or a $\sim$\,$2\times$ metallicity difference between these components. Nevertheless, both components were found to contribute to the galaxy red sequence (colour-magnitude) trend.

Evolutionary pathways will not necessarily preserve the archetypal S0 morphology. Furthermore, the simple exponential model (Type I; \citealp{Freeman1970}) adopted in most decomposition studies does not fully represent the observed range of S0 outer disc structures. `Broken' disks have been observed for S0 and spiral galaxies \citep{Freeman1970,Erwin2008}, wherein surface brightness profiles beyond a break radius deviate either downwards (i.e. fainter; `Type II') and upwards (i.e. brighter; `Type III') relative to a simple exponential (`Type I') profile. Such profiles result from the redistribution of stars due to evolutionary processes. For example, truncated discs may be formed when stars are physically removed from a galaxy's outer regions (e.g. during tidal interaction), while anti-truncated discs may result from merger events \citep{Younger2007,Borlaff2014}. Thus, investigation of galaxies with a wider range of structural morphologies provides a more complete picture of the ETG formation mechanisms.

Previous investigations of multi-component ETG structures (e.g. \citealp{Michard1985,Capaccioli1991,Laurikainen2005,Janz2012,Huang2013a,Janz2014,Weinzirl2013}) and disc breaks (e.g. \citealp{Erwin2008,Erwin2012,Roediger2012,Laine2014}) are typically limited by (relatively) small galaxy samples from narrow fields of view or 1D profile analyses. As noted in \cite{Dullo2014}, care must be taken to report ``real" structural components, rather than overfitting galaxies with unnecessarily complex models. 

Here, we build upon these pioneering studies by characterising the multi-component internal structures of galaxies within a wide radial area ($0<r_{\rm cluster}<1.3$\,$r_{200}$) of the Coma cluster (and an absolute magnitude range $-17>M_g>-22$) using deep Canada-France Hawaii Telescope (CFHT) $i$ band imaging data. 

The decomposition analysis reported in Paper I is extended by using a wider suite of candidate models (including 2- and 3-component broken disc galaxies) in order to explore the diversity of galaxy structure in the Coma cluster. Thus, we reinvestigate the structural morphologies of all Coma cluster galaxies investigated in Paper I, including the $\sim$\,$\frac{2}{3}$ previously removed from analysis as not well-described by an archetypal bulge + disc model. While a primary goal of this analysis is the investigation of Type I, II, and III discs galaxy structures, the extended range of (multi-component) models is necessary to avoid mis-classification of additional component structures (e.g. bars or rings) as surface brightness profile breaks. Bayesian model selection and sample filtering are applied to avoid overfitting, and to ensure that best fit models are reliable representations of the underlying galaxy structures.

We investigate four main questions regarding galaxy evolution: Does the multi-component structure of giant ellipticals suggest the `puffing-up' of a compact progenitor, or the accumulation of additional structures around a compact spheroid? Are broken discs structures (truncated or anti-truncated) correlated with the properties of the bulge/bar components? Do the structures of Freeman Type II galaxies indicate physical truncation of discs? Does such a truncation scenario explain the apparent size offset of S0 discs relative to star-forming spirals reported in Paper I?

The structure of this paper is as follows: first, in Section 2 we summarise the MegaCam imaging data and galaxy sample selection criteria used in this work. Secondly, Section 3 describes the multi-component decomposition methodology, highlighting differences from the bulge-disc decomposition pipeline previously-described in Paper I. Thirdly, in Section 4 we present the resulting galaxy morphology (model) fractions, including a census of disc types. Furthermore, we explore the properties of galaxies comprising multiple distinct \sersic structures, and galaxies containing disk breaks. Finally, a discussion of possible formation pathways for broken discs is presented in Section 5.

Throughout this paper, we make use of the following notation conventions: Fitted model structures (see Section \ref{anal}) are indicated in italics (e.g. `\emph{S}' for pure \Sersic) to distinguish them from morphological classifications (e.g. `S0'). Disk break types (i.e. Freeman types; untruncated, truncated, anti-truncated) are denoted with Roman numerals (e.g. `Type II'), and galaxies containing such structures are referred to as Type I, Type II, or Type III galaxies. Conversely, galaxy types using Arabic numerals (e.g. `Type 2') refer to \cite{Allen2006} surface brightness profile types (see also Section \ref{anal}). The Type 1 profile is a special case describing a central bulge and an outer (exponential or broken exponential) disk, and is referred to as an `archetypal S0' profile (`archetypal', or `S0' as shorthand). All other \cite{Allen2006} types are referred to as `atypical S0' profiles (or simply `atypical'). 

We use the WMAP7 cosmology: $H_0=70.4\rm{km}$\,$\rm{s}^{-1}\rm{Mpc}^{-1}$ (i.e. $h_{70} = 1.01$), $\Omega_m=0.272$ and $\Omega_{\Lambda}=0.728$ \citep{WMAP7}. Using $z_{\rm{CMB}}\rm{(Coma)}=0.024$, the luminosity distance for the Coma cluster is 104.1 Mpc, and the distance modulus, $m - M = 35.09$. At this distance, $1\minute$ corresponds to 28.9 kpc. Taking a value for velocity dispersion of $\sigma_{{\rm Coma}} = 1008$ km s$^{-1}$ \citep{Struble1999} and virial mass, $M_{200} = 5.1\times10^{14} h^{-1}_{70}$\Msol \citep{Gavazzi2009}, the virial radius, $r_{200}$, for Coma is 2.2 Mpc ($\sim75\minute$).

\section{Data and Initial Sample}\label{samp}
This study makes use of the data as previously described in Paper I. To recap: optical imaging covering a total of 9 deg$^2$ of the Coma cluster in the $i$ band was acquired using the MegaCam instrument on the 3.6 m CFHT during March - June 2008 (Run ID 2008AC24, PI: M. Hudson). Total (coadded) exposure times of 300 s were obtained for each observed field, yielding $\sim12\times$ deeper imaging data (from $D^2t_{\rm exp}$) compared to SDSS (2.5 m telescope, 53 s exposures). The MegaCam frames were sky-subtracted during pre-processing using a 64 pixel mesh. A point spread function (psf) full-width half-maximum (fwhm) of between $0.65\second$ and $0.84\second$ was typical. The pixel scale was $\sim0.186$ arcseconds/pixel.

The initial sample for analysis was selected from SDSS (DR9) catalogue galaxies in the 3 deg $\times$ 3 deg ($\equiv5.2$ Mpc $\times 5.2$ Mpc, 2 $r_{200} \times$ 2 $r_{200}$) area covered by the MegaCam observations. A limit of $-17.1>M_g$\footnote{Note that while the sample is defined based on $g$ band photometry, we analyse the $i$ band data in this paper.} was applied to ensure sufficient signal-to-noise (S/N) for reliable measurement of galaxy bulge and disc structures. These targets were limited to the redshift range $0.015 < z < 0.032$ (heliocentric $v_{\rm{Coma}} \pm2.5\sigma_{\rm{1D}}$) to ensure that only cluster members were included. Unlike Paper I, no colour cut is made during sample selection. Thus, an additional $\sim$\,60 blue galaxies ($g-r \leq 0.5$) are included in the present work, yielding an initial sample of 631 Coma cluster members.

\begin{figure*}
\centering
\includegraphics[width=155mm]{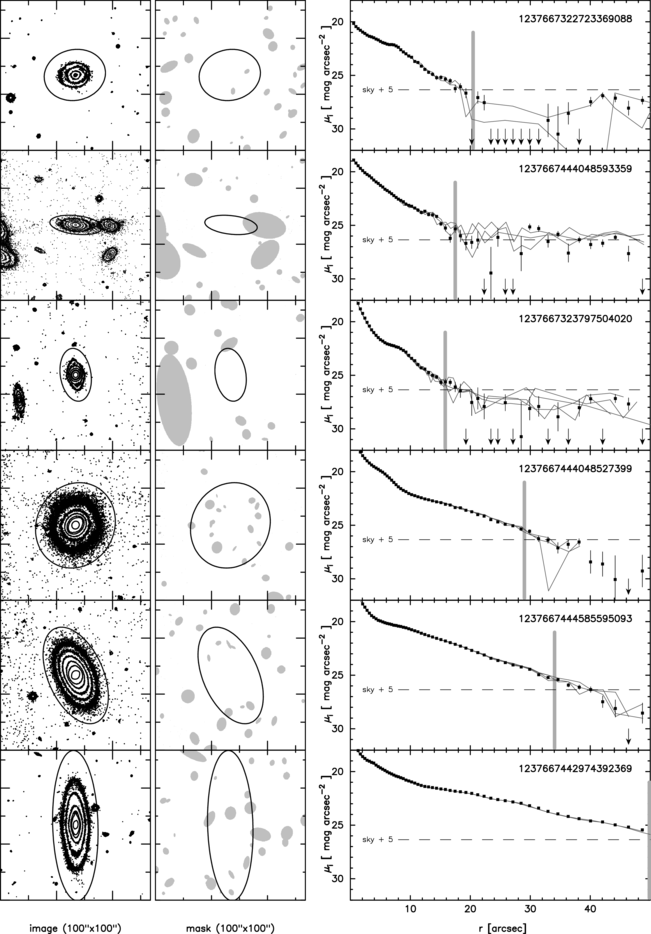}
\caption{
Diversity of surface brightness profiles observed in the Coma cluster galaxies. Six representative galaxies are shown. The first panel of each row is the galaxy thumbnail image contour plot (100$''$\,$\times$\,100$''$). The second panel shows the areas masked in the fitting procedure. The ellipses in these two panels denote the extent of the galaxy area fitted. The third panel shows the major axis surface brightness profiles (30$^{\circ}$ width wedges) and errors as derived from the stack MegaCam images; labelled by the SDSS {\tt ObjID} number. The measured profiles derived from the individual MegaCam exposures are shown as grey lines.  The one per cent level of the sky brightness  is shown as the horizontal dashed line. The vertical grey bar shows the radius limit used in the fitting.}
\label{fig:six_profiles}
\end{figure*}

To illustrate the initial image analysis undertaken, and the diversity of photometric structures observed, we show in Figure \ref{fig:six_profiles} the major axis surface brightness profiles for six representative Coma galaxies. As well as displaying the profile derived from the stacked image data {\it (black points)},
we show the results from the individual MegaCam exposures {\it (grey lines)}. The radial limit used in our profile analysis is shown {\it (grey vertical bars)};
this corresponds to $1-3\%$ of the sky brightness level ($i$ band $\simeq$\,25\,mag\,arcsec$^{-2}$). Within the fitted area, there is very good agreement between the stack and individual image profiles. This demonstrates that the stacking process, including the choice of sky-grid mesh size, has little or no effect on the derived profiles and hence on the 2D surface fitting analysis reported below.

Here we briefly note some of the key features apparent in the surface brightness profiles for these six galaxies. Galaxy 1237667322723369088 has a distinct break in profile at $\sim8\second$ with a second, downward-bending exponential-like outer shape. 1237667444048593359 has a slight upward-bending outer structure, although this may be affected by the nearby contaminating galaxy. 1237667323797504020 displays a relatively weak intermediate exponential-like structure, and an outer downward-bending shape. 1237667444048527399 has a very distinctive upward-bending, exponential-like outer structure. 1237667444585595093 has a weak, downward-bending exponential-like outer structure. 1237667442974392369 has a weak, upward-bending outer structure.

In Section 4 the model surface brightness profiles for these six galaxies derived from the multi-component fits are shown over-plotted on the data points. Work in progress will provide a detailed comparison of the surface brightness profiles of Coma cluster galaxies derived from a wide variety of independent imaging sources (MegaCam, HST ACS, SDSS, Pan-STARRS, etc.).

\section{Analysis}\label{anal}
\subsection{Decomposition}
\begin{table}
\begin{center}
\begin{tabular}{ccl}
\hline
\hline
Component&Parameter&Description\\
\hline
{\it D}&$m_i$&Total $i$ band magnitude\\
&$R_s$&Exponential scale length\\
&$q$&Axis ratio ($b/a$)\\
&$\Phi$&Position angle\\
\hline
{\it B} or {\it S}&$m$&Total $i$ band magnitude\\
&$R_e$&Effective half-light radius\\
&$n$&\sersic index\\
&$q$&Axis ratio ($b/a$)\\
&$\Phi$&Position angle\\
\hline
{\it C}&$m$&Total $i$ band magnitude\\
&$R_e$&Effective half-light radius\\
&$n$&\sersic index\\
&$q$&Axis ratio ($b/a$)\\
&$\Phi$&Position angle\\
&$C_0$&Boxiness/Diskiness parameter\\
\hline
{\it Dd}&$\mu_{\rm brk}$&Surface brightness at $r_{\rm brk}$\\
&$R_{s, {\rm in}}$&Inner exponential scale length\\
&$R_{s, {\rm out}}$&Outer exponential scale length\\
&$r_{\rm brk}$&Break radius\\
&$q$&Axis ratio ($b/a$)\\
&$\Phi$&Position angle\\
\hline
\hline
\end{tabular}
\end{center}
\caption{Table of the model components used during decomposition analysis, including descriptions of their free parameters. Note that the `bulge' label ({\it B}) is used to describe the central \sersic component. All components in a model share centroid position parameters ($x$, $y$). $C_0$ is defined in \protect\cite{GALFIT}.}
\label{multi_modC}
\end{table}

\begin{table}
\begin{center}
\begin{tabular}{lccc}
\hline
\hline
Model&Label&$k$&$n_{\rm comp}$\\
\hline
\sersic&{\it S}&7&1\\
\sersic + exponential&{\it BD}&11&2\\
Boxy \sersic + exponential&{\it CD}&12&2\\
Double \sersic&{\it BS}&12&2\\
\sersic + broken exponential&{\it BDd}&13&2\\
Boxy \sersic + broken exponential&{\it CDd}&14&2\\
\sersic + double exponential&{\it BDD}&15&3\\
Double \sersic + exponential&{\it BSD}&16&3\\
Triple \sersic&{\it BSS}&17&3\\
Double \sersic + broken exponential&{\it BSDd}&18&3\\
\hline
\hline
\end{tabular}
\end{center}
\caption[The multi-component models used during decomposition analysis, including the number of independent structural components, and number of free parameters.]{Table of the multi-component models used during decomposition analysis, including the number of independent structural components, $n_{\rm comp}$, and number of free parameters, $k$.}
\label{multi_modT}
\end{table}

Galaxy decomposition was carried out using {\footnotesize{\tt GALFIT}} (version 3.0.4; \citealp{GALFIT}) with an automated {\footnotesize{\tt python}} wrap-around derived from {\footnotesize{\tt AGONII}} (Automated Galfitting of Optical and Near Infra-red Imaging; Paper I). Details of this fitting procedure (including description of the extraction and calibration of input MegaCam data products) can be found in Appendix \ref{GALF} and Paper I.

In the present work, we fit galaxies with a range of analytical models in order to thoroughly explore the diversity of internal galaxy structures. These candidate models are comprised of 1 to 3 structural components, each described by one of four functional forms (see Table \ref{multi_modC}): exponential `discs' (`{\it D}'), general \sersics (`{\it S}'), boxy \sersics (`{\it C}'), and broken discs (`{\it Dd}'). Note that the central {\it S} component in any model is referred to as the bulge, and labeled as `{\it B}'. Conversely, non-central {\it S} components in models containing a disc are referred to here as `bars'. However, this convention does not explicitly require a stellar bar structure. As such, a `bar' may also correspond to a lens or oval structure. The broken disc component (see Appendix \ref{brk_disk}) comprises inner and outer exponential discs (with differing scale lengths, $R_{s{\rm , in}}$ and $R_{s{\rm , out}}$) connected by a smooth transition at a break radius ($r_{\rm brk}$).

The ten candidate multi-component models considered in this work are catalogued in Table \ref{multi_modT}. \Sersic-only (hereafter `{\it S}') and bulge + disc (hereafter `{\it BD}') models are unchanged from those presented in Paper I. In addition, we present the decomposition results when boxy bulge + disc (hereafter `{\it CD}'), double \sersic (hereafter `{\it BS}'), bulge + double disc (hereafter `{\it BDD}'), bulge + bar + disc (hereafter `{\it BSD}'), bulge + double \sersic (hereafter `{\it BSS}') models are also considered. Three further models variants implement the `broken disc' profile: bulge + broken disc (hereafter `{\it BDd}'), boxy bulge + broken disc (hereafter `{\it CDd}') and bulge + bar + broken disc (hereafter `{\it BSDd}'). In order to avoid fitting bias due to the choice of initial parameter values, model inputs are based on the best fit parameters of simpler model types (e.g. {\it BSD} input derived from best {\it BD} fit). This iterative build up of model complexity significantly improves reliability of the measured galaxy properties, particularly for highly-degenerate multi-component models.

We use \cite{Allen2006} types to describe the relationship between the bulge (i.e. innermost \Sersic) and (exponential or broken) disc profiles (see Appendix \ref{Allen+}). This convention is also used for 3-component model systems, as fewer constraints are place on bar/disc or bulge/bar morphology. The only exception in which attention is paid to these profile interactions is where profile inversion implies incorrect interpretation of the model components (i.e. `Type 4' bulge/bar or bar/disc structures; see also Appendix \ref{app_filter}).

\subsection{Model Selection and Results Filtering}\label{filt_intro}
Sample filtering is applied to the fitting results (similar to Paper I) in order to isolate a sample of accurately-fit galaxies. A key step in this process is the selection of best-fit models which are meaningful descriptions of each galaxy's underlying morphological structure, ensuring that all structural components are statistically justified. Galaxy models are assessed on both goodness-of-fit (i.e. ensuring that a galaxy is neither under-fitted nor over-fitted), and suitability of component structures (i.e. rejecting components with unrealistic parameters, or which do not measure the intended target substructure). By removing such instances of dissonance between the galaxy and model stuctures, the reliability of multi-component analysis results is vastly improved. A detailed description of the galaxy filtering conditions, and a flow chart illustrating the overall filtering process is presented in Appendix \ref{app_filter}.

Galaxies are initially assessed for asymmetry (via the $A$ parameter; \citealp{Homeier2006}) and contamination (via image mask fraction, $f_{\rm mask}$) to ensure robust measurements of galaxy properties. Highly asymmetric galaxies, or galaxies strongly contaminated by neighbouring sources cannot be reliably fit by smooth, symmetric models, and are thus removed from consideration. Due to high parameter uncertainty, galaxies are also removed if their best-fit models are poorly-fitted (high $\chi^2_{\nu}$), highly inclined to the line of sight (from the axis ratio of the outer component), or if a model component contributes less than 5\% of the total galaxy luminosity (component fraction, C/T$<0.05$). Additional filtering conditions are placed on broken disc galaxies to ensure that both the inner and outer disc contribute significantly to the overall galaxy profile, and to avoid erroneous regions of parameter space. 

Selection between alternative candidate models is made using the Bayesian Information Criterion (BIC; \citealp{BIC}), calculated over independent resolution elements (see details in Paper I). This is defined as

\begin{equation}\label{BICeq}
{\rm{BIC}}_{\rm{res}} = \frac{\chi^2}{A_{\rm{psf}}} + k\cdotp{\rm{ln}}\left(\frac{n_{\rm{pix}}}{A_{\rm{psf}}}\right),
\end{equation}
 
where $\chi^2$ is the standard (un-reduced) fitting chi-squared, $k$ is the number of model parameters (degrees of freedom), $n_{\rm pix}$ is the number of image pixels used during fitting, and $A_{\rm psf}$ is the area of a resolution element (in pixels). Here, $A_{\rm psf}$ is calculated as the area within two standard deviations ($\sigma$) of the psf image centre, as measured by fitting a Gaussian model. For a set of candidate models, the model with the lowest BIC$_{\rm res}$ maximises goodness-of-fit without introducing unnecessary free parameters (hereafter `best-fit' model). This ensures that each of the best-fit model components provide a statistically significant improvement to $\chi^2$.

Measurement error in $A_{\rm psf}$ leads to an associated uncertainty in BIC$_{\rm res}$ ($\sigma_{\rm res}$). Therefore, a  3$\sigma_{\rm res}$ reduction in BIC$_{\rm res}$ is required before a more complex (higher $k$) model is accepted as a statistical improvement over a simpler model. This 3$\sigma$ selection condition is based on comparison with by-eye classification, and is discussed in further detail in Paper I, and \cite{HeadThesis}. Here, $\sigma_{\rm res}$ is based on the scatter in $A_{\rm psf}$ (typically $\sim$\,$3\%$), as measured across multiple star images. For example, a galaxy fit by {\it S} and {\it BD} models yields BIC$_{\rm res}$ values of 3500 and 3450 (respectively) with an associated $\sigma_{\rm res}$ of 10. Since $\Delta {\rm BIC_{ res}} = 50 > 3 \sigma_{\rm res}$, the addition of the exponential disc component is a statistically significant improvement to the fit, and hence measures a distinct photometric structure.

\section{Results}
\subsection{Best-fit Models}\label{res_frac}
\begin{figure*}
\begin{center}
	\includegraphics[width=0.8\linewidth,clip=true]{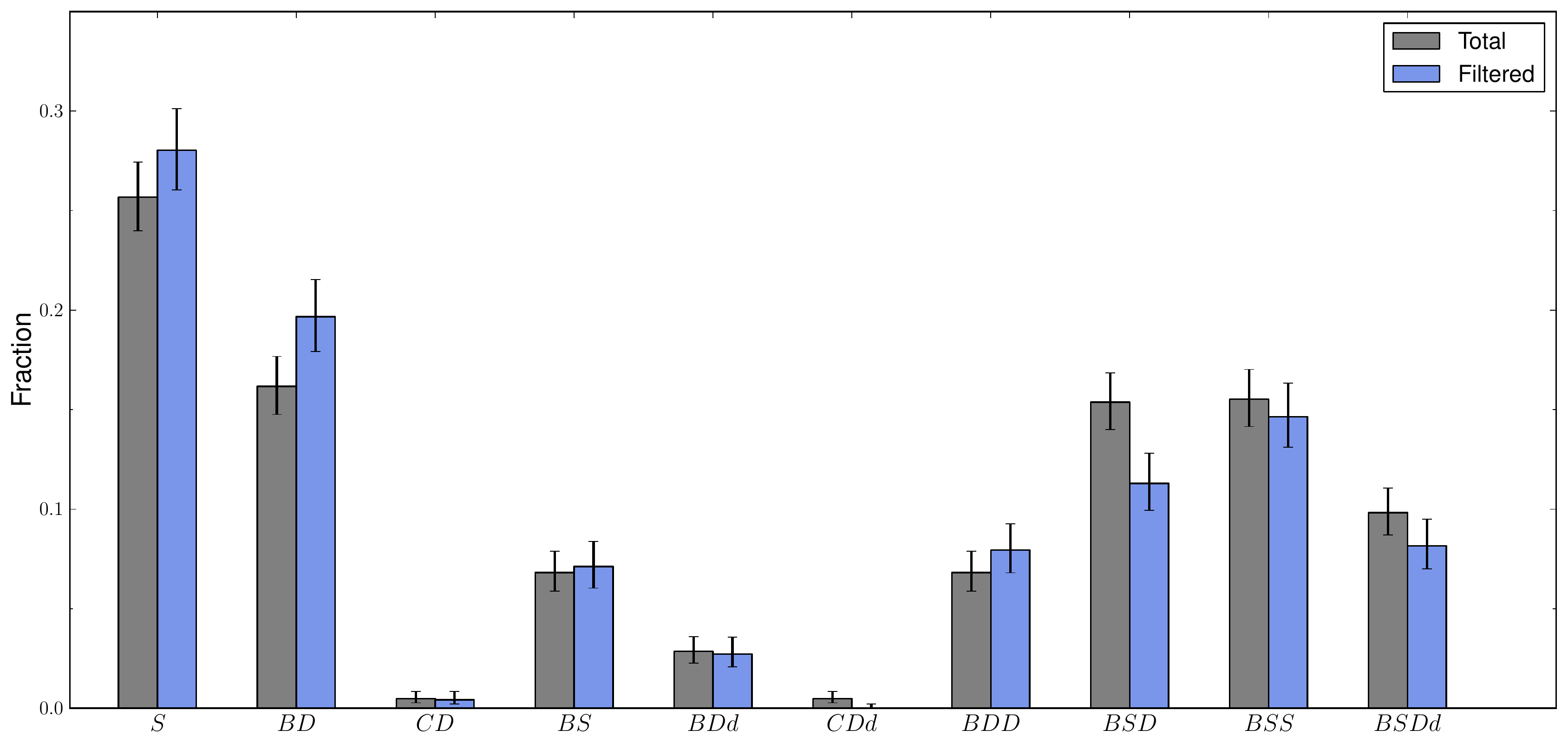}
\end{center}
\caption[Histogram of best-fit model type fractions for the initial and filtered samples.]{Histogram of best-fit model type fractions for the initial ($N=631$) and filtered ($N=478$) samples. Error bars are $68\%$ confidence limits.}% \protect\citep{Wilson1927}.}
\label{multisamp_hist}
\end{figure*}

A wide mix of best-fit model morphologies are found for the 631 galaxies (570 Coma sample + 61 blue Coma galaxies) investigated (see example plots in Appendix \ref{mod_eg}). The fractions of galaxies best described by each candidate multi-component model are illustrated in Figure \ref{multisamp_hist}. From the initial sample ($N=631$), 162 are best fitted by a \Sersic-only model ($26\%$), 102 are best fitted by {\it BD} ($16\%$), 43 are best fitted by {\it BS} ($7\%$), three are best fitted by {\it CD} ($<1\%$), 18 are best fitted by {\it BDd} ($3\%$), three are best fitted by {\it CDd} ($<1\%$), 43 are best fitted by {\it BDD} ($7\%$), 97 are best fitted by {\it BSD} ($15\%$), 98 are best fitted by {\it BSS} ($16\%$), and 62 are best fitted by {\it BSDd} ($10\%$). Thus, the majority of Coma cluster galaxies (58\%) have morphologies more complex than the simple {\it S} and {\it BD} models. 

Many of these complex structure galaxies were considered to be (archetypal S0) bulge + disc systems in Paper I. In total, 51 `archetypal' galaxies (from $N=200$; 25.5\%) remain best-fit by a {\it BD} model, while 129 (64.5\%) require more complex models and 20 (10.0\%) are demoted to a single-\sersic model (due to the more stringent model selection tests in the present work). The fractions of prior `atypical bulge + discs' adequately fit by a {\it BD} model (21.1\% of 137), requiring more complex models (73.0\%), and demoted to {\it S} (5.8\%) are similar. Note that many galaxies classed as `unstable' previously are best-fit here by complex 3-component models (68\% of 128). This is due to significant reductions in $\chi^2_{\nu}$ as additional structural components are accounted for. 

After sample filtering (see Section \ref{filt_intro}), 478 galaxies from the sample of 631 ($76\%$) remain. The 153 galaxies removed by filtering comprise: 80 galaxies removed due to asymmetry or contamination, 23 galaxies with high $\chi^2$, 13 highly-inclined galaxies, three galaxies with anomalous outer discs due to $R_{s,{\rm out}} > 0.1$\,$r_{\rm brk}$, two galaxies with anomalous inner discs due to $r_{\rm brk} < 5\second$, and 32 galaxies removed due to inverted \Sersic/disc components (i.e. disc-dominated at low radii, \Sersic-dominated at large radii). From the remaining filtered sample, 134 galaxies are best fitted by a \Sersic-only model ($28.0\%$), 94 are best fitted by {\it BD} ($20\%$), 34 are best fitted by {\it BS} ($7\%$), two are best fitted by {\it CD} ($<1\%$), 13 are best fitted by {\it BDd} ($3\%$), none are best fitted by {\it CDd}, 38 are best fitted by {\it BDD} ($8\%$), 54 are best fitted by {\it BSD} ($11\%$), 70 are best fitted by {\it BSS} ($15\%$), and 39 are best fitted by {\it BSDd} ($8\%$). 

Note that in total, 93 galaxies ($20\pm 2\%$) are well-described by `barred' models ({\it BSD}, {\it BSDd}). This barred fraction for Coma is significantly lower than the value reported in \cite{Lansbury2014} from either decomposition ($72^{+5}_{-6}\%$) or ellipse ($48\pm 6\%$) analyses. This difference cannot be reconciled, even if {\it BSS} models are included in the `barred' sample (yielding $34^{+3}_{-2}\%$ bar fraction). However, if the present sample is restricted to only contain galaxies with D80 morphological classifications (as in \citealp{Lansbury2014}), then the barred fraction (including {\it BSS} galaxies) rises to $63\pm 4\%$. This fraction rises further if only D80 S0s (including S0/a, E/S0) galaxies are considered, yielding bars in $71\pm 5\%$ of galaxies. As the D80 catalogue only covers the bright end of the Coma sample ($M_g \lesssim -18$), the bar fraction increase for D80 galaxies indicates a significantly decreasing bar detection rate for faint galaxies. However, the lower bar detection rate relative to \cite{Lansbury2014}, particularly if {\it BSS} galaxies are not considered `barred', reflects the more stringent conditions for accepting a more complex model in the present work.

Structural biases in decomposition studies with overly-simplistic galaxy models can be quantified by artificially limiting the range of candidate models considered during model selection. If model selection were repeated {\it without} considering double/triple \sersic models, 94 out of 631 galaxies ($20\%$) would be identified as best fitted by models including a broken disc ({\it BDd}, {\it CDd}, {\it BSDd}). Of these, 42 galaxies are better dsecribed by a double or triple \sersic model. Thus, a 2D decomposition analysis falsely reports broken disc models $45\%$ of the time if only models with exponential discs are considered. Alternatively, if 3-component models are excluded from consideration for 2D analysis, a 2-component model is preferred in 342 galaxies ($72\%$ of total). However, 199 of these galaxies would be better fit by a 3-component model\footnote{Disparity between this value and the 201 3-component models reported above is due to two galaxies which would be better fit by a single \sersic model, if 3-component models are excluded.}. Thus, 2D analysis selects an overly-simplistic 2-component model $58\%$ of the time if 3-component models are not considered.

Of all 141 galaxies with 2-component structures, 82 galaxies ($58\%$) exhibit Type 1 (i.e. `archetypal' inner + outer component) profiles, while 44 Type 3 profile galaxies (recurrent bulge; $31\%$) make up the second most common structural type. For 3-component systems ($N=201$), a larger proportion of galaxies (130; $65\%$) are characterised by Type 1 bulge/disc structures, while only 55 galaxies ($27\%$) had Type 3 bulge/discs. Thus, 2- and 3-component galaxies have archetypal bulge + disc structures in the majority of cases, with recurrent bulges (dominant over their discs at large radii) being the second most common structure.

\begin{table*}
\begin{center}
\begin{tabular}{clccc}
\hline
\hline
Model&Parameter&Comp. 1&Comp. 2& Comp. 3\\
\hline
\hline
{\it S}&$n$&$1.90\pm0.05$&-&-\\
&$R_e$ [kpc]&$1.99\pm0.07$&-&-\\
$N=134$&$q$& $0.63\pm0.02$&-&-\\
$m_i=16.58\pm0.03$&C/T&1.0&-&-\\
\hline
{\it BS}&$n$&$2.12\pm0.30$&$\phantom{1}0.66\pm0.10$&-\\
&$R_e$ [kpc]&$5.27\pm0.90$&$14.57\pm0.90$&-\\
$N=34$&$q$&$0.66\pm0.04$&$0.79\pm0.04$&-\\
$m_i=15.76\pm0.12$&C/T&$0.35\pm0.04$&$\phantom{1}0.65\pm0.04$&-\\
\hline
{\it BSS}
&$n$&$2.17\pm0.23$&$\phantom{1}0.43\pm0.13$&$\phantom{1}0.55\pm0.04$\\
&$R_e$ [kpc]&$4.20\pm0.55$&$12.68\pm0.76$&$25.80\pm1.75$\\
$N=70$&$q$&$0.71\pm0.03$&$0.48\pm0.04$&$0.58\pm0.04$ \\
$m_i=14.45\pm0.08$&C/T&$0.42\pm0.02$&$\phantom{1}0.26\pm0.02$&$\phantom{1}0.27\pm0.02$\\
\hline
\hline
\end{tabular}
\end{center}
\caption[The median structural parameter values for multi-\sersic model galaxies.]{Table of the median structural parameter values for multi-\sersic model galaxies ({\it S}, {\it BS}, {\it BSS}), indicating the half-light radii, \sersic indices, component axis ratios ($q$) and component fraction (C/T) of each model component. In addition, the median total apparent magnitude ($m_i$), and number of galaxies ($N$) are given for each model type. }
\label{msers_tab}
\end{table*}

If a {\it BD} model is forced on the 478 galaxies in the filtered sample, 214 ($62\%$) yield a Type 1 (archetypal) profile, and 72 galaxies ($21\%$) correspond to a Type 3 (recurrent bulge) profile. These profile fractions do not change significantly if galaxies best-fit by a 2 or 3-component model are considered separately (Type 1/3: $66\%/20\%$ for 2-component galaxies, $59\%/22\%$ for 3-component galaxies). Thus, Type 3 {\it BD} profiles do not intrinsically represent underfit galaxy structures, but rather a structural morphology distinct from archetypal bulge + disc systems.

In summary, thorough 2D decomposition analysis reveals a rich range of galaxy structures in the Coma cluster, with 3(+) structural components required in $42\%$ of galaxies. 2+ component structure systems were well-represented by archetypal (central) bulge + (outer) disc morphologies in the majority of cases ($N=202$), including 52 galaxies which exhibited broken disc profiles. This broken disc fraction would be overestimated, however, if multi-component models (including double/triple \sersic systems) were not considered during model selection.

\subsection{Multi-\sersic structures}
\begin{figure*}
\begin{center}
	\includegraphics[width=0.8\linewidth,clip=true]{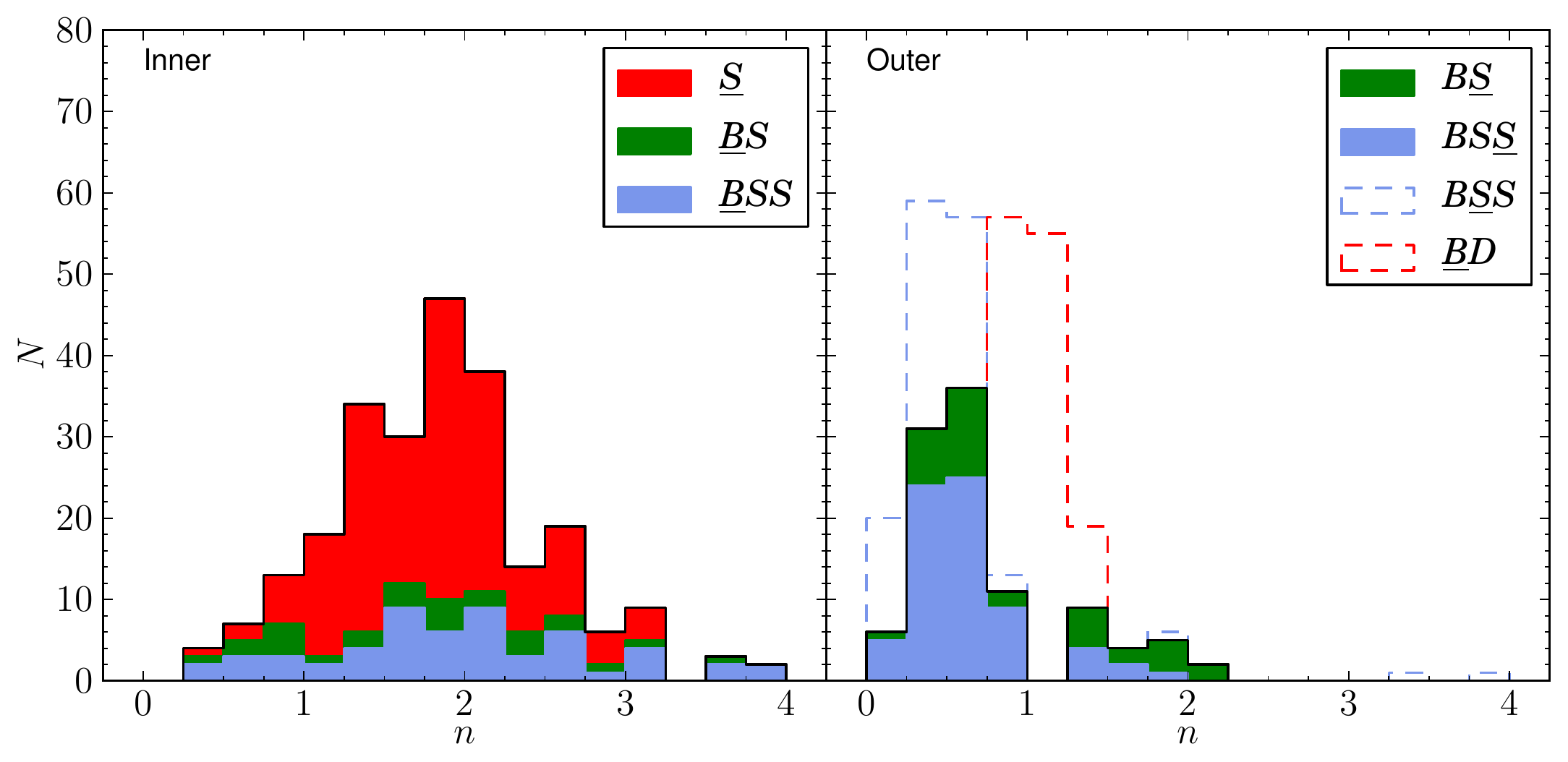}
\end{center}
\caption[Histograms of inner/outer \sersic structure $n$ in single, double, and triple \sersic galaxies.]{Histograms of \sersic index, $n$, for multi-component \sersic galaxies. {\bf Left:} the $n$ distribution for inner-dominant \sersic structures ({\textit {\textbf S}}, {\textit {\textbf B}S}, {\textit {\textbf B}SS}), divided by best-fit model type. {\bf Right:} the $n$ distribution of outer-dominant \sersic structures ({\textit B{\textbf S}}, {\textit BS{\textbf S}}) divided by best-fit model type. The equivalent distributions including middle \sersic structures ({\it B{\rm S}S}) or outer \sersic $n$ for {\it BS} fits to galaxies best-fit by a {\it BD} model are included as dashed blue and red histograms respectively.}
\label{sern_hist}
\end{figure*}

From the initial sample of 478 galaxies (filtered to exclude badly-fit galaxies; see Section \ref{filt_intro}), $\sim$\,50\% were best described by a model comprising one or more \sersic components (28\% {\it S}; 7\% {\it BS}; 15\% {\it BSS})\footnote{Note that outer \sersic $n\neq1$ for these models. Thus, we exclude the {\it BD} and {\it BSD} models as special cases of {\it BS} and {\it BSS}.}. In this section, we briefly discuss the structural results for these multi-component \sersic galaxies (Table \ref{msers_tab}). Note that average galaxy luminosity increases with number of model components, highlighting the strong S/N and spatial size dependence of multi-component structure detection. 

Figure \ref{sern_hist} illustrates the distributions of $n$ for inner (left panel) and outer (right panel) \sersic components. In all \sersic model variants, the central structure is compact and has a `pseudobulge-like' ($n\sim$\,$2$) profile. For {\it S} model galaxies, the \sersic structure is equivalent to a `naked' bulge for {\it BD} galaxies, albeit a factor of $2\times$ larger ($R_e\sim$\,$2$ kpc). The central `bulges' of both {\it BS} and {\it BSS} galaxies are consistent in size, but larger on average than a single \sersic ($R_e\sim$\,$4$ kpc). Note that very few inner structures refer to a classic $n=4$ (de Vaucouleur's) profile. If {\it BSS} (or other 3-component) galaxies were force fit by a single \sersic structure, however, the resulting $n$ distribution would extend to $n\sim$\,$8$, peaking strongly for $n = 3$-$4$. Hence, de Vaucouleur's profile may arise from underfitting more structurally complex systems.

Outer \sersic structures have Gaussian-like profiles ($n\sim$\,$0.5$) on average, although a weak tail exists in the $n$ distribution towards higher values (Figure \ref{sern_hist}, right panel). Since an outer component with $n=1$ would be described by a {\it BD} model, the $1.00<n<1.25$ bin is empty for outer structures. If the disc $n$ is allowed to vary for these bulge + disc galaxies (i.e. fitting a {\it BS} model), then a continuous distribution of outer structure \sersic index becomes apparent (red dashed histogram in Figure \ref{sern_hist}). The resulting `disc' $n$ distribution covers the range $0.5<n<1.5$, but peaks strongly at $n=1$ (median value: 1.00, standard deviation: 0.24)\footnote{Recall however, that these changes in outer profile $n$ do not yield statistically significant improvements to the goodness-of-fit relative to fixing $n=1$.}. Hence, a subset of the outer structures considered in this section may represent ($n\neq1$) discs. This is supported by the detection of rapid galaxy rotation ($V_c > 100$ km s$^{-1}$) for 11 of our 70 {\it BSS} galaxies in \citeauthor{Rawle2013} (\citeyear{Rawle2013}; $\sim$\,$\frac{1}{3}$ of their S0 sample). However as a practical choice, only $n=1$ discs will be considered in discussion of disc structure in later sections due to the uncertain nature of outer \sersic structures.

The outer and middle \sersic structures of {\it BS} and {\it BSS} galaxies are both $\gtrsim$\,$10$ kpc larger than `bulges', but represent drastically different fractions of their parent galaxy's total luminosity ($\sim$\,$\frac{2}{3}$ and $\sim$\,$\frac{1}{4}$ respectively). Conversely, the outer structure of {\it BSS} galaxies is comparable in luminosity to the middle \sersic, but is an additional 10 kpc larger. As such, {\it BSS} galaxies are structurally equivalent to {\it BS} galaxies with the addition of an outer \sersic structure. The outer \sersic structures may be the remnants of past merger events. As such, the distinction between {\it BS} and {\it BSS} may be a difference in the number of major merger events experienced in the past.

By comparison, the triple \sersic structures measured by \citeauthor{Huang2013a} (\citeyear{Huang2013a}; H13) in a small sample of nearby (visually-selected) ellipticals consist of a faint, compact central object ($R_e <1$ kpc), a middle component ($R_e\sim$\,$2.5$ kpc), and a dominant outer envelope ($R_e\sim$\,$10$ kpc). If the compact components are neglected, the H13 structures are comparable with the multi-\sersic models in the present work, albeit with smaller bulges, and more centrally-concentrated outer profiles ($n\sim$\,$1$-$2$ in H13). This discrepancy in outer component $n$ may indicate that the outer profiles in H13 encompass multiple distinct Gaussian structures (e.g. both outer \sersic components in our {\it BSS} galaxies). Alternatively, given the low local environment density of galaxies in the H13 sample, the increased detection rate of weak additional outer \sersic structures in the present work may instead reflect a more active merger history of present-day Coma cluster galaxies. This is supported by the higher average bulge size in the present work, as mergers will also increase bulge $R_e$.

\begin{figure*}
\begin{center}
	\includegraphics[width=0.85\linewidth,clip=true]{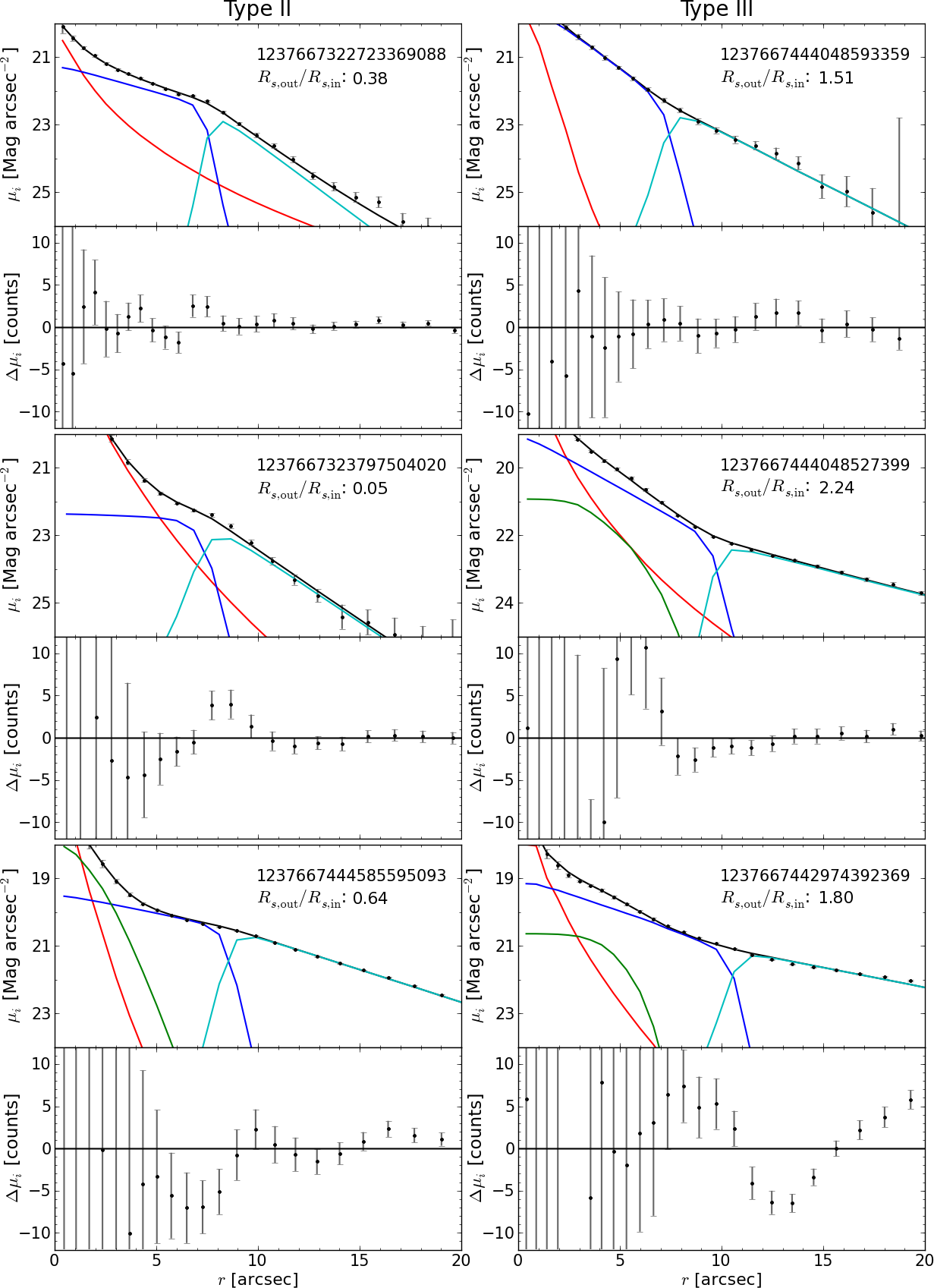}
\end{center}
\caption{Example (major axis) surface brightness profiles for six galaxies (3 {\it BDd}, 3 {\it BSDd}) best fitted by broken disc models (3 Type II, 3 Type III). {\bf Upper panels:} The $i$ band surface brightness as measured from the galaxy thumbnail (black points) in wedges of elliptical annuli. The corresponding model components are indicated as solid lines (black/red/green/blue/cyan: total/bulge/bar/inner disc/outer disc). {\bf Lower panels:} Model residual (in counts). Error bars in both plots (gray) are the standard error on the mean surface brightness in each wedge. All examples include the ratio of the outer and inner disc scale lengths ($R_{s,{\rm out}}/R_{s,{\rm in}}$).}% \protect\citep{Wilson1927}.}
\label{multi_demo}
\end{figure*}

\begin{figure*}
\begin{center}
	\includegraphics[width=0.8\linewidth,clip=true]{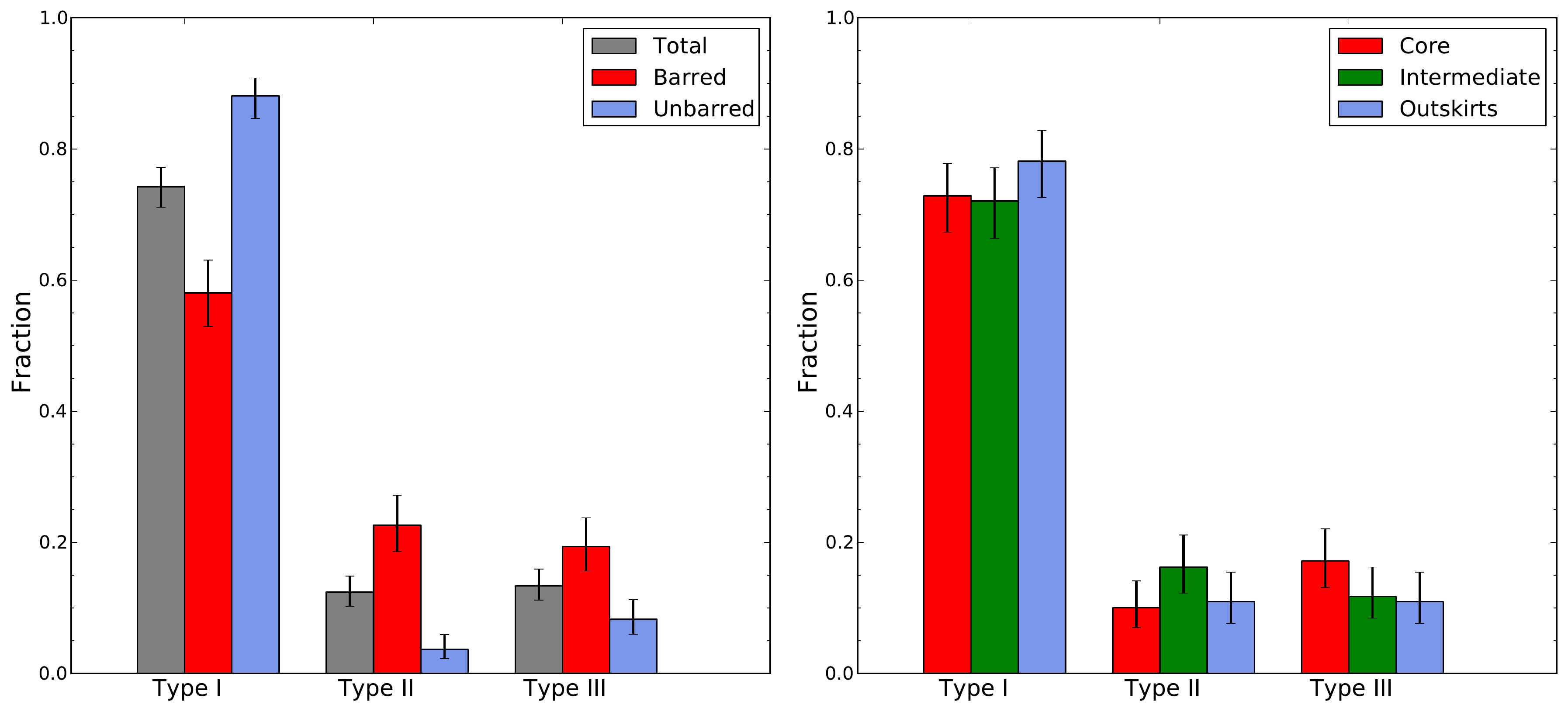}
\end{center}
\caption[Histograms of outer disc type fractions for barred/unbarred galaxies, and cluster inner/intermediate/outskirts galaxies.]{Histogram of outer disc type fractions for {\bf Left:} barred ({\it BSD}, {\it BSDd}) and unbarred ({\it BD}, {\it BDd}) bulge + disc models, {\bf Right:} galaxies in the core ($r_{\rm cluster} < 800$ kpc), intermediate ($0.8 < r_{\rm cluster} < 1.6$ Mpc), and outskirts ($r_{\rm cluster} > 1.6$ Mpc) cluster samples. Types I, II, and III refer to unbroken, truncated, and anti-truncated discs respectively. Error bars are $68\%$ confidence limits.} %\protect\citep{Wilson1927}.}
\label{diskty_hist}
\end{figure*}

If the {\it BS} and {\it BSS} models considered in this section represent the multi-component structures of traditional elliptical galaxies, then such galaxies comprise a (relatively) compact pseudobulge ($\sim$\,5 kpc) around which large ($\sim$\,10-20 kpc) outer (Gaussian) structures have been assembled. These compact central structures are $2$-$3\times$ larger than `red nugget' objects ($\sim$\,1-2 kpc) detected at high redshift ($z\sim$\,$2$; \citealp{Damjanov2009}). Thus, if the multi-component \sersic galaxies observed in Coma in the present work evolved from red nuggets, then their bulge structures must have experienced significant size growth (`puffing up'). However, the {\emph{total}} effective radii for {\it BS} and {\it BSS} galaxies (estimated from the combined luminosities of all model components, assuming alignment of component PAs) is $\sim$\,10-11 kpc on average\footnote{Even if the outermost structures in {\it BSS} galaxies were dismissed as fitting artefacts, the total $R_e$ of such systems would remain in excess of 7 kpc.}, suggesting an even more drastic growth mechanism ($\sim$\,$6\times$, consistent with \citealp{vDokkum2014}). 

In summary, galaxies comprising multiple \sersic structures (with outer $n\neq1$) resemble a compact central psuedobulge (reminiscent of single \sersic systems; $n\sim$\,2, $R_e \sim$\,4 kpc) embedded in extended Gaussian ($n\sim$\,0.5) envelopes. The combined effective half-light radii of these systems typically exceeds 10 kpc. Thus, if multi-\sersic galaxies evolved from compact `red nuggets' as detected at high redshift, then such systems must have experienced a $\gtrsim 6\times$ increase in size. 

\subsection{Freeman Disc Type Fractions}\label{disc_frac}
Galaxies with single disc-like outer profiles ({\it BD}, {\it CD}, {\it BDd}, {\it CDd}, {\it BSD}, {\it BSDd}) were categorised by their disc types (i.e. Freeman Type I, II, or III). In total, 202 valid disc galaxies are contained within the sample after filtering. Of these, 150 galaxies ($74\pm 3\%$) have Type I (untruncated) discs, 25 galaxies ($12^{+3}_{-2}\%$) have Type II (truncated) discs, and 27 galaxies ($13^{+3}_{-2}\%$) have Type III (anti-truncated) discs (Figure \ref{diskty_hist}, left panel). 
Compared to the disc type fractions reported in the Virgo cluster (Type I: $46\pm 10\%$, Type II: $0^{+4}_{-0}\%$, Type III: $54\pm 10\%$; \citealp{Erwin2012}), we detect significantly more Type I and II discs in Coma, but fewer Type III discs. By comparison, the field S0 sample in \cite{Erwin2012} yields significantly fewer Type I discs ($26^{+7}_{-6}\%$), but greater Type II ($28^{+7}_{-6}\%$) and III ($46\pm 7\%$) fractions than the Coma sample.

If considered separately, Type I discs are found more frequently in unbarred ({\it BD}, {\it CD}, {\it BDd}, {\it CDd}; $88\pm 3\%$) galaxies than those containing bars ({\it BSD}, {\it BSDd}; $58\pm 5\%$). Consequently, barred galaxies have a greater fraction of Type II and III discs ($23^{+5}_{-4}\%$ and $19\pm 4\%$) than galaxies without bars ($4^{+2}_{-1}\%$ and $8^{+3}_{-2}\%$). \citeauthor{Erwin2012} also reported a decrease in the Type I fraction for barred Virgo galaxies ($23^{+9}_{-7}\%$), however the increased barred Type II fraction in this work only widens the disparity between Coma and Virgo Type II disc detection. Note that no strong correlation (Pearson's $\rho \sim$\,0.3) is detected between bar and broken disc axis ratios ($q$) of {\it BSDd} galaxies, indicating that these model components are structurally distinct. Thus, the detection of a large number of broken discs in barred galaxies is not an artefact of of overfitting (i.e. via coupling of the inner disc to the \sersic bar profile).

To test the variation of disc type with environment, the filtered Coma sample was sub-divided into core, intermediate, and outskirts samples based on galaxy distance from the cluster centre ($r_{\rm cluster} < 0.8$ Mpc, $0.8 < r_{\rm cluster} < 1.6$ Mpc, and $r_{\rm cluster} > 1.6$ Mpc respectively; Figure \ref{diskty_hist}, right panel). These clustercentric radial ranges are selected such that each sample has approximately equal occupancy ($N=70$, 68, and 64). In all three samples, Type I discs form the vast majority, with a slightly increased Type I disc fraction for outskirt galaxies ($73^{+5}_{-6}\%$, $72^{+5}_{-6}\%$, and $78^{+5}_{-6}\%$ for core, intermediate, and outskirt galaxies). Type II and III disc fractions are consistent across all radial samples (Type II: $10^{+4}_{-3}\%$, $16^{+5}_{-4}\%$, $11^{+5}_{-3}\%$; Type III: $17^{+5}_{-4}\%$, $12^{+4}_{-3}\%$, $11^{+5}_{-3}\%$) although slight peaks in Type II and Type III disc fractions are apparent in the intermediate and core samples (respectively).

In summary, greater fractions of Freeman Type I (untruncated; $74\%$) and Type II (truncated; $12\%$) discs were detected in the present work than have been reported previously in the Virgo cluster. Conversely, the measured fraction of Type III (anti-truncated) discs in Coma ($13\%$) was lower than Virgo. The majority of galaxies with Type II or III discs also contain galaxy bars (Type II: $89\%$; Type III: $71\%$), compared to less than half of galaxies with unbroken discs (Type I: $42\%$). No significant variation in Type I/II/III fraction was detected with local environment within Coma.

\begin{table*}
\begin{center}
\begin{tabular}{clccc}
\hline
\hline
Disc Type&Parameter&{\it B}&{\it S}&{\it D}/{\it Dd}\\
\hline
\hline
Type I&$n$&$1.52\pm0.09$&$0.44\pm0.04$&$1.0$\\
&$R_e$ [kpc]&$0.62\pm0.04$&$1.94\pm0.16$&$3.68\pm0.18$\\
$N=97$&$q$&$0.65\pm0.02$&$0.54\pm0.04$&$0.67\pm0.02$\\
$m_i=15.82\pm0.10$&C/T&$0.27\pm0.02$&$0.11\pm0.02$&$0.62\pm0.02$\\
\hline
Type II&$n$&$2.32\pm0.14$&$0.44\pm0.04$&$1.0$\\
&$R_e$ [kpc]&$0.96\pm0.07$&$2.91\pm0.22$&$9.28\pm1.44\phantom{0}$/$\phantom{0}4.12\pm0.20$\\
$N=18$&$q$&$0.69\pm0.05$&$0.49\pm0.08$&$0.56\pm0.05$\\
$m_i=14.65\pm0.12$&C/T&$0.33\pm0.04$&$0.21\pm0.03$&$0.46\pm0.04$\\
\hline
Type III&$n$&$1.74\pm0.19$&$0.43\pm0.03$&$1.0$\\
&$R_e$ [kpc]&$0.54\pm0.05$&$2.82\pm0.49$&$3.58\pm0.20\phantom{0}$/$\phantom{0}5.62\pm0.34$\\
$N=24$&$q$&$0.70\pm0.04$&$0.39\pm0.07$&$0.54\pm0.05$\\
$m_i=14.49\pm0.17$&C/T&$0.24\pm0.04$&$0.14\pm0.02$&$0.62\pm0.03$\\
\hline
\hline
\end{tabular}
\end{center}
\caption{Table of the average structural parameter values for Type I/II/III archetypal disc galaxies with simple exponential discs ({\it BD}, {\it CD}, {\it BSD}) and broken exponential discs ({\it BDd}, {\it BSDd}). The average \sersic indices ($n$), half-light radii ($R_e$), component axis ratio ($q$), and component fractions (C/T) are indicated for each model component. For Type II and III galaxies, both the inner and outer $R_e$ ($=1.678R_s$) are included. In addition, the median total apparent magnitude ($m_i$), and number of galaxies ($N$) are given for each disc type.}
\label{mdisk_tab}
\end{table*}

\subsection{Freeman Type I, II, and III galaxy structures}
\begin{figure}
\begin{center}
	\includegraphics[width=\linewidth,clip=true]{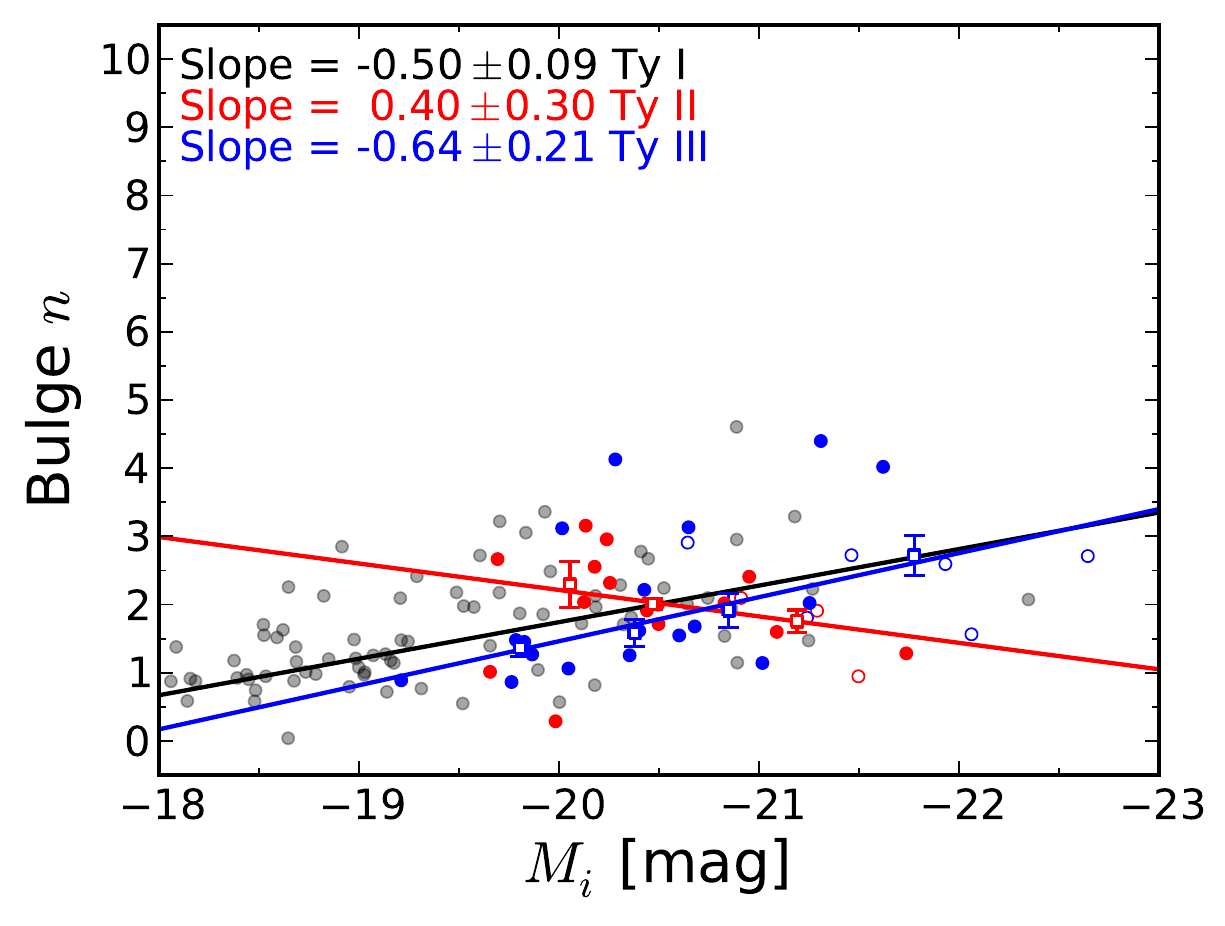}
	\includegraphics[width=\linewidth,clip=true]{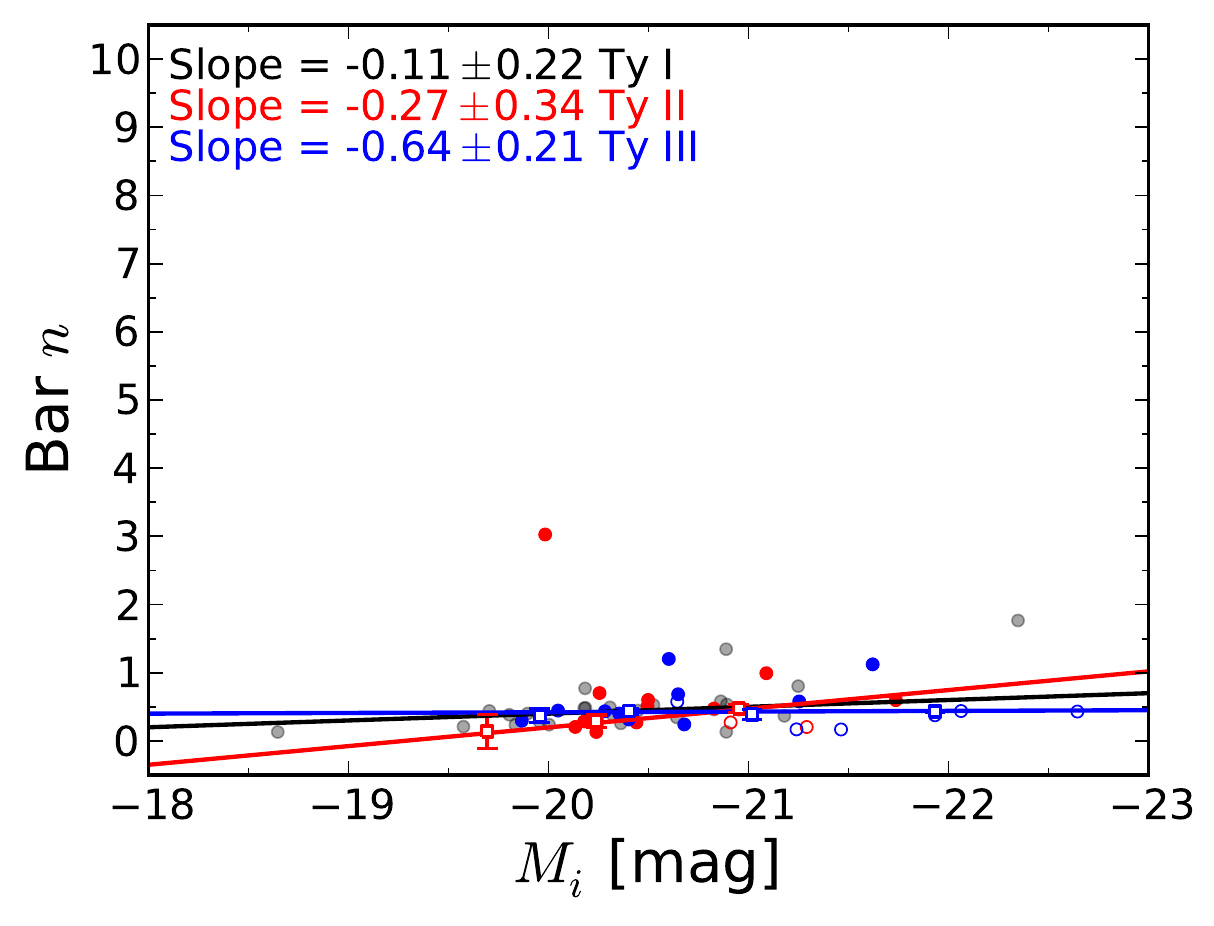}
\end{center}
\caption[Bulge and bar \sersic indices as a function of total model magnitude for galaxies with Type I, II, and III discs.]{Bulge and bar \sersic indices ($n$) for {\it BD}, {\it CD}, {\it BDd}, {\it CDd}, {\it BSD}, and {\it BSDd} model galaxies with Type I (black), II (red), and III (blue) discs as a function of total absolute $i$ band model magnitude. {\bf Upper plot:} Bulge \sersic index. {\bf Lower plot:} Bar \sersic index. Unfilled data points indicate flagged galaxies. Large square points are median parameter values in bins of $M_i$, to which a linear trend has been fit. Type I galaxies are indicated by small grey points for clarity.}
\label{brkparams5}
\end{figure}

In this section, the galaxy sample is divided by Freeman type to investigate differences in internal structure for galaxies with untruncated (Type I), truncated (Type II), or anti-truncated (Type III) discs. We consider all models with a single (exponential) disc component which dominates (relative to the bulge) at large galaxy-centric radii ({\it BD}, {\it CD}, {\it BDd}, {\it BSD}, {\it BSDd}). Here, the distributions and trends in structural parameters for galaxies with (single) disc-dominated outer regions ({\it BD}, {\it CD}, {\it BDd}, {\it BSD}, {\it BSDd}) are investigated. In order to ensure that model parameters are measured from consistent structures (i.e. exponential/broken exponential components measure galaxy disc properties), we only consider galaxies with archetypal bulge/disc models (i.e. Type 1; central bulge + outer disc). This reduces the sample of analysed galaxies to 146 (67 2-component galaxies, 79 3-component galaxies), of which 97 galaxies have Type I discs, 18 have Type II discs, and 24 have Type III discs. The average structural properties and total magnitudes of archetypal galaxies containing discs of each type are summarised in Table \ref{mdisk_tab}. Example surface brightness profiles for ({\it BDd} and {\it BSDd}) galaxies with Type II and Type III discs are presented in Figure \ref{multi_demo}. As a convenient shorthand, we hereafter use the phrase `Type I/II/III galaxy' to refer to galaxies containing Freeman Type I/II/III discs.

\subsubsection{Central components of Type I/II/III galaxies}
The bulge and bar \sersic indices for galaxies of each disc type are presented in Figure \ref{brkparams5}. Bulge $n$ is smaller (on average) in Type I ($1.89\pm0.08$) galaxies those with Type II ($2.32\pm0.14$) broken discs, and consistent with the bulges of Type III galaxies ($1.74\pm0.19$). By comparison, {\emph{bar}} \sersic index is consistent across all Freeman types ($0.44\pm0.04$ for Type I, $0.44\pm0.04$ for Type II, and $0.43\pm0.03$ for Type III). Thus, while consistent bar profiles are measured independent of disc type, the bulge profile shape depends on disc structure. Note that the Type I averages are calculated from galaxies in the magnitude range $-19<M_i<-22$ for consistency with the range of Type II and III galaxy luminosities.

With increasing galaxy luminosity, no significant variation in bar $n$ is detected for any galaxy type. However, the bulges of both Type I and Type III galaxies become more centrally-concentrated (higher $n$) for more luminous galaxies. Similar $n$-$M_i$ slopes are measured for both galaxy types (consistent with the equivalent trend measured previously for archetypal {\it BD} models in Paper I). The reverse trend (lower $n$ for higher galaxy luminosity) is measured for Type II galaxies. While this trend is not significant ($\sim$\,$1.5\sigma$), it remains discrepant with the measured Type I/III trends at a $3\sigma$ level. Thus, the bulges of galaxies with truncated discs are structurally distinct from those found in galaxies with untruncated, or anti-truncated discs. This is analogous to the distinct $n$-luminosity trends measured in the previous section for barred and unbarred galaxies. However, as barred galaxies comprise approximately equal numbers of Type II ($23^{+5}_{-4}\%$) and III discs ($19\pm4\%$), this apparent bulge $n$ bimodality is not strongly related to the presence of a bar component. 

\begin{figure}
\begin{center}
	\includegraphics[width=\linewidth,clip=true]{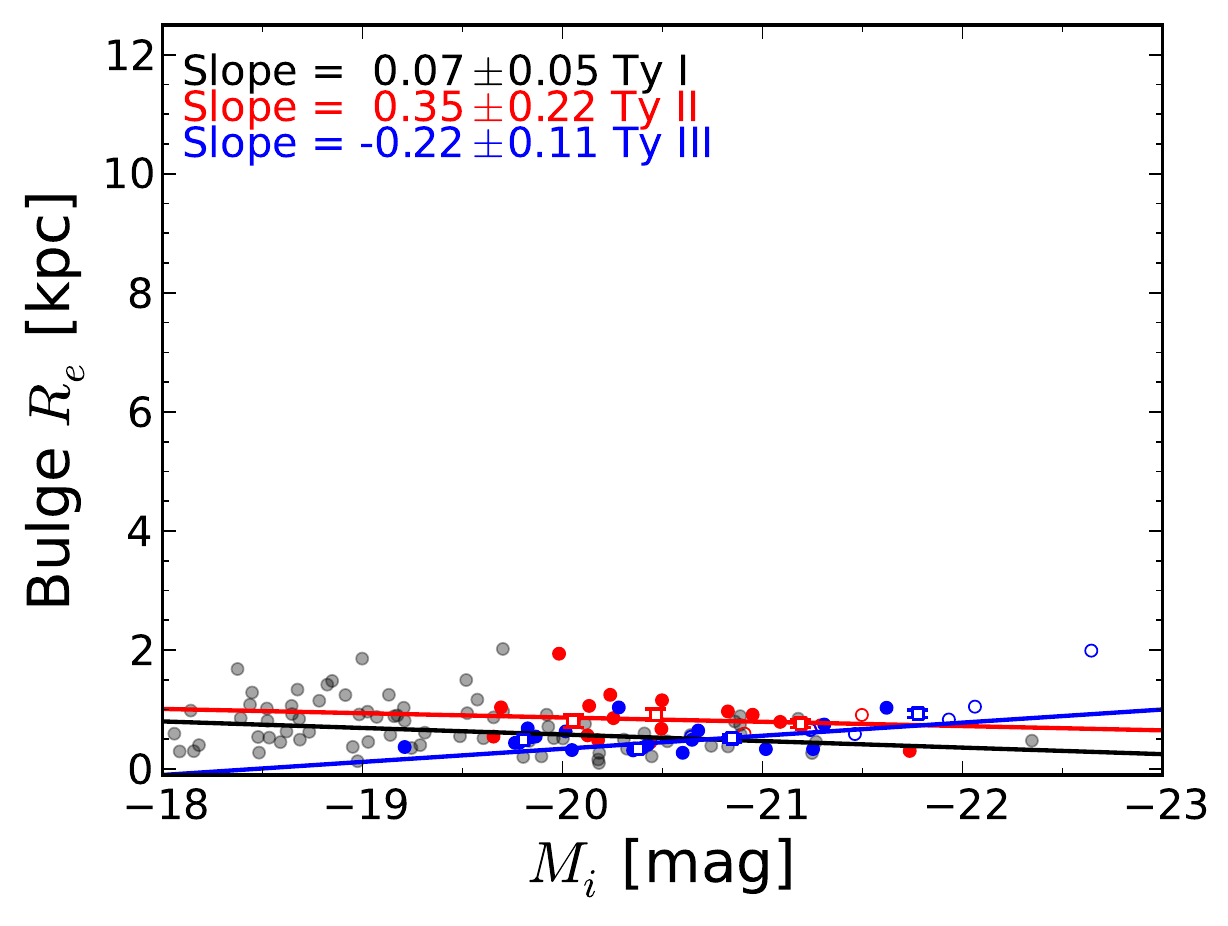}
	\includegraphics[width=\linewidth,clip=true]{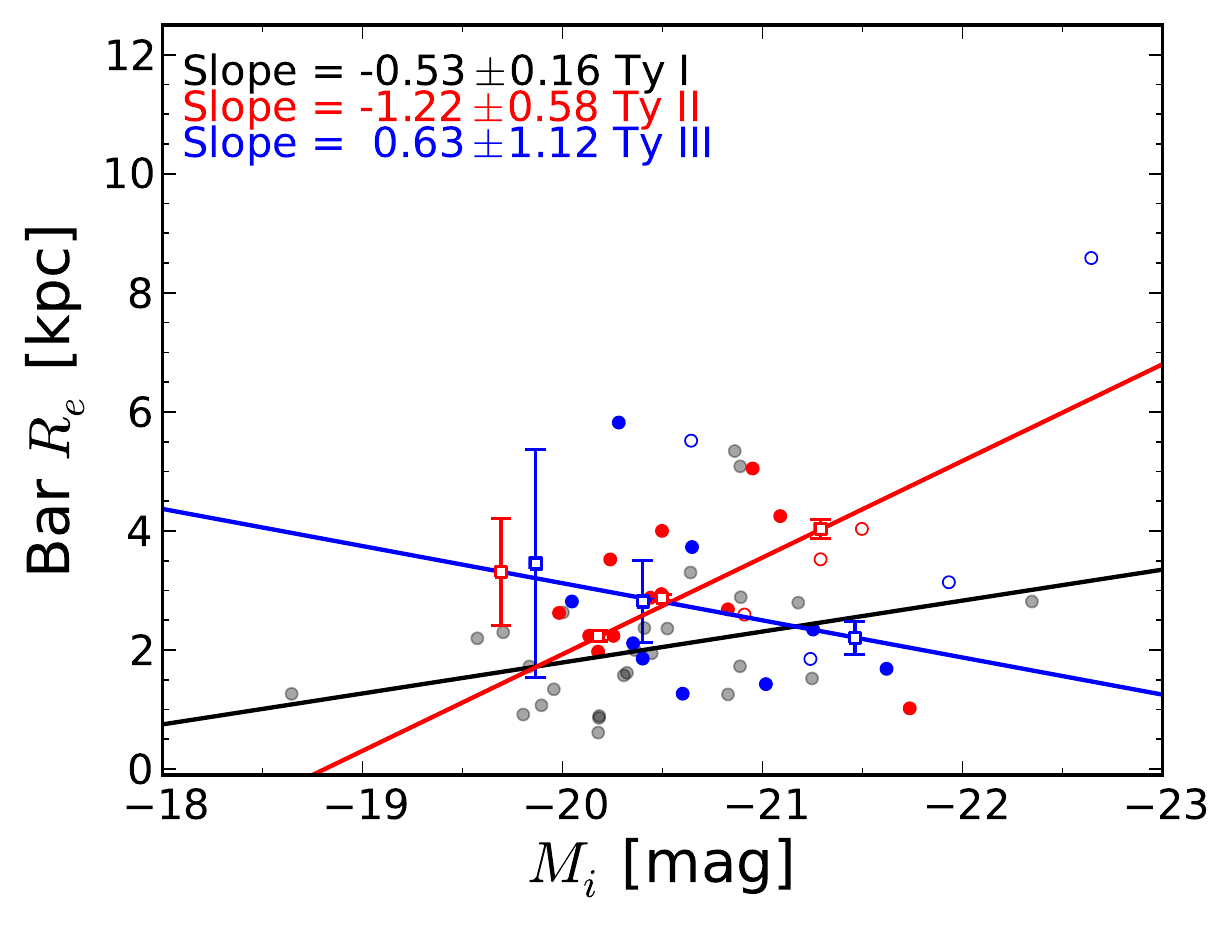}
\end{center}
\caption[Bulge and bar effective half-light radii as a function of total model magnitude for galaxies with Type I, II, and III discs.]{Bulge and bar effective half-light radii ($R_e$) for {\it BD}, {\it CD}, {\it BDd}, {\it CDd}, {\it BSD}, and {\it BSDd} model galaxies with Type I (black), II (red), and III (blue) discs as a function of total absolute $i$ band model magnitude. {\bf Upper plot:} Bulge $R_e$. {\bf Lower plot:} Bar $R_e$. Unfilled data points indicate flagged galaxies. Large square points are median parameter values in bins of $M_i$, to which a linear trend has been fit. Type I galaxies are indicated by small grey points for clarity.}
\label{brkparams6}
\end{figure}

Half light radii for the bulges and bars of Type I, II, and III galaxies are presented in Figure \ref{brkparams6}. The bulges of Type I and III galaxies show no significant size difference on average (Type I: $0.57\pm0.05$ kpc, Type III: $0.54\pm0.05$ kpc), while Type II galaxies have systematically larger bulges ($0.96\pm0.07$ kpc). No notable trends in bulge size with galaxy luminosity is noted for galaxies of any Freeman type.

The bars in Type II galaxies are systematically larger on average ($2.91\pm0.22$ kpc) than those found in Type I galaxies ($1.95\pm0.18$ kpc), but similar to the bars of Type III galaxies ($2.82\pm0.49$ kpc). As with bulge components, no significant size-luminosity trends are noted for galaxy bars. Thus, large galaxy bars are found more frequently in galaxies with broken (truncated/anti-truncated) discs, regardless of total galaxy luminosity. 

In summary, the bulges of galaxies with Type II discs have systematically larger $n$ and $R_e$ than the bulges of Type I or III galaxies. In addition, no significant bulge $n$-luminosity trend is detected for Type II galaxies. Thus, the bulges of galaxies with truncated discs are distinct in structure and origin from the equivalent components in galaxies with either untruncated or anti-truncated discs. Galaxy bars are consistent in profile shape across all Freeman types, but have systematically larger $R_e$ in galaxies with Type II or III broken discs.  

\subsubsection{Type I/II/III Disc Properties}
\begin{figure}
\begin{center}
	\includegraphics[width=\linewidth,clip=true]{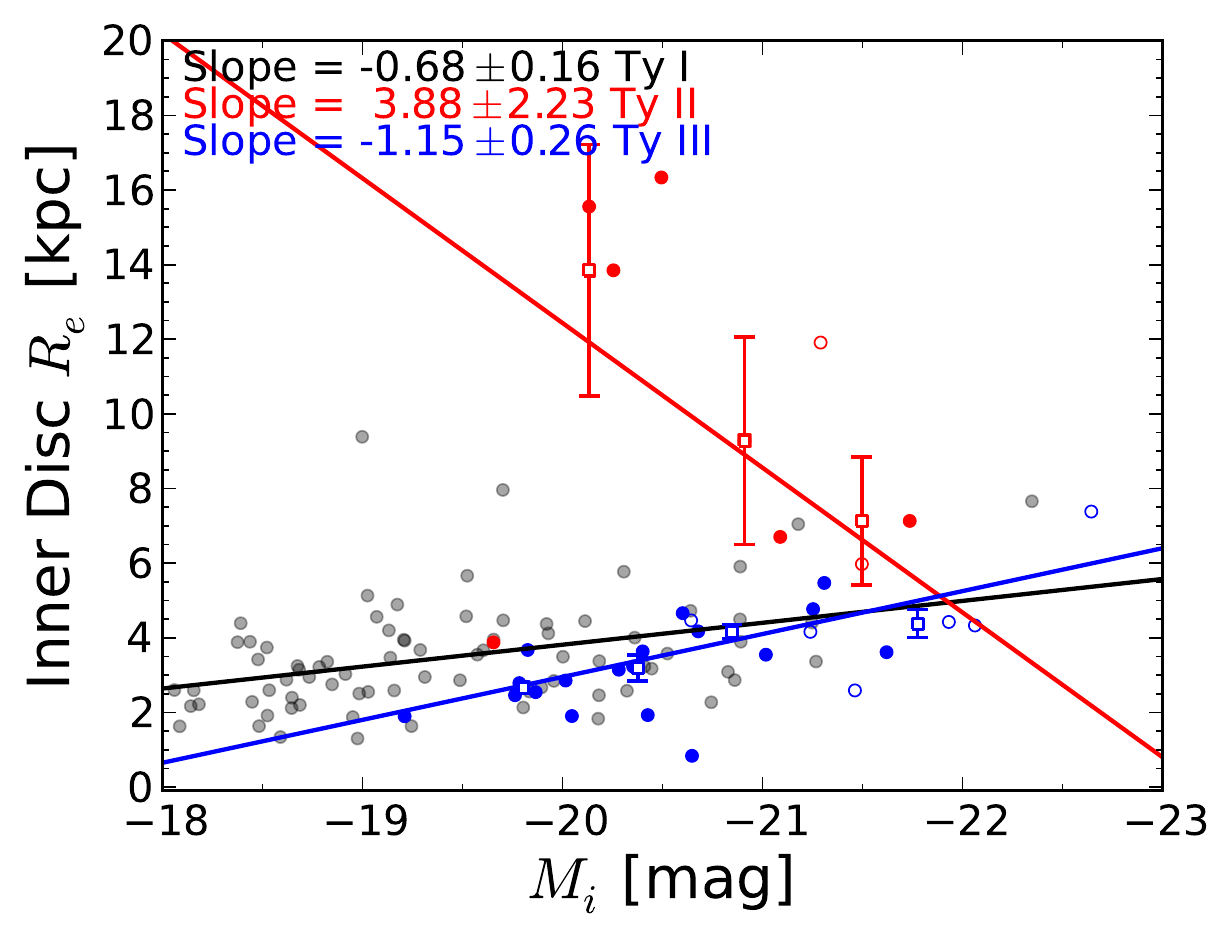}
	\includegraphics[width=\linewidth,clip=true]{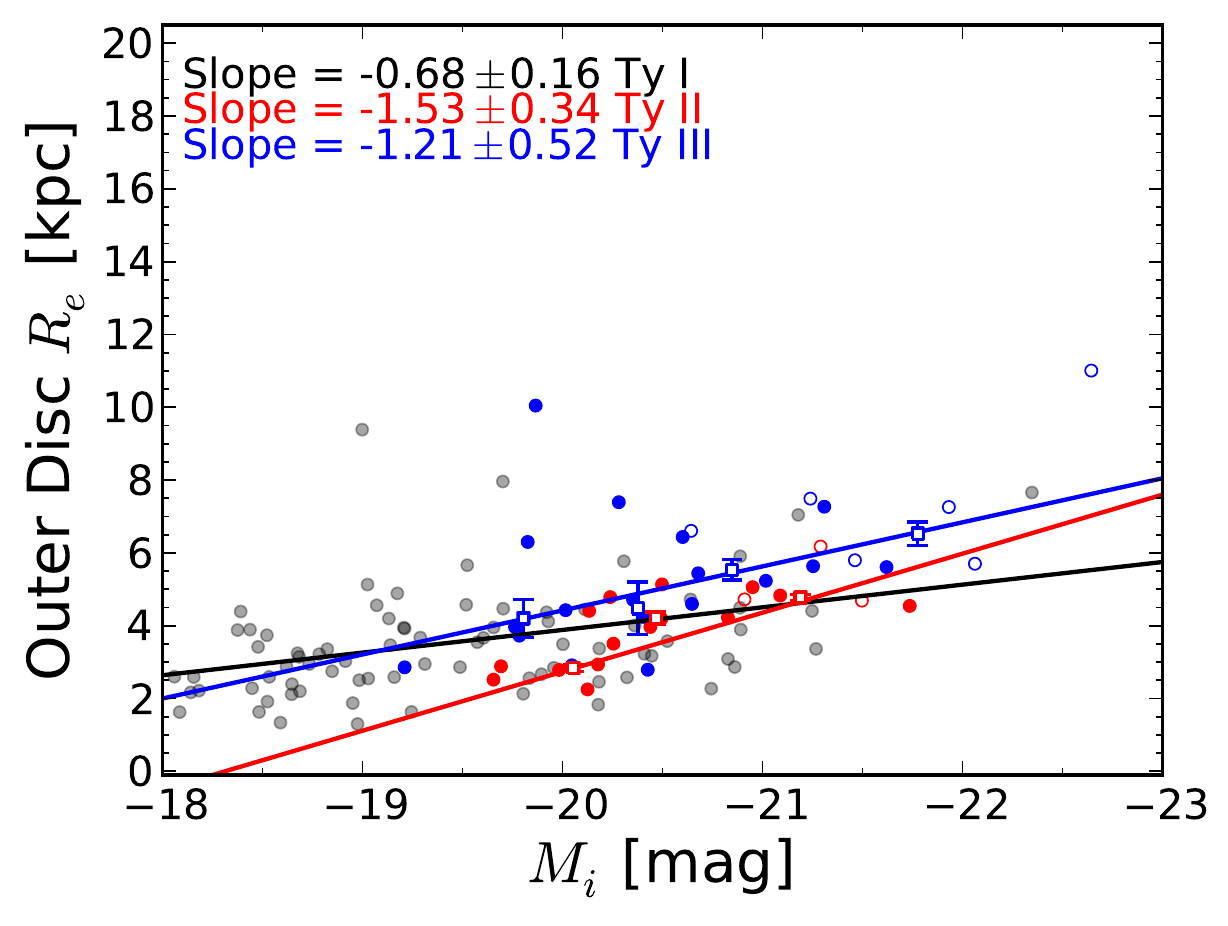}
\end{center}
\caption[Inner/outer disc effective half-light radii as a function of total model magnitude for Type I, II, and III discs.]{Inner and outer disc effective half-light radii ($R_e$) for {\it BD}, {\it CD}, {\it BDd}, {\it CDd}, {\it BSD}, and {\it BSDd} model galaxies with Type I (black), II (red), and III (blue) discs as a function of total absolute $i$ band model magnitude. {\bf Upper plot:} Inner disc $R_e$. {\bf Lower plot:} Outer disc $R_e$. Unfilled data points indicate flagged galaxies. Large square points are median parameter values in bins of $M_i$, to which a linear trend has been fit. Type I galaxies are indicated by small grey points for clarity.}
\label{brkparams7}
\end{figure}

The effective half-light radii for the inner and outer discs (i.e. the discs internal and external to the break radius, $r_{\rm brk}$) of Type II (truncated) and III (anti-truncated) galaxies are presented in Figure \ref{brkparams7}, with the disc $R_e$ for Type I galaxies included in both panels. Note that by definition, the inner $R_e$ of Type II/III galaxies is larger/smaller than the outer $R_e$, yielding a shallower disc surface brightness profile within/beyond $r_{\rm brk}$. On average, Type III inner discs are consistent in size ($3.58\pm0.20$ kpc) with Type I discs ($3.68\pm0.18$ kpc), and have a consistent size-luminosity relation (despite a $\sim$\,$2\times$ difference in slope). By contrast, Type II inner discs are {\emph{substantially}} larger (than Type I discs) on average ($9.28\pm1.44$ kpc), with an extremely steep trend ($4.76\pm3.09$ kpc per mag) of {\emph{decreasing}} inner disc size with increasing galaxy luminosity, albeit at low significance ($\sim$\,$1.5\sigma$). 

The outer discs of Type II galaxies have scale lengths ($4.12\pm0.20$ kpc) consistent with Type I discs (for galaxies in the range $-19<M_i<-22$) on average, while Type III outer discs are systematically larger ($5.62\pm0.34$ kpc). Outer disc size-luminosity relations are similar for both Type II and III discs, yielding size increases for more luminous galaxies a factor of approximately two times greater than the measured trend for Type I discs. However, this difference relative to Type I discs is only significant (at a $\sim$\,$2.5\sigma$ level) for Type II galaxies. The detection of consistent scale lengths (and similar size-luminosity relations) for Type I discs, Type II outer discs, and Type III inner discs (in agreement with \citealp{Laine2014}) suggests that the outer/inner structures of Type II/III discs preserve the structural properties of their progenitor discs. 

\begin{figure}
\begin{center}
	\includegraphics[width=\linewidth,clip=true]{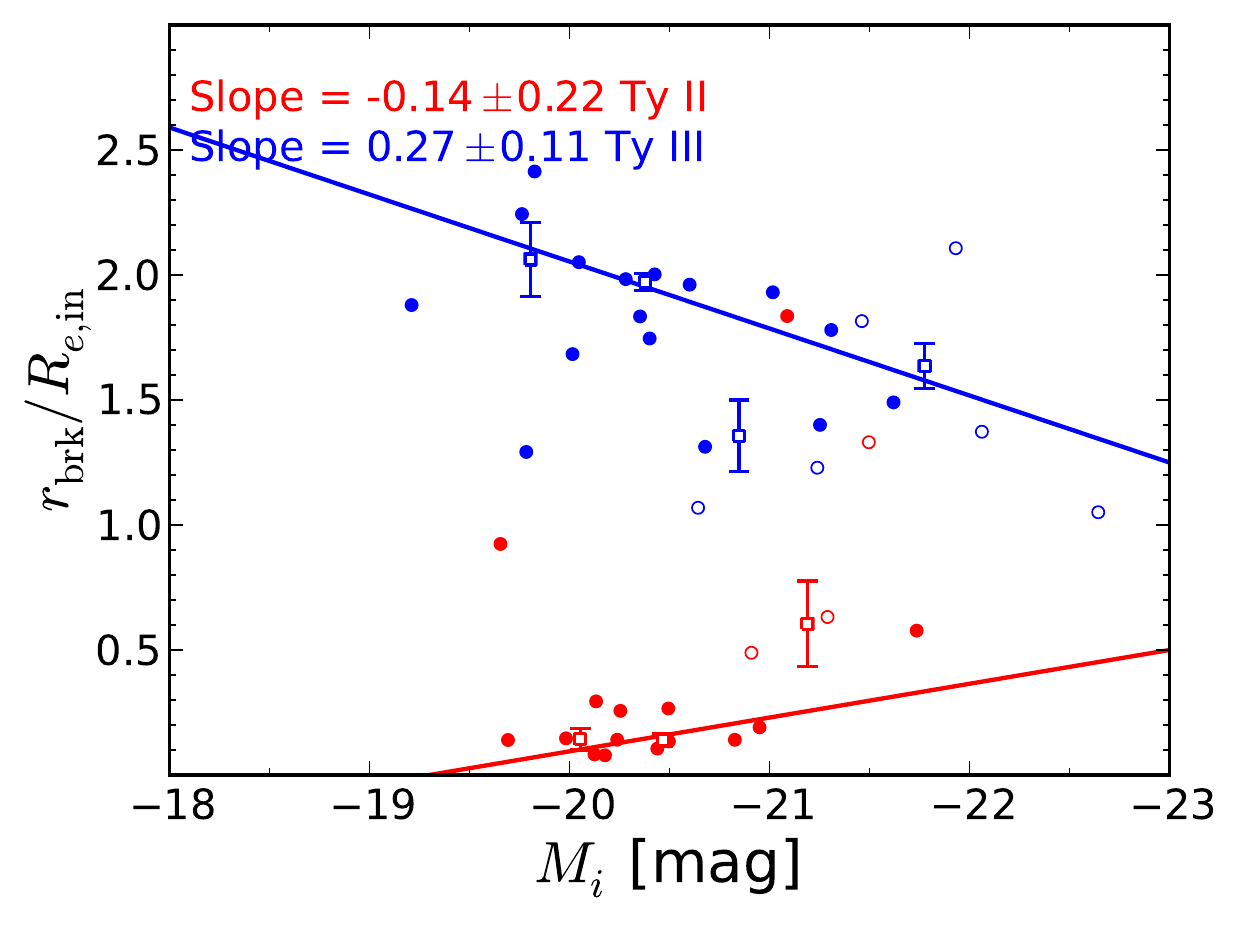}
	\includegraphics[width=\linewidth,clip=true]{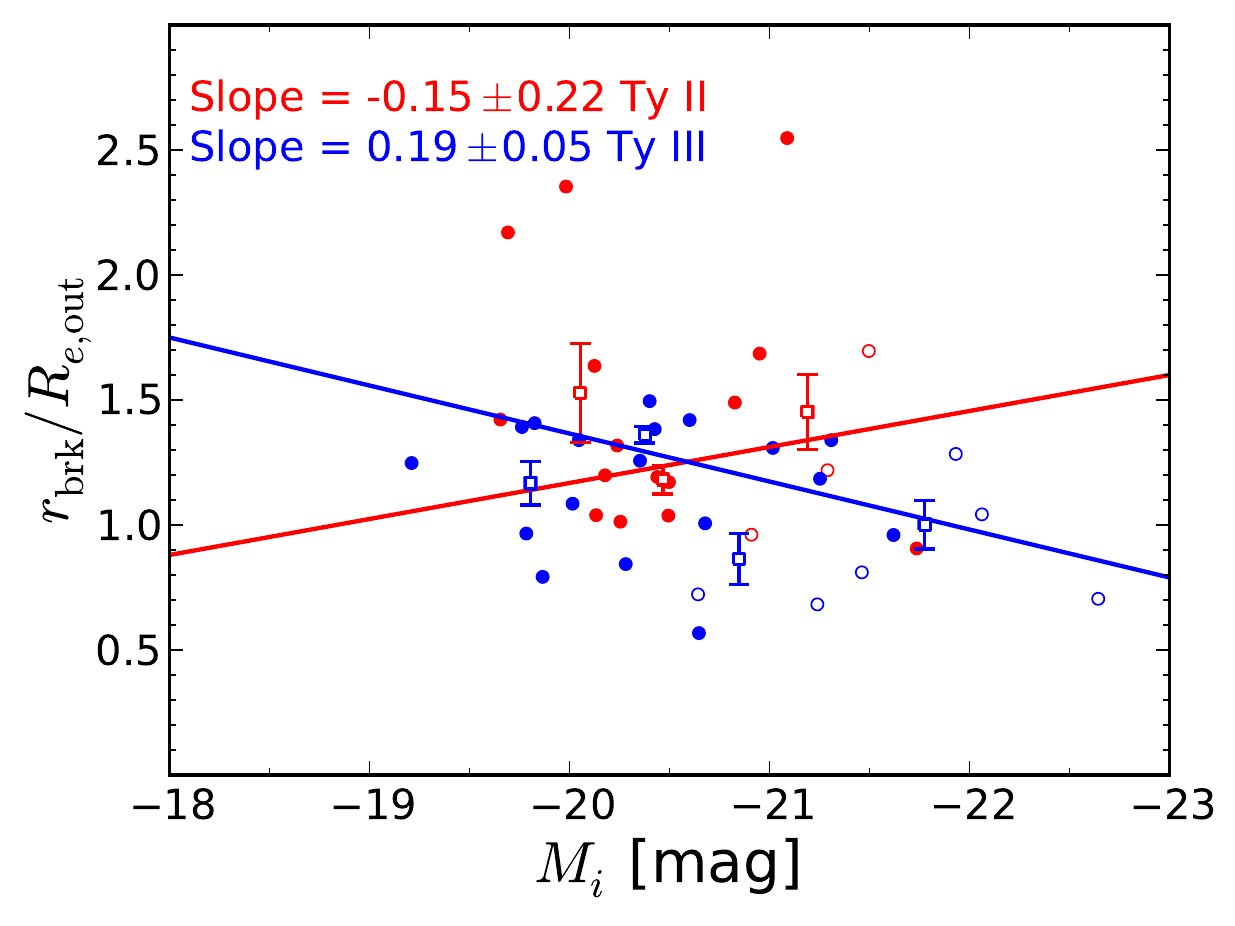}
\end{center}
\caption[Break radius for galaxies with Type II and III discs as a function of the total model magnitude.]{Break radius, $r_{\rm brk}$, as a function of the total absolute $i$ band model magnitude for galaxies with Type II (red) and III (blue) disc models relative to the inner ({\bf top}) and outer ({\bf bottom}) disc $R_e$. Unfilled data points indicate flagged galaxies. Large square points are median parameter values in bins of $M_i$, to which a linear trend has been fit.}
\label{rbk1}
\end{figure}

The break radius, $r_{\rm brk}$, is plotted in Figure \ref{rbk1} for Type II and III discs as a fraction of both inner and outer disc $R_e$. For Type II discs, the break radius is a small fraction of the inner disc size ($\sim$\,$0.25$). However, since $R_e$ is large for these structures, the contribution of the (flat) inner disc to the total disc luminosity is non-negligible. By comparison, Type III disc break radii are significantly beyond the inner disc half-light radius ($r_{\rm brk}\sim$\,$2R_e$), indicating that only the outer wings of Type III inner disc structures are modified by the profile break. Alternatively, both Type II and III profile breaks are comparable in size to the {\emph{outer}} disc $R_e$. Thus, in both cases, the outer structure of broken discs contribute $\sim$\,$50\%$ of the light of an equivalently-sized untruncated disc. 

In comparison to either disc, Type III breaks occur at smaller fractions of disc $R_e$ for increasingly luminous galaxies. Note, however, that this correlation is significant at a $>3\sigma$ level for outer disc $R_e$, but only significant at a $\sim$\,$2.5\sigma$ level for inner disc $R_e$. A decreased fractional break radius indicates that a Type III disc contains a smaller proportion of the primordial disc. Conversely, Type II break radii exhibit a non-significant increase (as fractions of both $R_e$) with galaxy luminosity. Thus, the break radius of a Type II disc is approximately the same fraction of the inner/outer disc size for any galaxy. Note that if Type II discs where $R_{e,{\rm in}}$ reaches the {\footnotesize{\tt GALFIT}} limit are excluded, the trend in $r_{\rm brk}$ relative to $R_{e,{\rm out}}$ is made considerably shallower. Hence, $r_{\rm brk}$ increases in size at a similar rate to outer disc $R_e$ with total galaxy luminosity.

\begin{table*}
\begin{center}
\begin{tabular}{clcccc}
\hline
\hline
&&$N$&$\Delta_{\rm B/T}$&$\Delta_{\rm S/T}$&$\Delta_{\rm D/T}$\\
\hline
&Type I&56&$-0.01\pm0.03$&-&$\phantom{-}0.01\pm0.03$\\
2-comp.&Type II&4&$-0.10\pm3.65$&-&$\phantom{-}0.10\pm3.65$\\
&Type III&7&$-0.12\pm0.11$&-&$\phantom{-}0.12\pm0.11$\\
\hline
&Type I&41&$-0.04\pm0.03$&$-0.10\pm0.03$&$\phantom{-}0.12\pm0.04$\\
3-comp.&Type II&21&$\phantom{-}0.11\pm0.07$&$\phantom{-}0.01\pm0.05$&$-0.22\pm0.06$\\
&Type III&17&$-0.14\pm0.02$&$\phantom{-}0.07\pm0.03$&$\phantom{-}0.05\pm0.03$\\
\hline
\hline
\end{tabular}
\end{center}
\caption[Table of best-fit trend slopes for component luminosity fraction varying with increased total luminosity.]{Table of best-fit component light fraction-luminosity trends ($\Delta_{\rm C/T}$; C/T per magnitude galaxy luminosity) measured for 2-component ({\bf top}) and 3-component ({\bf bottom}) galaxies. Here, a negative value indicates increasing C/T with luminosity.}
\label{dC_tab}
\end{table*}

In summary, the inner discs of anti-truncated (Type III) galaxies are consistent in size with the discs of unbroken (Type I) galaxies. Conversely, Type III outer discs are systematically larger than unbroken discs. Both inner and outer Type III discs exhibit a size-luminosity relation consistent with Type I discs. Thus, the inner discs of Type III galaxies preserve the properties of the unbroken progenitor disc. The inner discs of truncated (Type II) galaxies are not consistent in size or size-luminosity trend with unbroken discs. This rules out a formation scenario in which physical truncation preserves the primordial disc within the break radius. Conversely, the outer discs of Type II galaxies have sizes (and size-luminosity relations) consistent with untruncated disc structures. 

\subsubsection{Component fractions (C/T) of Type I/II/III galaxies}
In this section, we discuss the component flux fractions (B/T, S/T, and D/T for bulges, bars, and discs respectively) for galaxies with Type I, II, and III discs.

Measured across all (2- and 3-component) galaxies, Type I galaxies are strongly disc-dominated (median D/T $=0.62\pm0.02$), with (sub-dominant) bulges (B/T $=0.28\pm0.02$) and weak bar components (S/T $=0.10\pm0.01$). The corresponding component fractions for Type III galaxies are measured to be consistent with Type I galaxies (D/T$=0.62\pm0.03$, B/T $=0.24\pm0.02$, S/T $=0.14\pm0.03$). By contrast, Type II galaxies have a diminished disc light fraction on average (D/T$=0.46\pm0.04$), with corresponding increases in bulge (B/T$=0.33\pm0.04$) and bar (S/T$=0.21\pm0.03$) fractions. Note, however, that these averages are heavily biased by the lack of a bar (i.e. S/T$=0.00$) in 2-component galaxies. If the average is calculated from only 3-component galaxies, then bar light fraction (S/T) increases significantly for all three disc types (Type I: $0.24\pm0.02$; Type II: $0.25\pm0.02$; Type III: $0.20\pm0.03$). The corresponding disc light fractions (D/T) decrease on average for 3-component galaxies (Type I: $0.51\pm0.03$; Type II: $0.43\pm0.04$; Type III: $0.55\pm0.03$), while average bulge fractions (B/T) are not significantly changed (Type I: $0.25\pm0.02$; Type II: $0.31\pm0.04$; Type III: $0.20\pm0.03$).

The best-fit component light fraction trends with galaxy luminosity (2-component and 3-component galaxies considered separately) are presented in Table \ref{dC_tab}. For 2-component galaxies, no significant trends are noted in Type I galaxy B/T or D/T, while no conclusions can be drawn for Type II and III galaxies due to small sample sizes. However, with increasing luminosity, 3-component Type I galaxies become significantly more bar-dominated ($-0.10\pm0.03$), and less disc-dominated ($0.12\pm0.04$). Conversely, 3-component Type II galaxy disc light fraction and Type III galaxy bulge light fraction increases with galaxy luminosity ($-0.22\pm0.06$ and $-0.14\pm0.02$ respectively). 

These component light fraction-luminosity trends can be used to estimate whether the distinction between faint and bright galaxies is dominated by the luminosity difference of one particular component. This can characterise, for example, whether the difference between an average galaxy and an equivalent galaxy one magnitude brighter is primarily due to an increase in bulge or disc luminosity. Hence, we will determine whether the apparent differences in C/T trends between Freeman Types corresponds to intrinsically different component light scaling relations. 

For two galaxies separated in total galaxy luminosity by one magnitude ($M_0-M=1.0$), the fractional difference in the luminosity of a particular component ({\it C}) can be parametrised as $x_C= L_C/L_{C,0}$. For example, if the galaxy luminosity difference in Type I galaxies at $M_{i,0} = -20$ and $M_i = -21$ was caused by the bulge and disc components being $3\times$ brighter at $M_i = -21$ (but bars being as luminous in both cases), then $x_B = 3$, $x_S=1$, and $x_D=3$.

The reported C/T slopes ($\Delta_{\rm C/T}$; Table \ref{dC_tab}) can be expressed as:
\begin{equation}
\Delta_{\rm C/T} = {\rm C}_0{\rm /T}_0 - {\rm C/T} = {\rm C}_0{\rm /T}_0\left(1-\frac{x_C}{x_T}\right),
\label{CTeqn}
\end{equation}
\noindent
where fractional difference in total luminosity, $x_T=2.5$ across one magnitude. Note that $x_C/x_T$ is greater than unity if component luminosity increases at a greater rate than galaxy luminosity. 

\begin{figure}
\begin{center}
	\includegraphics[width=0.9\linewidth,clip=true]{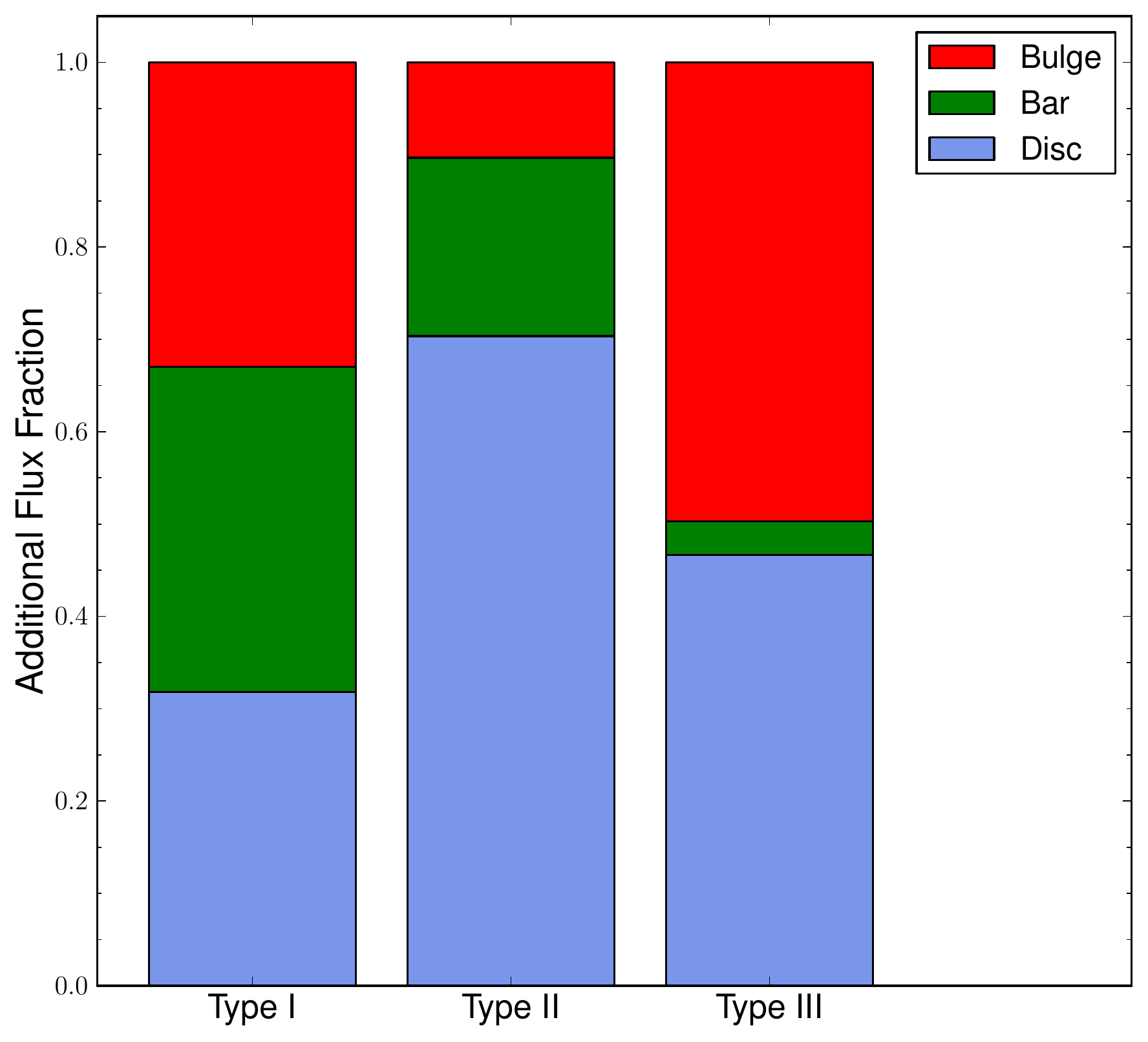}
\end{center}
\caption[`Additional' component light fractions for average Type I, II, or III galaxies.]{`Additional' component light fractions (i.e. the proportions of the luminosity difference for each component structure per unit galaxy luminosity) for average galaxies with Freeman Type I, II and III disc components. Indicates the fraction of light added to the bulge (red), disc (blue) or bar (green) per unit galaxy luminosity increase. Type I galaxies - bulge: $33\pm3\%$, disc: $35\pm3\%$, bar: $32\pm4\%$; Type II galaxies - bulge: $10\pm2\%$, bar: $19\pm2\%$, disc: $70\pm12\%$; Type III galaxies - bulge: $50\pm4\%$, bar: $4\pm1\%$, disc: $47\pm6\%$.}
\label{add_frac}
\end{figure}

Table \ref{xC_tab} presents $x_C$ values relative to a galaxy of average luminosity and C/T (i.e. C$_0$/T$_0 = \langle$C/T$\rangle$). For a galaxy brighter than the average by an arbitrary magnitude difference ($M_{\rm tot} = M_{\rm tot,0} + \Delta M_{\rm tot}$), the proportion of the total luminosity difference ($\Delta L_{\rm tot}$; where $L_{\rm tot,0} = L_{\rm tot} + \Delta L_{\rm tot}$) attributed to each photometric component ($\Delta L_C$) can be estimated using:

\begin{equation}
\frac{\Delta L_C}{\Delta L_{\rm tot}} = {\rm C/T}\frac{(x_T- 1)-x_T\Delta_{\rm C/T}}{x_T - 1}.
\label{eq3}
\end{equation}
\noindent
The resulting component fractions of the additional galaxy luminosity is illustrated in Figure \ref{add_frac} for Type I, II and III galaxies. 

For Type I galaxies, bars become more luminous at a significantly greater rate than the overall galaxy luminosity (bar luminosity doubles for a 42\% increase in total luminosity), while discs increase in luminosity at a slower rate (52\% increase in disc luminosity as galaxy luminosity doubles). However, the luminosity difference for Type I galaxies arbitrarily brighter than the average is distributed equally between all three structural components (from Equation \ref{eq3}; Figure \ref{add_frac}). For example, an average Type I galaxy ($\langle M_i\rangle =-20.3$) has a total luminosity of $8.6\times10^9$ \Lsol. Relative to this average, a $9.6\times10^9$ \Lsol\, Type I galaxy (i.e. $10^9$ \Lsol\, brighter) would have a bulge more luminous by $(3.3\pm0.4) \times 10^8$\Lsol, a bar more luminous by $(3.5\pm0.3) \times 10^8$ \Lsol, and a disc $(3.2\pm0.4) \times 10^8$\Lsol\, more luminous.

\begin{table}
\begin{center}
\begin{tabular}{lcccc}
\hline
\hline
&$N$&$x_B/x_T$&$x_S/x_T$&$x_D/x_T$\\
\hline
Type I&41&$1.16\pm0.12$&$1.41\pm0.13$&$0.76\pm0.08$\\
Type II&21&$0.65\pm0.23$&$0.96\pm0.20$&$1.51\pm0.15$\\
Type III&17&$1.56\pm0.09$&$0.66\pm0.16$&$0.91\pm0.06$\\
\hline
\hline
\end{tabular}
\end{center}
\caption[Table of fractional component luminosity changes with increased total luminosity.]{Table of approximate fractional component luminosity changes for 3-component Type I, II, and III galaxies as total galaxy luminosity increases ($x_T=2.5$).}
\label{xC_tab}
\end{table}

For Type II galaxies, the disc component is the dominant contribution to luminosity growth ($70\pm12\%$ of $\Delta L_{\rm tot}$), doubling in luminosity for each 32\% increase in galaxy luminosity. Hence, the disc-total luminosity trend is significantly steeper for Type II discs than Type I, indicating a larger difference in disc luminosity between faint and bright Type II galaxies than for Type I galaxies. This implies that fainter Type II galaxies have experienced a greater truncation (of light) than intrinsically more luminous galaxies.

For Type III galaxies, the bulge is the dominant component (doubling in luminosity for a 28\% increase in global luminosity). For an average Type III galaxy, the bulge component's contribution to galaxy luminosity is approximately equal to the (intrinsically more luminous) disc. The corresponding bar light contribution is minimal ($3.7\pm0.4$\% of $\Delta L_{\rm tot}$), indicating approximately equally-luminous bars in all Type III galaxies, independent of total galaxy luminosity. 
 
No other components (bulges in Type I galaxies, bulges and bars in Type II galaxies, and bars and discs in Type III galaxies) differ significantly from increasing in luminosity proportional to the galaxy ($x_C\sim$\,1). Note that since $x_T = 2.5$, no component in Type I, II, or III galaxies decreases in luminosity in brighter galaxies. The proportions of added galaxy luminosity contributed by each model component are illustrated in .

In summary, 3-component archetypal (central bulge + outer-dominant disc + any bar) galaxies are disc-dominated on average, with approximately equal bulge and bar light fractions, independent of Freeman disc/galaxy type. The measured trends in component light fraction with total magnitude were used to quantify the contributions of each structural component to galaxy luminosity. All three structural components contribute equally on average to the increasing total luminosity in galaxies unbroken discs (Type I). However, the bar component exhibits the largest fractional increase in luminosity. discs were found to dominate truncated (Type II) galaxy luminosities. The corresponding disc-total luminosity trend is steeper than for Type I galaxies, which may indicate disc (luminosity) truncation. Increasing anti-truncated disc (Type III) galaxy luminosities correlate strongly with both their bulges and discs. Hence, bar luminosity in Type III galaxies is independent of galaxy luminosity. 

\begin{figure}
\begin{center}
	\includegraphics[width=\linewidth,clip=true]{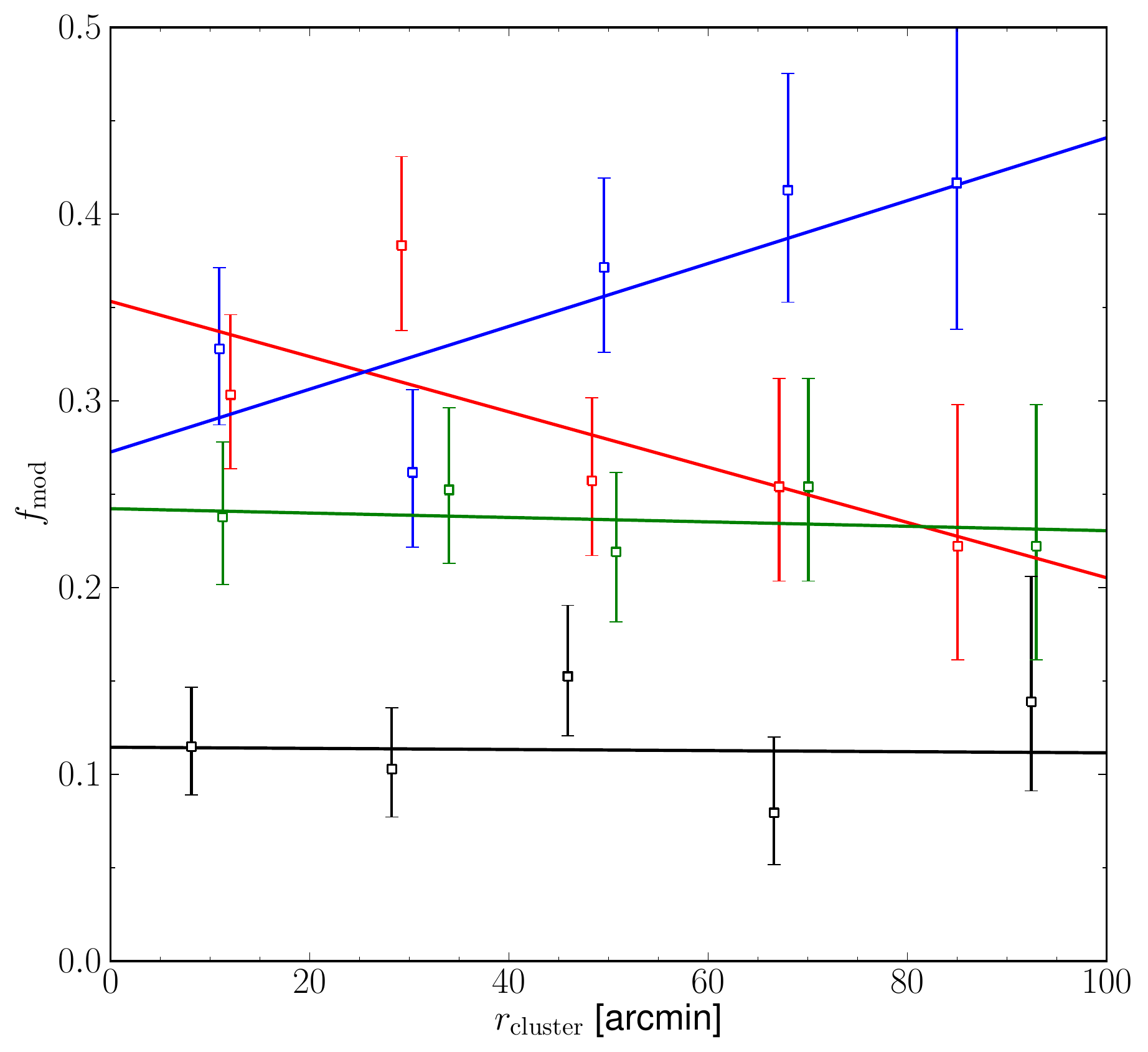}
\end{center}
\caption{The model fractions, $f_{\rm mod}$ (i.e. number of each model type $\div$ total number of galaxies per bin) as a function of radial position in the Coma cluster, $r_{\rm cluster}$. Single \sersic objects ({\it S}) are plotted in red, galaxies with outer exponential discs ({\it BD, CD, BSD}) are plotted in blue, galaxies with broken discs ({\it BDd, BSDd}) are plotted in black, and multi-\sersic systems({\it BS, BSS}) are plotted in green.}
\label{mdr}
\end{figure}

\subsection{Structural Trends with Environment}
In this section, we investigate variation in the multi-component structures of galaxies as a function of the (projected) distance from the Coma cluster centre, $r_{\rm{cluster}}$. Observed $r_{\rm cluster}$ correlates with the time at which a galaxy first entered the cluster environment \citep{Gao2004,Smith2012,DeLucia2012,Taranu2012}, albeit with substantial scatter. A cluster-centric radial analysis therefore highlights the environment-mediated processes that have acted on these multi-component systems, and hence the cluster environment's role in their formation. 

The morphological mix of galaxies varies with position in the cluster (Figure \ref{mdr}). Neither the fraction of multi-\sersic models, nor the fraction of galaxies with broken discs vary significantly with cluster-centric radius. Note that, this would not change if Type II and III galaxies were considered separately (see Section \ref{disc_frac}). The fraction of pure \sersic systems increases towards the cluster centre, while the fraction of (exponential) outer disc galaxies decreases. However, neither of these radial morphology trends are significant. 

With increasing distance from the cluster centre, no significant variation ($>3\sigma$) is detected in the structural properties of Type I, II, or II galaxies. However, weakly significant radial trends ($\sim$\,$2\sigma$) are detected in (barred) Type I galaxy bar size ($R_e$; $1.3\pm0.6$ kpc per $r_{200}$), Type II outer disc size ($R_{e, {\rm out}}$; $-1.5\pm0.8$ kpc per $r_{200}$)\footnote{Note that the sign of this trend indicates {\it increasing} outer disc size towards the cluster centre, and hence does not correspond to environmental truncation.}, and Type III galaxy bulge profile ($n$; $0.9\pm0.5$ per $r_{200}$). Thus, the strucural properties of galaxies with disc-dominated outer regions are independent of their radial position in the cluster. Therefore, the formation of disc breaks may result from secular, rather than environment-mediated processes.     

\section{Discussion: Disc break formation scenarios}
Here, we briefly investigate the evolutionary origins of archetypal broken disc (truncated/anti\-truncated; Freeman Type II/III) galaxies through comparison of their structural and component photometric properties to unbroken (Type I) galaxies. As a working hypothesis, we assume a break formation scenario in which Type II and III galaxies had Type I discs at some point in the past. While the observed {\it present-day} Type I discs are not necessarily the progenitors of present-day broken discs, all three Freeman type galaxies are assumed to have evolved from a common population of primordial galaxies with (Type I) discs. Thus, characteristics of the structural/photometric distributions unique to galaxies of a particular Freeman type can be used to constrain their evolutionary pathways. %that acted exclusively on them.  

The absence of any strong cluster-radial trends in galaxy structure disfavours a (cluster) environment-driven origin for disc breaks. Furthermore, bar structures appear to be strongly related to the formation of Type II and III discs: while one half of all Type I galaxies contain a bar ($42^{+5}_{-5}\%$ barred, $58^{+5}_{-5}\%$ unbarred), the bar fraction is considerably higher for Type II ($89^{+5}_{-9}\%$) and Type III ($71^{+8}_{-10}\%$) galaxies. Galaxy bars are also significantly larger if their host galaxy has a truncated/anti-truncated disc than if the galaxy disc remains unbroken. This implies that either the formation mechanism induces bar growth, or that bars stabilise discs during truncation/anti-truncation, such that the detection of a disc break for bright galaxies is more likely if a bar is present. 

The detection in galaxy simulations of inner (and outer) disc evolution with time \citep{Debattista2006,Minchev2012} supports a scenario of stellar (or gas) redistribution. In particular, the radial angular momentum transfer mechanism proposed in \cite{Minchev2012} would explain the apparent importance of a bar component, as such a structure would induce significant gravitational torques in disc gas. The significant increase in bar size for more luminous Type II galaxies may therefore suggest a period of enhanced star formation in the bar due to gas inflows, or the migration of disc stars into the bar.  

Systematically larger Type II inner disc scale lengths (and inconsistent size-luminosity relations) compared to untruncated Type I galaxies indicates that the inner discs of Type II galaxies are not structures equivalent to Type I discs. This disfavours a scenario in which Type II discs represent a truncated system in which the outer disc is suppressed relative to the surviving primordial inner disc. Furthermore, while D/T is systematically lower in Type II galaxies compared to Type I, the fractional change in disc light does not differ significantly from unity ($x_D/x_T=0.9\pm0.1$ as in Equation \ref{CTeqn}, where $x_T$ is the galaxy luminosity change between Type I and Type II galaxies). Hence, assuming an evolutionary scenario in which Type II galaxies evolve from Type I, disc luminosity increases proportional to the $\sim$\,$40\%$ increase in total galaxy luminosity (i.e. Type I: $m_T=14.8\pm0.1$ vs. Type II: $m_T=14.5\pm0.1$). Intrinsically brighter Type II discs rule out a formation mechanism in which Type I discs are physical truncated. This conclusion is not compromised by the comparison of present-day truncated and untruncated discs unless evolution from primordial to present-day Type I galaxies also involves reduction of disc luminosity while preserving their untruncated profiles.

Beyond $r_{\rm brk}$, Type II discs represent structures reminiscent of their primordial Type I discs (see also \citealp{Foyle2008}). Conversely, in the inner region ($r<r_{\rm brk}$) disc light has been redistributed such that the profile is flattened relative to a Type I profile. Bulges and bars in Type II galaxies are systematically larger than those in untruncated galaxies, implying that secular bulge/bar enhancement effects are significant for the formation of Type II galaxies. Thus, disc stars within $r_{\rm brk}$ may have been redistributed to form a bar and/or grow the galaxy bulge (see \citealp{Valenzuela2003}). Alternatively, the break formation mechanism may be enhanced via interaction with an existing bar, resulting in steeper inner/outer disc size trends with luminosity due to the strong bar size - galaxy luminosity relation.

Consistency in component scale-lengths (and size-luminosity trends) between Type III and Type I galaxies implies that the Type III inner discs may correspond to undisturbed primordial (Type I) discs. Conversely, the (significantly larger) outer disc may represent an additional extended structure. Nevertheless, this outer structure maintains a disc-like size-luminosity relation. An evolutionary scenario from Type I (or Type I {\emph{progenitors}}) to Type III is supported by the consistent bulge and disc component light fractions for both disc types, despite Type III galaxies being a factor of $1.9\times$ brighter on average. Hence, bulge and disc luminosities increase proportional to the galaxy luminosity difference between Type I and III galaxies.

Bulge and bar sizes in anti-truncated galaxies are significantly larger than those in Type I galaxies, while bulge luminosity increases strongly in more luminous Type III galaxies ($x_B=1.5x_T$; see Table \ref{xC_tab}). Thus, similar to Type IIs, the formation of Type III galaxies involves bulge/bar enhancement. However, unlike Type II discs, inner anti-truncated discs do not appear to be structurally disturbed relative to Type I discs. Therefore, the additional bulge and outer disc light does not appear to result from restructuring of inner disc stars. 

The transfer of angular momentum from a bar structure into the disc would cause an increase in disc scale length outside a break radius (i.e. outer disc stars are redistributed to higher radii; \citealp{Minchev2012}). However, this mechanism does not explain the intrinsic {\it increase} in disc luminosity relative to Type I discs. Instead, it would be necessary to invoke additional star formation to build this additional stellar mass. If the progenitor disc was gas rich, outward angular momentum transfer from the bar could lead to an increased gas density at larger radii, and hence yield heightened star formation in the outer disc. Additionally, if disc gas {\it within} the break radius was simultaneously driven inwards by the bar, then the resulting central burst of star formation could explain the increased bulge luminosity. Central or outer starburst scenarios would be easily confirmed via the optical colours of these structural components (i.e. systematically bluer, indicating recent star formation). However, such a multi-band analysis is beyond the scope of the present work.

If secular angular momentum transfer due to bar components is the primary mechanism of disc break formation, then the distinction between Type II and III galaxies may be due to the absence/presence of cold gas in the progenitor disc: a gasless (i.e. quenched) progenitor would result in the redistribution of disc stars, and hence form a Type II disc, while the bar in a gas-rich progenitor may interact primarily with gas, yielding a Type III disc. %However, the difference between Type II and Type III progenitor galaxies may instead be caused by differences

Alternatively, anti-truncated disc formation scenario via merger events has been proposed in \cite{Borlaff2014}. Such a merger event would add mass ($\equiv$luminosity) to the galaxy, and would grow the bulge component ($\propto M^{1{\rm -} 2}$; \citealp{BKolchin2005,vDokkum2010,Hilz2012}). In this paradigm, the outer Type III disc corresponds to a merger remnant structure, while the inner disc represents the surviving progenitor disc (potentially stabilised by the presence of a bar). If brighter galaxies assembled more mass via mergers, then the observation of decreasing $r_{\rm brk}/R_{e,{\rm in}}$ with increasing Type III galaxy luminosity can be understood as a decreasing fraction of the primordial disc surviving increasing mass ratio mergers.

\section{Summary and Conclusions}
In this paper, we have presented detailed decomposition analyses (both bulge-disc and more complex, multi-component models) of $\sim$\,630 Coma cluster galaxies (in the luminosity range $-17>M_g>-22$) using CFHT $i$ band imaging data. As this data is $\gtrsim12\times$ deeper than SDSS, fitting accuracy and reliability was substantially improved relative to studies based on SDSS imaging data. This work focused on early-type galaxies (notably those with outer discs, i.e. S0s). 

The \sersic bulge + exponential disc decomposition analysis previously presented in \citeauthor{Head2014} (\citeyear{Head2014}; `Paper I'), has been extended to a wider range of candidate models including 3-component and/or broken disc models. This has allowed a detailed re-investigation of the $\sim$400 Coma cluster galaxies previously considered to be poorly-described by an archetypal (central) bulge + (outer) disc morphology, in addition to the $\sim$\,200 archetypal S0s in Paper I's analysis sample.

Rigorous model selection testing was implemented to ensure no dissonance exists between galaxy and (best-fit) model structure. We have investigated the structural properties beyond the simple bulge + (exponential) disk morphology, the multi-component structure of classic ellipticals, and the role of galaxy bars in the evolution of disk-dominated galaxies. Furthermore, the properties of broken disk structures (Freeman Types II and III) have been contrasted with the previously-considered (unbroken) exponential disk (Freeman Type I), allowing investigation of the formation mechanisms (and hence evolutionary history) of galaxies containing such structures. 

\noindent
The key conclusions drawn from our analysis sample of 478 reliably-fit Coma galaxies are as follows:

\begin{enumerate}[label=\roman{*}),ref=(\roman{*})]
	\item $48\pm3\%$ of galaxies ($N=230$) are well-described by a simple \Sersic, or \sersic + exponential model, while 3(+) component models are required to describe $42\pm3\%$ of galaxies ($N=201$). Hence, a wide range of complex structures are found for ETGs in Coma.
	\item Disc breaks are detected in $26\pm4\%$ of archetypal (central bulge + outer disc) galaxies, with equal numbers of ‘truncated’ (Freeman Type II; $12^{+3}_{-2}\%$) and ‘anti-truncated’ (Freeman Type III; $13^{+3}_{-2}\%$) discs. This corresponds to a significantly higher truncated disc fraction, and lower anti-truncated disc fraction than has previously been detected for Virgo cluster galaxies.
	\item Multi-component \sersic galaxies were resolved into a compact core (with $n\sim$\,$2$), surrounded by large Gaussian-like structures. The total (combined) half-light radii for these multi-component \sersic galaxies are typically $\sim$\,$11$ kpc. Thus, if these galaxies formed from the compact `red nuggets' detected at high redshifts, then these objects require a factor of $\sim$\,$6\times$ growth in size. 
	\item No significant variation in galaxy morphology or multi-component structure was detected with projected distance from the Coma cluster centre. Therefore, secular processes are responsible for the {\it structural} changes responsible for the formation of broken disc galaxies.
	\item Disc breaks are found overwhelmingly in barred galaxies (Type II: $89^{+5}_{-9}\%$ contain bars; Type III: $71^{+8}_{-10}\%$ contain bars), while the minority of galaxies with unbroken discs also contain bars ($42\pm5\%$). In addition, broken discs (of both types) are structurally correlated with bar size. Galaxy bars therefore play an important role in the formation or stabilisation of Type II and Type III broken discs.
	\item Type II discs may not be physically truncated. Rather, inner disc surface brightness may be suppressed in these structures, while the outer disc approximately preserves the progenitor disc properties. However, Type II disc luminosity trends are steeper than untruncated discs, suggesting luminosity truncation in fainter galaxies. 
	\item Significant growth of bulge size and luminosity implies a bulge enhancement origin (e.g. mergers, starbursts) for Type III galaxies, while the inner disc ($r<r_{\rm brk}$) remains structurally consistent with that of untruncated galaxies. Thus, `anti-truncated' discs are likely to result from either radial redistribution of disc gas due to bars, or (disc-preserving) merger events.
\end{enumerate}

Model selection techniques are biased by the assumption that the set of considered models contains the `true' representation of the underlying data. Here, the detection of genuine broken disc galaxies would have been significantly distorted if only a narrow a range of models are considered. False positive broken disc detection (i.e. the fraction of reported `broken disc' galaxies revealed to have more complex, {\emph{unbroken}} structures via a more detailed analysis) can exceed $50\%$ if 3-component and/or multi-\sersic models are not also considered. Thus, decomposition analyses require a sufficiently broad range of candidate models in order to ensure meaningful results. Accordingly, consideration of models inculding more varied structural components (e.g. Ferrer bars, core-\sersic bulges; \citealp{Graham2003b}) may provide additional insight into the galaxies analysed in the present work. However, this does not compromise the results of this study, which has explored the diversity of galaxy structures via the best fits {\it from the considered range of models}.

\section*{Acknowledgements}
%We thank the anonymous referee for their comments, which have improved the paper.
We thank Russell Smith for helpful discussion during the development of this paper. 
JTCGH was supported by an STFC studentship (ST/I505656/1). JRL is supported by STFC Rolling Grant ST/I001573/1. MJH acknowledges support from NSERC (Canada). 

This work is based on observations obtained with MegaPrime/MegaCam, a joint project of CFHT and CEA/DAPNIA, at the Canada-France-Hawaii Telescope (CFHT) which is operated by the National Research Council (NRC) of Canada, the Institute National des Sciences de l'Univers of the Centre National de la Recherche Scientifique of France, and the University of Hawaii. This work is based in part on data products produced at TERAPIX with the expert assistance of Partick Hudelot and Yannick Mellier. Observational data used in this paper are available from the CFHT archive http://www3.cadc-ccda.hia-iha.nrc-cnrc.gc.ca/cfht/cfht.html

This work uses data from SDSS-III. Funding for SDSS-III has been provided by the Alfred P. Sloan Foundation, the Participating Institutions, the National Science Foundation, and the U.S. Department of Energy Office of Science. The SDSS-III web site is http://www.sdss3.org/.

SDSS-III is managed by the Astrophysical Research Consortium for the Participating Institutions of the SDSS-III Collaboration including the University of Arizona, the Brazilian Participation Group, Brookhaven National Laboratory, Carnegie Mellon University, University of Florida, the French Participation Group, the German Participation Group, Harvard University, the Instituto de Astrofisica de Canarias, the Michigan State/Notre Dame/JINA Participation Group, Johns Hopkins University, Lawrence Berkeley National Laboratory, Max Planck Institute for Astrophysics, Max Planck Institute for Extraterrestrial Physics, New Mexico State University, New York University, Ohio State University, Pennsylvania State University, University of Portsmouth, Princeton University, the Spanish Participation Group, University of Tokyo, University of Utah, Vanderbilt University, University of Virginia, University of Washington, and Yale University.

%-----------------------------------------------------
\bibliographystyle{mn2e}
\bibliography{jtcgh.bib}

\begin{thebibliography}{}

\bibitem[\protect\citeauthoryear{{Allen}, {Driver}, {Graham}, {Cameron},
  {Liske} \& {de Propris}}{{Allen} et~al.}{2006}]{Allen2006}
{Allen} P.~D.,  {Driver} S.~P.,  {Graham} A.~W.,  {Cameron} E.,  {Liske} J.,
  {de Propris} R.,  2006, \mnras, 371, 2

\bibitem[\protect\citeauthoryear{{Arag{\'o}n-Salamanca}}{{Arag{\'o}n-Salamanca}}{2008}]{Aragon2008}
{Arag{\'o}n-Salamanca} A.,  2008, in {M.~Bureau, E.~Athanassoula, \& B.~Barbuy}
  ed., IAU Symposium Vol.~245 of IAU Symposium, {Stellar populations in the
  bulges of S0s and the formation of S0 galaxies}.
pp 285--288

\bibitem[\protect\citeauthoryear{{Barr}, {Bedregal}, {Arag{\'o}n-Salamanca},
  {Merrifield} \& {Bamford}}{{Barr} et~al.}{2007}]{Barr2007}
{Barr} J.~M.,  {Bedregal} A.~G.,  {Arag{\'o}n-Salamanca} A.,  {Merrifield}
  M.~R.,    {Bamford} S.~P.,  2007, \aap, 470, 173

\bibitem[\protect\citeauthoryear{{Barway}, {Wadadekar}, {Kembhavi} \&
  {Mayya}}{{Barway} et~al.}{2009}]{Barway2009}
{Barway} S.,  {Wadadekar} Y.,  {Kembhavi} A.~K.,    {Mayya} Y.~D.,  2009,
  \mnras, 394, 1991

\bibitem[\protect\citeauthoryear{{Borlaff}, {Eliche-Moral},
  {Rodr{\'{\i}}guez-P{\'e}rez}, {Querejeta}, {Tapia}, {P{\'e}rez-Gonz{\'a}lez},
  {Zamorano}, {Gallego} \& {Beckman}}{{Borlaff} et~al.}{2014}]{Borlaff2014}
{Borlaff} A.,  {Eliche-Moral} M.~C.,  {Rodr{\'{\i}}guez-P{\'e}rez} C.,
  {Querejeta} M.,  {Tapia} T.,  {P{\'e}rez-Gonz{\'a}lez} P.~G.,  {Zamorano} J.,
   {Gallego} J.,    {Beckman} J.,  2014, ArXiv e-prints

\bibitem[\protect\citeauthoryear{{Boselli} \& {Gavazzi}}{{Boselli} \&
  {Gavazzi}}{2006}]{Boselli2006}
{Boselli} A.,  {Gavazzi} G.,  2006, \pasp, 118, 517

\bibitem[\protect\citeauthoryear{{Bower}, {Lucey} \& {Ellis}}{{Bower}
  et~al.}{1992}]{BowLucEll}
{Bower} R.~G.,  {Lucey} J.~R.,    {Ellis} R.~S.,  1992, \mnras, 254, 601

\bibitem[\protect\citeauthoryear{{Boylan-Kolchin}, {Ma} \&
  {Quataert}}{{Boylan-Kolchin} et~al.}{2005}]{BKolchin2005}
{Boylan-Kolchin} M.,  {Ma} C.-P.,    {Quataert} E.,  2005, \mnras, 362, 184

\bibitem[\protect\citeauthoryear{{Capaccioli}, {Vietri}, {Held} \&
  {Lorenz}}{{Capaccioli} et~al.}{1991}]{Capaccioli1991}
{Capaccioli} M.,  {Vietri} M.,  {Held} E.~V.,    {Lorenz} H.,  1991, \apj, 371,
  535

\bibitem[\protect\citeauthoryear{{Cappellari}}{{Cappellari}}{2013}]{Cappellari2013}
{Cappellari} M.,  2013, \apjl, 778, L2

\bibitem[\protect\citeauthoryear{{Cappellari}, {Emsellem}, {Krajnovi{\'c}},
  {McDermid}, {Serra}, {Alatalo}, {Blitz}, {Bois}, {Bournaud}, {Bureau},
  {Davies} \& et al.}{{Cappellari} et~al.}{2011}]{Cappellari2011}
{Cappellari} M.,  {Emsellem} E.,  {Krajnovi{\'c}} D.,  {McDermid} R.~M.,
  {Serra} P.,  {Alatalo} K.,  {Blitz} L.,  {Bois} M.,  {Bournaud} F.,  {Bureau}
  M.,  {Davies} R.~L.,    et al. 2011, \mnras, 416, 1680

\bibitem[\protect\citeauthoryear{{Carter}, {Goudfrooij}, {Mobasher},
  {Ferguson}, {Puzia}, {Aguerri}, {Balcells}, {Batcheldor}, {Bridges}, {Davies}
  \& et al.}{{Carter} et~al.}{2008}]{Carter2008}
{Carter} D.,  {Goudfrooij} P.,  {Mobasher} B.,  {Ferguson} H.~C.,  {Puzia}
  T.~H.,  {Aguerri} A.~L.,  {Balcells} M.,  {Batcheldor} D.,  {Bridges} T.~J.,
  {Davies} J.~I.,    et al. 2008, \apjs, 176, 424

\bibitem[\protect\citeauthoryear{{Chilingarian}, {Melchior} \&
  {Zolotukhin}}{{Chilingarian} et~al.}{2010}]{KCorr}
{Chilingarian} I.~V.,  {Melchior} A.-L.,    {Zolotukhin} I.~Y.,  2010, \mnras,
  405, 1409

\bibitem[\protect\citeauthoryear{{Chilingarian} \& {Zolotukhin}}{{Chilingarian}
  \& {Zolotukhin}}{2012}]{Kcorr2}
{Chilingarian} I.~V.,  {Zolotukhin} I.~Y.,  2012, \mnras, 419, 1727

\bibitem[\protect\citeauthoryear{{Damjanov}, {McCarthy}, {Abraham},
  {Glazebrook}, {Yan}, {Mentuch}, {Le Borgne}, {Savaglio}, {Crampton},
  {Murowinski}, {Juneau}, {Carlberg}, {J{\o}rgensen}, {Roth}, {Chen} \&
  {Marzke}}{{Damjanov} et~al.}{2009}]{Damjanov2009}
{Damjanov} I.,  {McCarthy} P.~J.,  {Abraham} R.~G.,  {Glazebrook} K.,  {Yan}
  H.,  {Mentuch} E.,  {Le Borgne} D.,  {Savaglio} S.,  {Crampton} D.,
  {Murowinski} R.,  {Juneau} S.,  {Carlberg} R.~G.,  {J{\o}rgensen} I.,  {Roth}
  K.,  {Chen} H.-W.,    {Marzke} R.~O.,  2009, \apj, 695, 101

\bibitem[\protect\citeauthoryear{{De Lucia}, {Weinmann}, {Poggianti},
  {Arag{\'o}n-Salamanca} \& {Zaritsky}}{{De Lucia} et~al.}{2012}]{DeLucia2012}
{De Lucia} G.,  {Weinmann} S.,  {Poggianti} B.~M.,  {Arag{\'o}n-Salamanca} A.,
    {Zaritsky} D.,  2012, \mnras, 423, 1277

\bibitem[\protect\citeauthoryear{{de Vaucouleurs}}{{de
  Vaucouleurs}}{1948}]{dVauc}
{de Vaucouleurs} G.,  1948, Annales d'Astrophysique, 11, 247

\bibitem[\protect\citeauthoryear{{Debattista}, {Mayer}, {Carollo}, {Moore},
  {Wadsley} \& {Quinn}}{{Debattista} et~al.}{2006}]{Debattista2006}
{Debattista} V.~P.,  {Mayer} L.,  {Carollo} C.~M.,  {Moore} B.,  {Wadsley} J.,
    {Quinn} T.,  2006, \apj, 645, 209

\bibitem[\protect\citeauthoryear{{Dressler}}{{Dressler}}{1980}]{Dressler1980}
{Dressler} A.,  1980, \apj, 236, 351

\bibitem[\protect\citeauthoryear{{Dressler} \& {Gunn}}{{Dressler} \&
  {Gunn}}{1983}]{Dressler1983}
{Dressler} A.,  {Gunn} J.~E.,  1983, \apj, 270, 7

\bibitem[\protect\citeauthoryear{{Driver}, {Popescu}, {Tuffs}, {Liske},
  {Graham}, {Allen} \& {de Propris}}{{Driver} et~al.}{2007}]{Driver2007}
{Driver} S.~P.,  {Popescu} C.~C.,  {Tuffs} R.~J.,  {Liske} J.,  {Graham} A.~W.,
   {Allen} P.~D.,    {de Propris} R.,  2007, \mnras, 379, 1022

\bibitem[\protect\citeauthoryear{{Dullo} \& {Graham}}{{Dullo} \&
  {Graham}}{2014}]{Dullo2014}
{Dullo} B.~T.,  {Graham} A.~W.,  2014, \mnras, 444, 2700

\bibitem[\protect\citeauthoryear{{Eliche-Moral}, {Gonz{\'a}lez-Garc{\'{\i}}a},
  {Aguerri}, {Gallego}, {Zamorano}, {Balcells} \& {Prieto}}{{Eliche-Moral}
  et~al.}{2013}]{EMoral2013}
{Eliche-Moral} M.~C.,  {Gonz{\'a}lez-Garc{\'{\i}}a} A.~C.,  {Aguerri} J.~A.~L.,
   {Gallego} J.,  {Zamorano} J.,  {Balcells} M.,    {Prieto} M.,  2013, \aap,
  552, A67

\bibitem[\protect\citeauthoryear{{Emsellem}, {Cappellari}, {Krajnovi{\'c}},
  {Alatalo}, {Blitz}, {Bois}, {Bournaud}, {Bureau}, {Davies}, {Davis}, {de
  Zeeuw} \& et al.}{{Emsellem} et~al.}{2011}]{Emsellem2011}
{Emsellem} E.,  {Cappellari} M.,  {Krajnovi{\'c}} D.,  {Alatalo} K.,  {Blitz}
  L.,  {Bois} M.,  {Bournaud} F.,  {Bureau} M.,  {Davies} R.~L.,  {Davis}
  T.~A.,  {de Zeeuw} P.~T.,    et al. 2011, \mnras, 414, 888

\bibitem[\protect\citeauthoryear{{Erwin}, {Guti{\'e}rrez} \& {Beckman}}{{Erwin}
  et~al.}{2012}]{Erwin2012}
{Erwin} P.,  {Guti{\'e}rrez} L.,    {Beckman} J.~E.,  2012, \apjl, 744, L11

\bibitem[\protect\citeauthoryear{{Erwin}, {Pohlen} \& {Beckman}}{{Erwin}
  et~al.}{2008}]{Erwin2008}
{Erwin} P.,  {Pohlen} M.,    {Beckman} J.~E.,  2008, \aj, 135, 20

\bibitem[\protect\citeauthoryear{{Foyle}, {Courteau} \& {Thacker}}{{Foyle}
  et~al.}{2008}]{Foyle2008}
{Foyle} K.,  {Courteau} S.,    {Thacker} R.~J.,  2008, \mnras, 386, 1821

\bibitem[\protect\citeauthoryear{{Freeman}}{{Freeman}}{1970}]{Freeman1970}
{Freeman} K.~C.,  1970, \apj, 160, 811

\bibitem[\protect\citeauthoryear{{Gao}, {De Lucia}, {White} \& {Jenkins}}{{Gao}
  et~al.}{2004}]{Gao2004}
{Gao} L.,  {De Lucia} G.,  {White} S.~D.~M.,    {Jenkins} A.,  2004, \mnras,
  352, L1

\bibitem[\protect\citeauthoryear{{Gavazzi}}{{Gavazzi}}{1989}]{Gavazzi1989}
{Gavazzi} G.,  1989, \apj, 346, 59

\bibitem[\protect\citeauthoryear{{Gavazzi}, {Fumagalli}, {Cucciati} \&
  {Boselli}}{{Gavazzi} et~al.}{2010}]{Gavazzi2010}
{Gavazzi} G.,  {Fumagalli} M.,  {Cucciati} O.,    {Boselli} A.,  2010, \aap,
  517, A73

\bibitem[\protect\citeauthoryear{{Gavazzi}, {Adami}, {Durret}, {Cuillandre},
  {Ilbert}, {Mazure}, {Pell{\'o}} \& {Ulmer}}{{Gavazzi}
  et~al.}{2009}]{Gavazzi2009}
{Gavazzi} R.,  {Adami} C.,  {Durret} F.,  {Cuillandre} J.-C.,  {Ilbert} O.,
  {Mazure} A.,  {Pell{\'o}} R.,    {Ulmer} M.~P.,  2009, \aap, 498, L33

\bibitem[\protect\citeauthoryear{{Graham}, {Erwin}, {Trujillo} \& {Asensio
  Ramos}}{{Graham} et~al.}{2003}]{Graham2003b}
{Graham} A.~W.,  {Erwin} P.,  {Trujillo} I.,    {Asensio Ramos} A.,  2003, \aj,
  125, 2951

\bibitem[\protect\citeauthoryear{{Guzman}, {Lucey}, {Carter} \&
  {Terlevich}}{{Guzman} et~al.}{1992}]{Guzman1992}
{Guzman} R.,  {Lucey} J.~R.,  {Carter} D.,    {Terlevich} R.~J.,  1992, \mnras,
  257, 187

\bibitem[\protect\citeauthoryear{{H{\"a}ussler}, {McIntosh}, {Barden}, {Bell},
  {Rix}, {Borch}, {Beckwith}, {Caldwell}, {Heymans}, {Jahnke}, {Jogee},
  {Koposov}, {Meisenheimer}, {S{\'a}nchez}, {Somerville}, {Wisotzki} \&
  {Wolf}}{{H{\"a}ussler} et~al.}{2007}]{Haussler2007}
{H{\"a}ussler} B.,  {McIntosh} D.~H.,  {Barden} M.,  {Bell} E.~F.,  {Rix}
  H.-W.,  {Borch} A.,  {Beckwith} S.~V.~W.,  {Caldwell} J.~A.~R.,  {Heymans}
  C.,  {Jahnke} K.,  {Jogee} S.,  {Koposov} S.~E.,  {Meisenheimer} K.,
  {S{\'a}nchez} S.~F.,  {Somerville} R.~S.,  {Wisotzki} L.,    {Wolf} C.,
  2007, \apjs, 172, 615

\bibitem[\protect\citeauthoryear{{Head}}{{Head}}{2014}]{HeadThesis}
{Head} J.~T.~C.~G.,  2014, PhD thesis, Durham University

\bibitem[\protect\citeauthoryear{{Head}, {Lucey}, {Hudson} \& {Smith}}{{Head}
  et~al.}{2014}]{Head2014}
{Head} J.~T.~C.~G.,  {Lucey} J.~R.,  {Hudson} M.~J.,    {Smith} R.~J.,  2014,
  \mnras, 440, 1690

\bibitem[\protect\citeauthoryear{{Hilz}, {Naab}, {Ostriker}, {Thomas},
  {Burkert} \& {Jesseit}}{{Hilz} et~al.}{2012}]{Hilz2012}
{Hilz} M.,  {Naab} T.,  {Ostriker} J.~P.,  {Thomas} J.,  {Burkert} A.,
  {Jesseit} R.,  2012, \mnras, 425, 3119

\bibitem[\protect\citeauthoryear{{Homeier}, {Postman}, {Menanteau},
  {Blakeslee}, {Mei}, {Demarco}, {Ford}, {Illingworth} \& {Zirm}}{{Homeier}
  et~al.}{2006}]{Homeier2006}
{Homeier} N.~L.,  {Postman} M.,  {Menanteau} F.,  {Blakeslee} J.~P.,  {Mei} S.,
   {Demarco} R.,  {Ford} H.~C.,  {Illingworth} G.~D.,    {Zirm} A.,  2006, \aj,
  131, 143

\bibitem[\protect\citeauthoryear{{Huang}, {Ho}, {Peng}, {Li} \&
  {Barth}}{{Huang} et~al.}{2013}]{Huang2013a}
{Huang} S.,  {Ho} L.~C.,  {Peng} C.~Y.,  {Li} Z.-Y.,    {Barth} A.~J.,  2013,
  \apj, 766, 47

\bibitem[\protect\citeauthoryear{{Hudson}, {Stevenson}, {Smith}, {Wegner},
  {Lucey} \& {Simard}}{{Hudson} et~al.}{2010}]{Hudson2010}
{Hudson} M.~J.,  {Stevenson} J.~B.,  {Smith} R.~J.,  {Wegner} G.~A.,  {Lucey}
  J.~R.,    {Simard} L.,  2010, \mnras, 409, 405

\bibitem[\protect\citeauthoryear{{Janz}, {Laurikainen}, {Lisker}, {Salo},
  {Peletier}, {Niemi}, {den Brok}, {Toloba}, {Falc{\'o}n-Barroso}, {Boselli} \&
  {Hensler}}{{Janz} et~al.}{2012}]{Janz2012}
{Janz} J.,  {Laurikainen} E.,  {Lisker} T.,  {Salo} H.,  {Peletier} R.~F.,
  {Niemi} S.-M.,  {den Brok} M.,  {Toloba} E.,  {Falc{\'o}n-Barroso} J.,
  {Boselli} A.,    {Hensler} G.,  2012, \apjl, 745, L24

\bibitem[\protect\citeauthoryear{{Janz}, {Laurikainen}, {Lisker}, {Salo},
  {Peletier}, {Niemi}, {Toloba}, {Hensler}, {Falc{\'o}n-Barroso}, {Boselli},
  {den Brok}, {Hansson}, {Meyer}, {Ry{\'s}} \& {Paudel}}{{Janz}
  et~al.}{2014}]{Janz2014}
{Janz} J.,  {Laurikainen} E.,  {Lisker} T.,  {Salo} H.,  {Peletier} R.~F.,
  {Niemi} S.-M.,  {Toloba} E.,  {Hensler} G.,  {Falc{\'o}n-Barroso} J.,
  {Boselli} A.,  {den Brok} M.,  {Hansson} K.~S.~A.,  {Meyer} H.~T.,  {Ry{\'s}}
  A.,    {Paudel} S.,  2014, \apj, 786, 105

\bibitem[\protect\citeauthoryear{{J{\o}rgensen}}{{J{\o}rgensen}}{1999}]{Jorgensen1999}
{J{\o}rgensen} I.,  1999, \mnras, 306, 607

\bibitem[\protect\citeauthoryear{{J{\o}rgensen} \& {Franx}}{{J{\o}rgensen} \&
  {Franx}}{1994}]{Jorgensen1994}
{J{\o}rgensen} I.,  {Franx} M.,  1994, \apj, 433, 553

\bibitem[\protect\citeauthoryear{{Kaviraj}, {Ting}, {Bureau}, {Shabala},
  {Crockett}, {Silk}, {Lintott}, {Smith}, {Keel}, {Masters}, {Schawinski} \&
  {Bamford}}{{Kaviraj} et~al.}{2012}]{Kaviraj2012}
{Kaviraj} S.,  {Ting} Y.-S.,  {Bureau} M.,  {Shabala} S.~S.,  {Crockett} R.~M.,
   {Silk} J.,  {Lintott} C.,  {Smith} A.,  {Keel} W.~C.,  {Masters} K.~L.,
  {Schawinski} K.,    {Bamford} S.~P.,  2012, \mnras, 423, 49

\bibitem[\protect\citeauthoryear{{Kent}}{{Kent}}{1985}]{Kent1985}
{Kent} S.~M.,  1985, \apjs, 59, 115

\bibitem[\protect\citeauthoryear{{Komatsu}, {Smith}, {Dunkley}, {Bennett},
  {Gold}, {Hinshaw}, {Jarosik}, {Larson}, {Nolta}, {Page}, {Spergel},
  {Halpern}, {Hill} \& {Kogut}}{{Komatsu} et~al.}{2011}]{WMAP7}
{Komatsu} E.,  {Smith} K.~M.,  {Dunkley} J.,  {Bennett} C.~L.,  {Gold} B.,
  {Hinshaw} G.,  {Jarosik} N.,  {Larson} D.,  {Nolta} M.~R.,  {Page} L.,
  {Spergel} D.~N.,  {Halpern} M.,  {Hill} R.~S.,    {Kogut} A.,  2011, \apjs,
  192, 18

\bibitem[\protect\citeauthoryear{{Kormendy} \& {Bender}}{{Kormendy} \&
  {Bender}}{2012}]{Kormendy2012}
{Kormendy} J.,  {Bender} R.,  2012, \apjs, 198, 2

\bibitem[\protect\citeauthoryear{{Laine}, {Laurikainen}, {Salo}, {Comer{\'o}n},
  {Buta}, {Zaritsky}, {Athanassoula}, {Bosma}, {Mu{\~n}oz-Mateos}, {Gadotti} \&
  et al.}{{Laine} et~al.}{2014}]{Laine2014}
{Laine} J.,  {Laurikainen} E.,  {Salo} H.,  {Comer{\'o}n} S.,  {Buta} R.~J.,
  {Zaritsky} D.,  {Athanassoula} E.,  {Bosma} A.,  {Mu{\~n}oz-Mateos} J.-C.,
  {Gadotti} D.~A.,    et al. 2014, \mnras, 441, 1992

\bibitem[\protect\citeauthoryear{{Lansbury}, {Lucey} \& {Smith}}{{Lansbury}
  et~al.}{2014}]{Lansbury2014}
{Lansbury} G.~B.,  {Lucey} J.~R.,    {Smith} R.~J.,  2014, \mnras, 439, 1749

\bibitem[\protect\citeauthoryear{{Laurikainen}, {Salo} \& {Buta}}{{Laurikainen}
  et~al.}{2005}]{Laurikainen2005}
{Laurikainen} E.,  {Salo} H.,    {Buta} R.,  2005, \mnras, 362, 1319

\bibitem[\protect\citeauthoryear{{Lucey}, {Guzman}, {Carter} \&
  {Terlevich}}{{Lucey} et~al.}{1991}]{Lucey1991}
{Lucey} J.~R.,  {Guzman} R.,  {Carter} D.,    {Terlevich} R.~J.,  1991, \mnras,
  253, 584

\bibitem[\protect\citeauthoryear{{Michard}}{{Michard}}{1985}]{Michard1985}
{Michard} R.,  1985, \aaps, 59, 205

\bibitem[\protect\citeauthoryear{{Minchev}, {Famaey}, {Quillen}, {Di Matteo},
  {Combes}, {Vlaji{\'c}}, {Erwin} \& {Bland-Hawthorn}}{{Minchev}
  et~al.}{2012}]{Minchev2012}
{Minchev} I.,  {Famaey} B.,  {Quillen} A.~C.,  {Di Matteo} P.,  {Combes} F.,
  {Vlaji{\'c}} M.,  {Erwin} P.,    {Bland-Hawthorn} J.,  2012, \aap, 548, A126

\bibitem[\protect\citeauthoryear{{M{\"o}llenhoff}, {Popescu} \&
  {Tuffs}}{{M{\"o}llenhoff} et~al.}{2006}]{Mollenhoff2006}
{M{\"o}llenhoff} C.,  {Popescu} C.~C.,    {Tuffs} R.~J.,  2006, \aap, 456, 941

\bibitem[\protect\citeauthoryear{{Pastrav}, {Popescu}, {Tuffs} \&
  {Sansom}}{{Pastrav} et~al.}{2013}]{Pastrav2013}
{Pastrav} B.~A.,  {Popescu} C.~C.,  {Tuffs} R.~J.,    {Sansom} A.~E.,  2013,
  \aap, 557, A137

\bibitem[\protect\citeauthoryear{{Peng}, {Ho}, {Impey} \& {Rix}}{{Peng}
  et~al.}{2010}]{GALFIT}
{Peng} C.~Y.,  {Ho} L.~C.,  {Impey} C.~D.,    {Rix} H.,  2010, \aj, 139, 2097

\bibitem[\protect\citeauthoryear{{Querejeta}, {Eliche-Moral}, {Tapia},
  {Borlaff}, {Rodr{\'{\i}}guez-P{\'e}rez}, {Zamorano} \& {Gallego}}{{Querejeta}
  et~al.}{2014}]{Querejeta2014}
{Querejeta} M.,  {Eliche-Moral} M.~C.,  {Tapia} T.,  {Borlaff} A.,
  {Rodr{\'{\i}}guez-P{\'e}rez} C.,  {Zamorano} J.,    {Gallego} J.,  2014,
  ArXiv e-prints

\bibitem[\protect\citeauthoryear{{Rawle}, {Lucey}, {Smith} \& {Head}}{{Rawle}
  et~al.}{2013}]{Rawle2013}
{Rawle} T.~D.,  {Lucey} J.~R.,  {Smith} R.~J.,    {Head} J.~T.~C.~G.,  2013,
  \mnras, 433, 2667

\bibitem[\protect\citeauthoryear{{Rix} \& {White}}{{Rix} \&
  {White}}{1990}]{Rix1990}
{Rix} H.-W.,  {White} S.~D.~M.,  1990, \apj, 362, 52

\bibitem[\protect\citeauthoryear{{Roediger}, {Courteau},
  {S{\'a}nchez-Bl{\'a}zquez} \& {McDonald}}{{Roediger}
  et~al.}{2012}]{Roediger2012}
{Roediger} J.~C.,  {Courteau} S.,  {S{\'a}nchez-Bl{\'a}zquez} P.,    {McDonald}
  M.,  2012, \apj, 758, 41

\bibitem[\protect\citeauthoryear{{Schlafly} \& {Finkbeiner}}{{Schlafly} \&
  {Finkbeiner}}{2011}]{Schlafly2011}
{Schlafly} E.~F.,  {Finkbeiner} D.~P.,  2011, \apj, 737, 103

\bibitem[\protect\citeauthoryear{Schwarz}{Schwarz}{1978}]{BIC}
Schwarz G.,  1978, The annals of statistics, 6, 461

\bibitem[\protect\citeauthoryear{{S{\'e}rsic}}{{S{\'e}rsic}}{1963}]{Sersic}
{S{\'e}rsic} J.~L.,  1963, Boletin de la Asociacion Argentina de Astronomia La
  Plata Argentina, 6, 41

\bibitem[\protect\citeauthoryear{{Shapley}}{{Shapley}}{1934}]{Shapley1934}
{Shapley} H.,  1934, Harvard College Observatory Bulletin, 896, 3

\bibitem[\protect\citeauthoryear{{Smith}, {Lucey} \& {Hudson}}{{Smith}
  et~al.}{2009}]{Smith2009}
{Smith} R.~J.,  {Lucey} J.~R.,    {Hudson} M.~J.,  2009, \mnras, 400, 1690

\bibitem[\protect\citeauthoryear{{Smith}, {Lucey}, {Price}, {Hudson} \&
  {Phillipps}}{{Smith} et~al.}{2012}]{Smith2012}
{Smith} R.~J.,  {Lucey} J.~R.,  {Price} J.,  {Hudson} M.~J.,    {Phillipps} S.,
   2012, \mnras, 419, 3167

\bibitem[\protect\citeauthoryear{{Struble} \& {Rood}}{{Struble} \&
  {Rood}}{1999}]{Struble1999}
{Struble} M.~F.,  {Rood} H.~J.,  1999, \apjs, 125, 35

\bibitem[\protect\citeauthoryear{{Taranu}, {Hudson}, {Balogh}, {Smith},
  {Power}, {Oman} \& {Krane}}{{Taranu} et~al.}{2014}]{Taranu2012}
{Taranu} D.~S.,  {Hudson} M.~J.,  {Balogh} M.~L.,  {Smith} R.~J.,  {Power} C.,
  {Oman} K.~A.,    {Krane} B.,  2014, \mnras, 440, 1934

\bibitem[\protect\citeauthoryear{{Valenzuela} \& {Klypin}}{{Valenzuela} \&
  {Klypin}}{2003}]{Valenzuela2003}
{Valenzuela} O.,  {Klypin} A.,  2003, \mnras, 345, 406

\bibitem[\protect\citeauthoryear{{van den Bergh}}{{van den
  Bergh}}{1976}]{vdBergh1976}
{van den Bergh} S.,  1976, \apj, 206, 883

\bibitem[\protect\citeauthoryear{{van den Bergh}}{{van den
  Bergh}}{1990}]{vdBergh1990}
{van den Bergh} S.,  1990, \apj, 348, 57

\bibitem[\protect\citeauthoryear{{van den Bergh}}{{van den
  Bergh}}{2009a}]{vdBergh2009a}
{van den Bergh} S.,  2009a, \apj, 702, 1502

\bibitem[\protect\citeauthoryear{{van den Bergh}}{{van den
  Bergh}}{2009b}]{vdBergh2009b}
{van den Bergh} S.,  2009b, \apjl, 694, L120

\bibitem[\protect\citeauthoryear{{van Dokkum}, {Abraham}, {Merritt}, {Zhang},
  {Geha} \& {Conroy}}{{van Dokkum} et~al.}{2014}]{vDokkum2014b}
{van Dokkum} P.,  {Abraham} R.,  {Merritt} A.,  {Zhang} J.,  {Geha} M.,
  {Conroy} C.,  2014, ArXiv e-prints

\bibitem[\protect\citeauthoryear{{van Dokkum}, {Bezanson}, {van der Wel},
  {Nelson}, {Momcheva}, {Skelton}, {Whitaker}, {Brammer}, {Conroy},
  {F{\"o}rster Schreiber} \& et al.}{{van Dokkum} et~al.}{2014}]{vDokkum2014}
{van Dokkum} P.~G.,  {Bezanson} R.,  {van der Wel} A.,  {Nelson} E.~J.,
  {Momcheva} I.,  {Skelton} R.~E.,  {Whitaker} K.~E.,  {Brammer} G.,  {Conroy}
  C.,  {F{\"o}rster Schreiber} N.~M.,    et al. 2014, \apj, 791, 45

\bibitem[\protect\citeauthoryear{{van Dokkum}, {Whitaker}, {Brammer}, {Franx},
  {Kriek}, {Labb{\'e}}, {Marchesini}, {Quadri}, {Bezanson}, {Illingworth},
  {Muzzin}, {Rudnick}, {Tal} \& {Wake}}{{van Dokkum}
  et~al.}{2010}]{vDokkum2010}
{van Dokkum} P.~G.,  {Whitaker} K.~E.,  {Brammer} G.,  {Franx} M.,  {Kriek} M.,
   {Labb{\'e}} I.,  {Marchesini} D.,  {Quadri} R.,  {Bezanson} R.,
  {Illingworth} G.~D.,  {Muzzin} A.,  {Rudnick} G.,  {Tal} T.,    {Wake} D.,
  2010, \apj, 709, 1018

\bibitem[\protect\citeauthoryear{{Weinzirl}, {Jogee}, {Neistein}, {Khochfar},
  {Kormendy}, {Marinova}, {Hoyos}, {Balcells}, {den Brok}, {Hammer},
  {Peletier}, {Kleijn}, {Carter}, {Goudfrooij}, {Lucey} \& et al.}{{Weinzirl}
  et~al.}{2014}]{Weinzirl2013}
{Weinzirl} T.,  {Jogee} S.,  {Neistein} E.,  {Khochfar} S.,  {Kormendy} J.,
  {Marinova} I.,  {Hoyos} C.,  {Balcells} M.,  {den Brok} M.,  {Hammer} D.,
  {Peletier} R.~F.,  {Kleijn} G.~V.,  {Carter} D.,  {Goudfrooij} P.,  {Lucey}
  J.~R.,    et al. 2014, \mnras, 441, 3083

\bibitem[\protect\citeauthoryear{{Wolf}}{{Wolf}}{1902}]{Wolf1902}
{Wolf} M.,  1902, Publikationen des Astrophysikalischen Instituts
  Koenigstuhl-Heidelberg, 1, 125

\bibitem[\protect\citeauthoryear{{Younger}, {Cox}, {Seth} \&
  {Hernquist}}{{Younger} et~al.}{2007}]{Younger2007}
{Younger} J.~D.,  {Cox} T.~J.,  {Seth} A.~C.,    {Hernquist} L.,  2007, \apj,
  670, 269

\end{thebibliography}
%----------------------------------------------------

\appendix
\section{Model Type Examples}\label{mod_eg}
\begin{figure}
\begin{center}
	\includegraphics[width=.99\linewidth,clip=true]{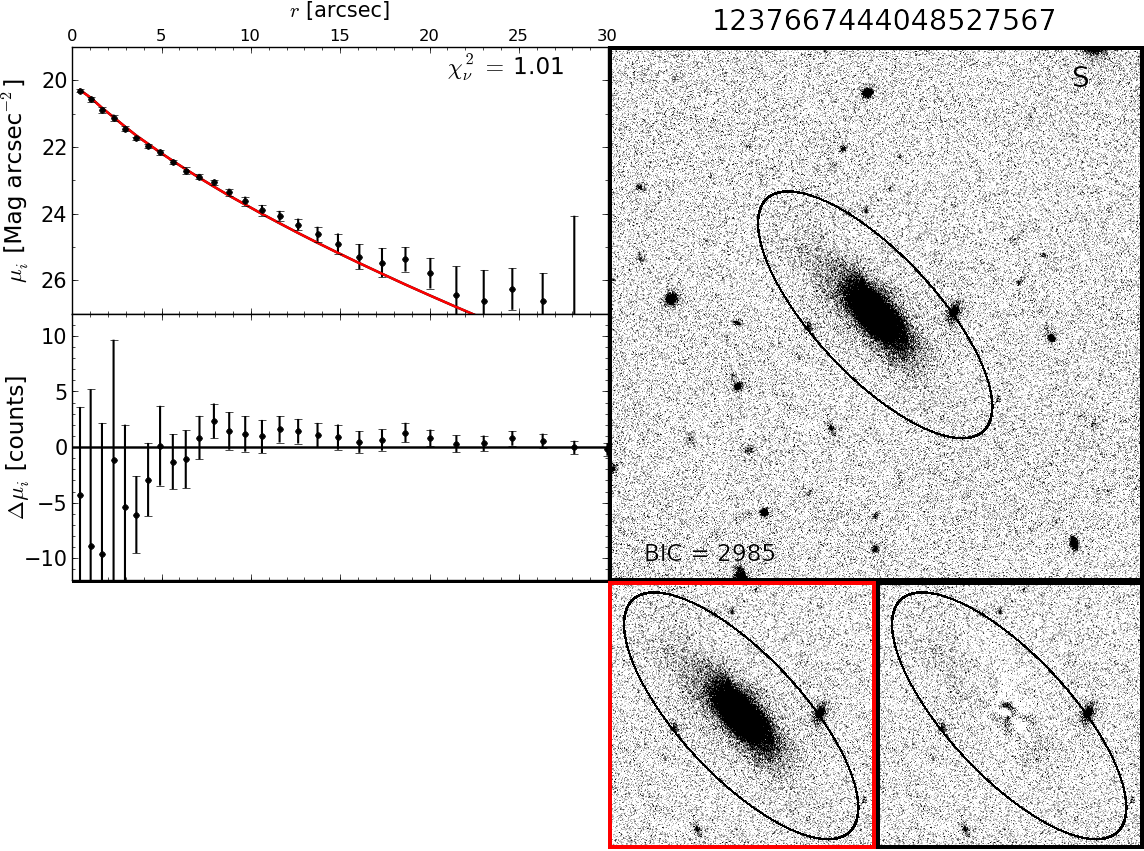}
	\includegraphics[width=.99\linewidth,clip=true]{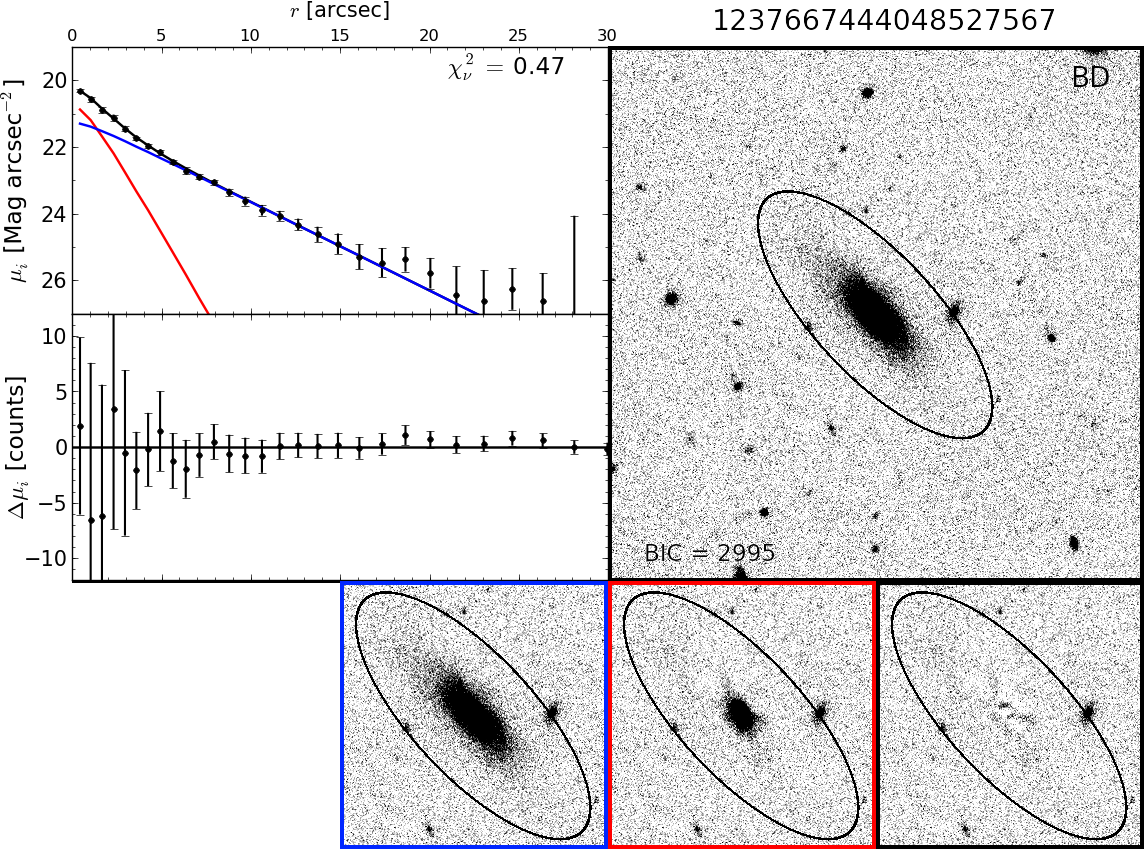}
\end{center}
\caption[Surface brightness profiles, residuals, and galaxy thumbnails for an example S galaxy.]{An example galaxy best fit by an {\it S} model (DR8 ObjID 1237667444048527567): Surface brightness profiles ($\mu_i$), residuals ($\Delta\mu_i$), and $i$ band thumbnails for the {\it S} model ({\bf top:} \sersic = red), and the corresponding {\it BD} model ({\bf bottom:} bulge = red, disc = blue). Small images depict isolated model components (border colours $\equiv\mu_i$ plot), and the total residual (black borders). The target ellipse is noted in black in all thumbnails, and 1D $\chi_{\nu}^2$ (major axis) and 2D BIC values are included for both models.}
\label{egB}
\end{figure}

Figure \ref{egB} presents an illustrative example galaxy best fit by a \Sersic-only model (upper panels), and the corresponding (overfit) bulge + disc model (lower panels). Surface brightness plots ($\mu_i$) and model residuals (image $-$ model in counts) are included for both model fits (top left), as measured from the galaxy and model thumbnails in wedges of elliptical annuli (angle ${\rm cos}^{-1}(e^2)$, where $e$ is the eccentricity of the galaxy's target ellipse). The $i$ band residual images (including only the central quarters) are presented in the bottom right corners (black border) for each model fit. In addition, component residual images (i.e. the residual image after all model components {\emph{except}} the target component are subtracted) are included along the bottom in panels bordered by their $\mu_i$ plot line colours (i.e. red and blue for \sersic and exponential components respectively). Here, the addition of a disc component improves the goodness of fit (lower $\chi_{\nu}^2$), but this improvement is not statistically significant given the increased number of fitting parameters (increased BIC).

\begin{figure}
\begin{center}
	\includegraphics[width=.99\linewidth,clip=true]{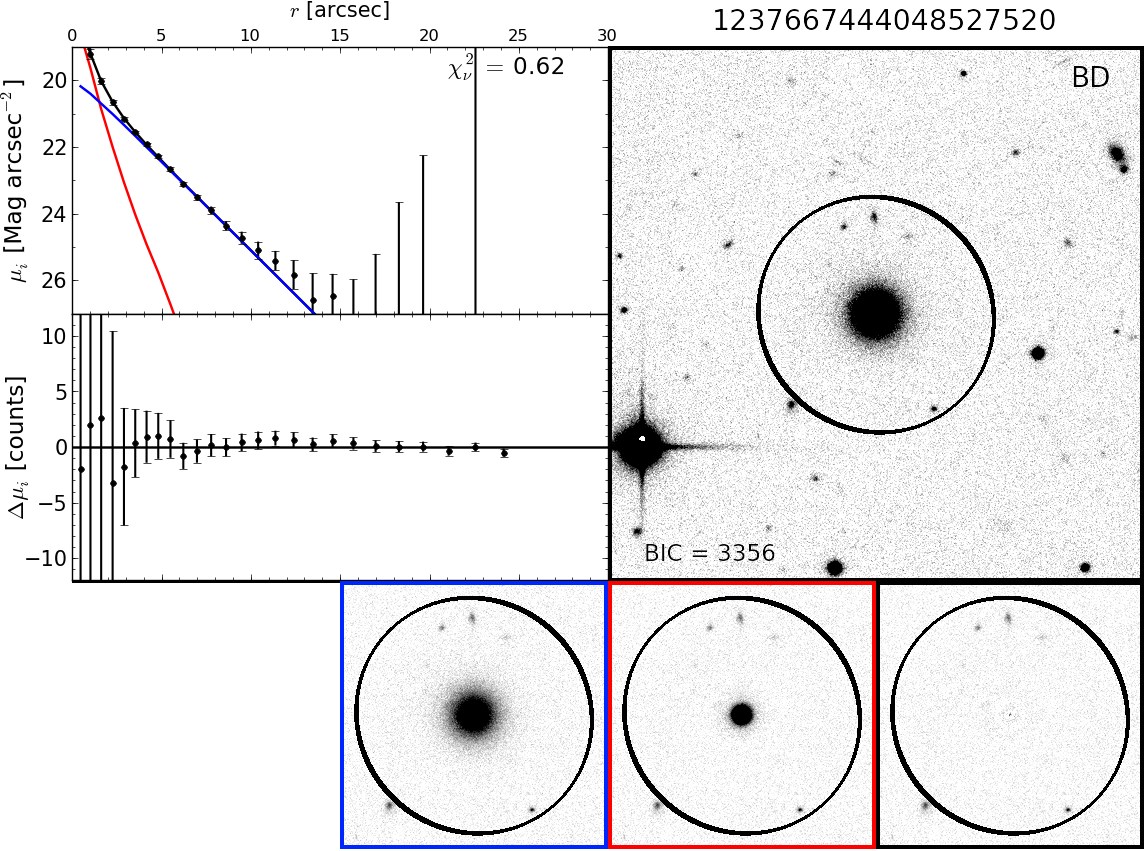}
	\includegraphics[width=.99\linewidth,clip=true]{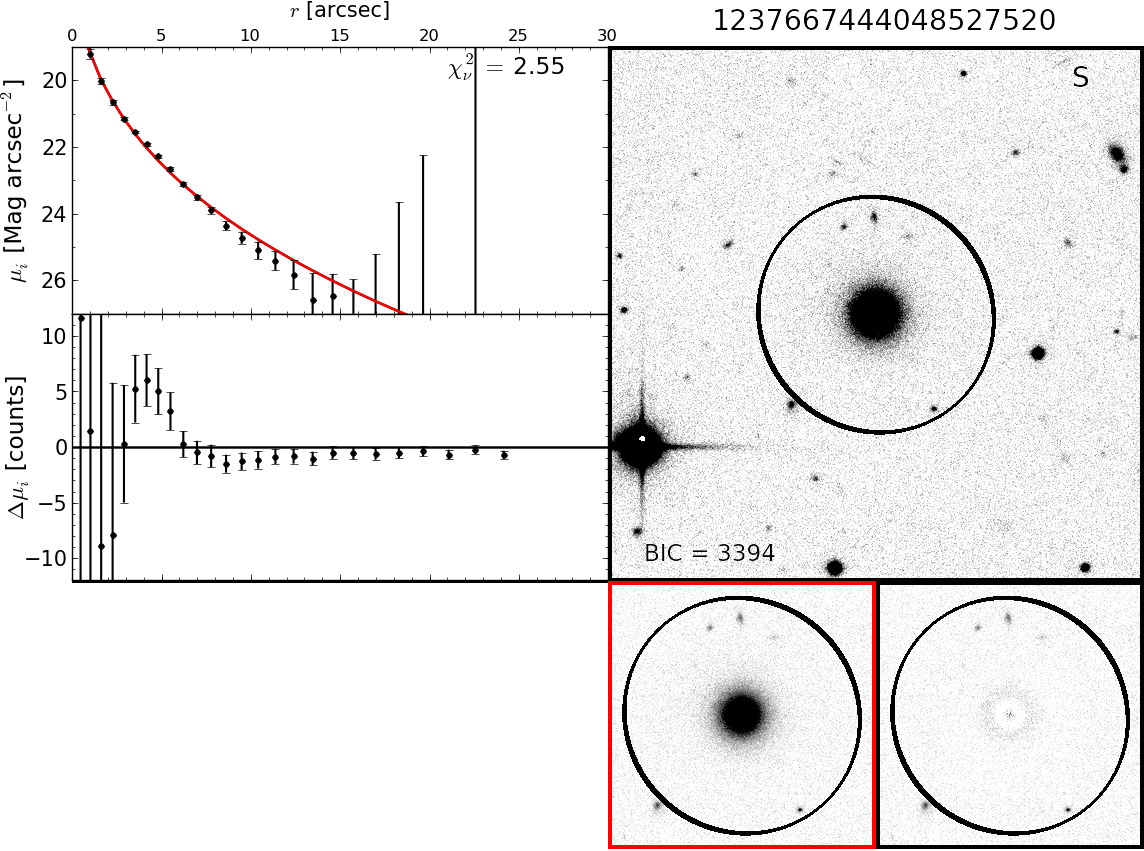}
\end{center}
\caption[Surface brightness profiles, residuals, and galaxy thumbnails for an example BD galaxy.]{An example galaxy best fit by a {\it BD} model (DR8 ObjID 1237667444048527520). As Figure \ref{egB} for a {\it BD} model ({\bf top:} bulge = red, disc = blue), and the corresponding {\it S} model ({\bf bottom:} \sersic = red).}
\label{egBD}
\end{figure}

\begin{figure}
\begin{center}
	\includegraphics[width=.99\linewidth,clip=true]{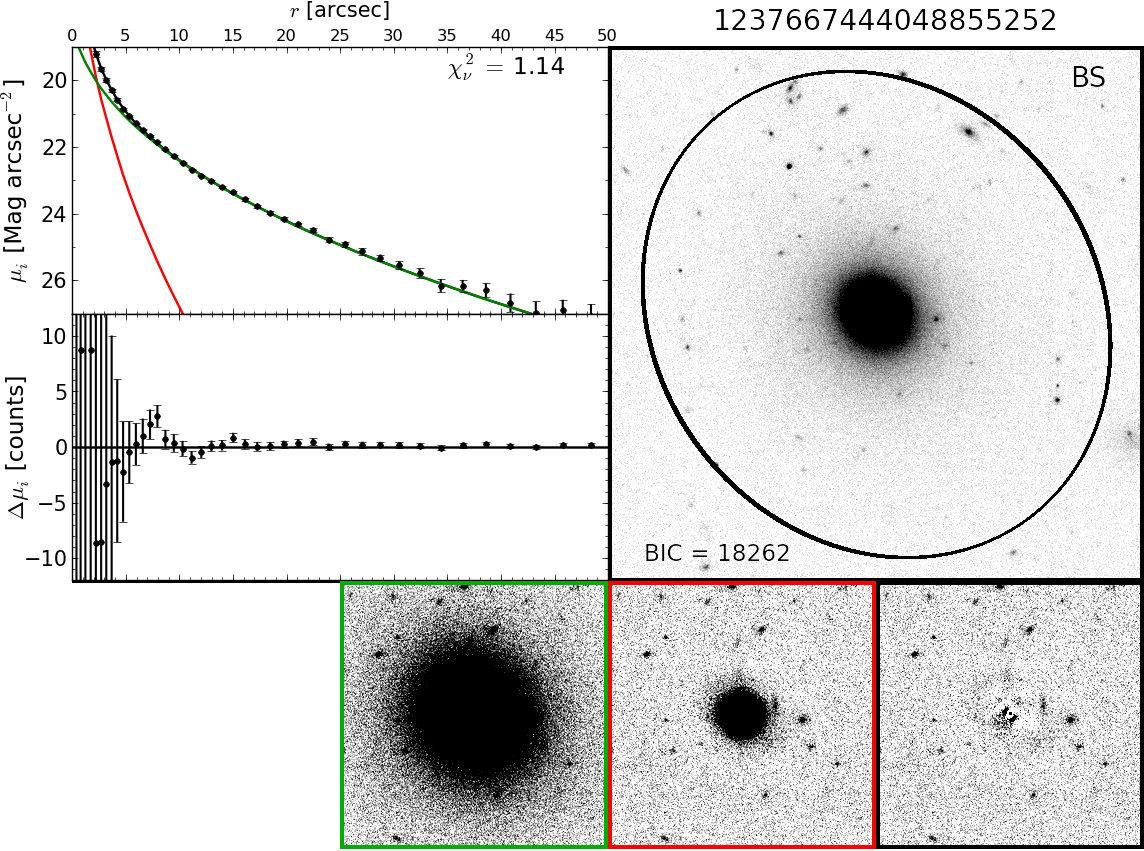}
	\includegraphics[width=.99\linewidth,clip=true]{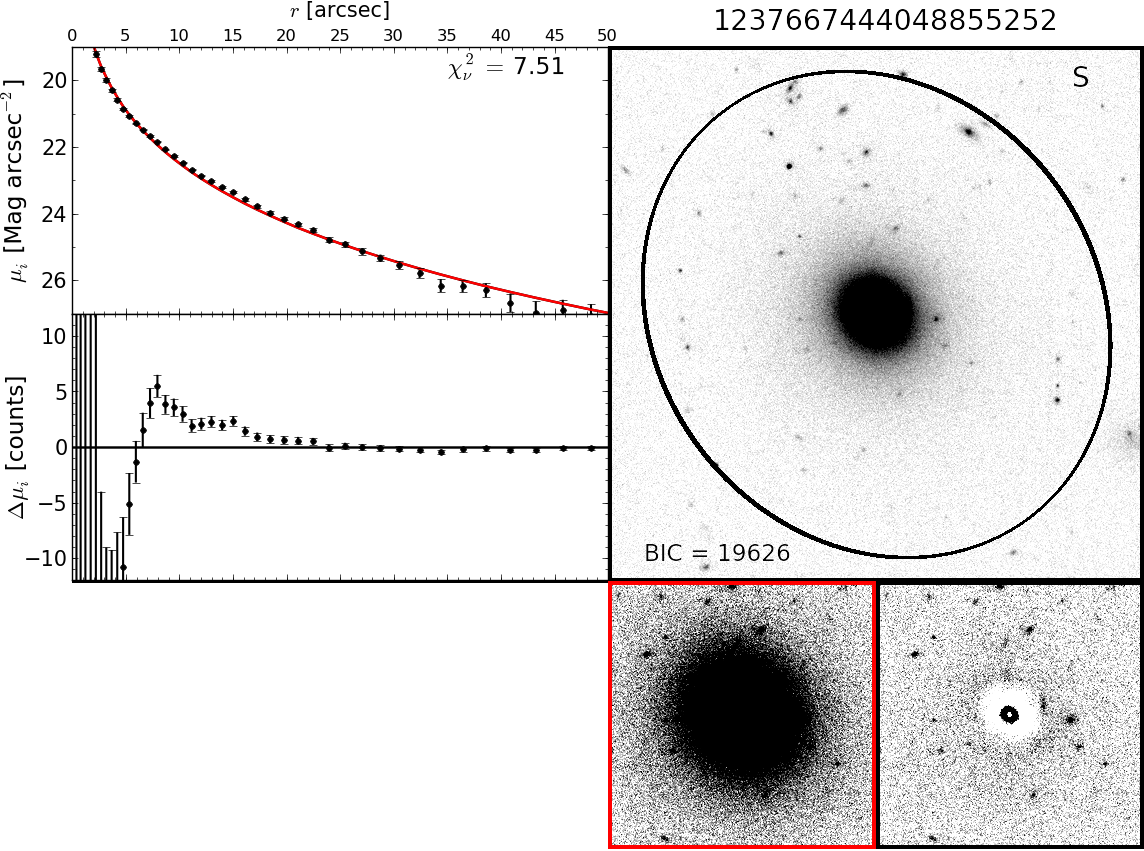}
\end{center}
\caption[Surface brightness profiles, residuals, and galaxy thumbnails for an example BS galaxy.]{An example galaxy best fit by a {\it BS} model (SDSS DR8 ObjID 1237667444048855252). As Figure \ref{egB} for a {\it BS} model ({\bf top:} bulge = red, \sersic = green), and the corresponding {\it S} model ({\bf bottom:} \sersic = red).}
\label{egBS}
\end{figure}

\begin{figure}
\begin{center}
	\includegraphics[width=.99\linewidth,clip=true]{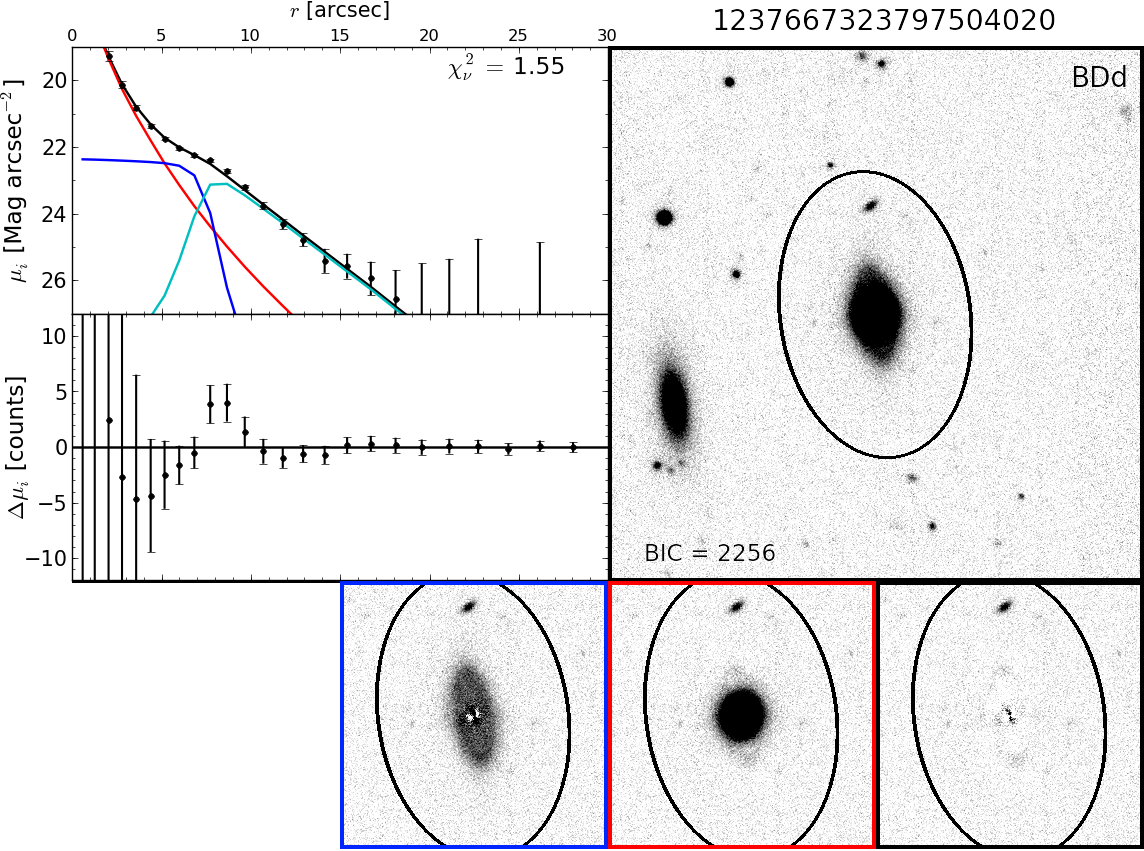}
	\includegraphics[width=.99\linewidth,clip=true]{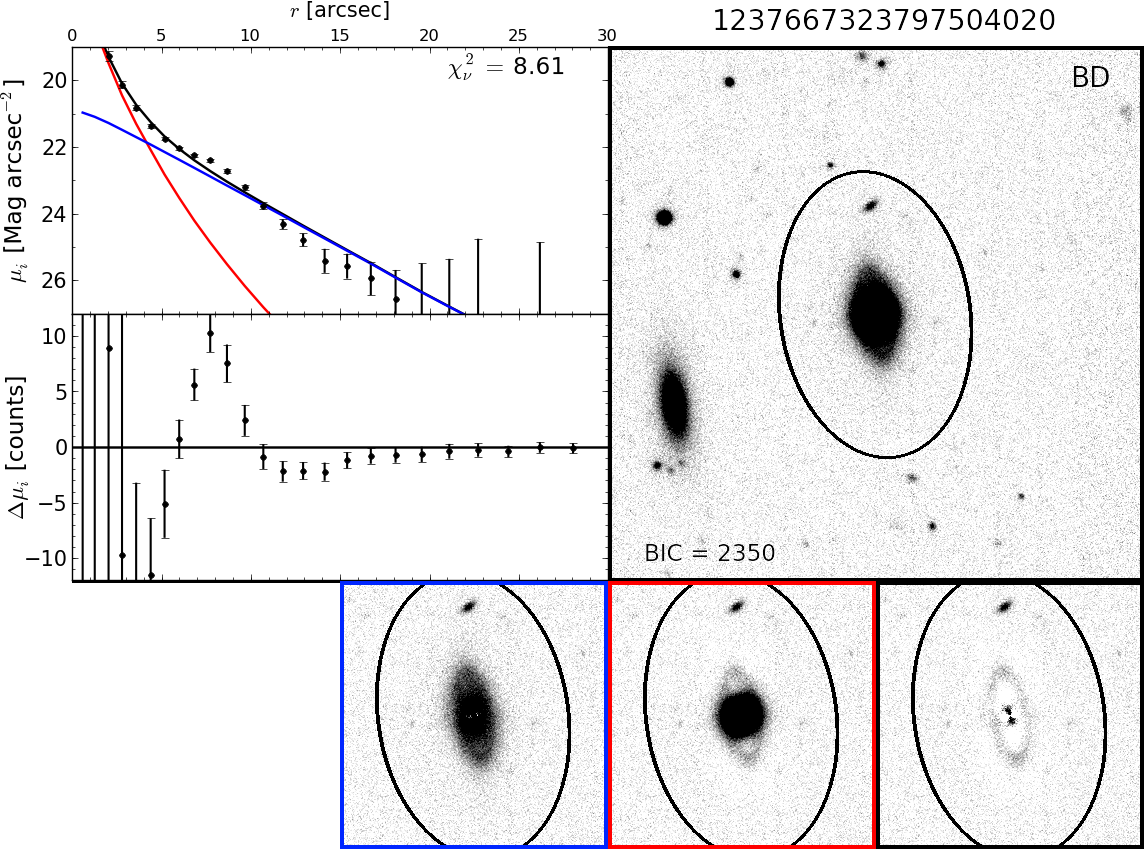}
\end{center}
\caption[Surface brightness profiles, residuals, and galaxy thumbnails for an example BDd galaxy.]{An example galaxy best fit by a {\it BDd} model (SDSS DR8 ObjID 1237667323797504020). As Figure \ref{egB} for a {\it BDd} model ({\bf top:} bulge = red, inner/outer disc = blue/cyan), and the corresponding {\it BD} model ({\bf bottom:} bulge = red, disc = blue).}
\label{egBDd}
\end{figure}

\begin{figure}
\begin{center}
	\includegraphics[width=.99\linewidth,clip=true]{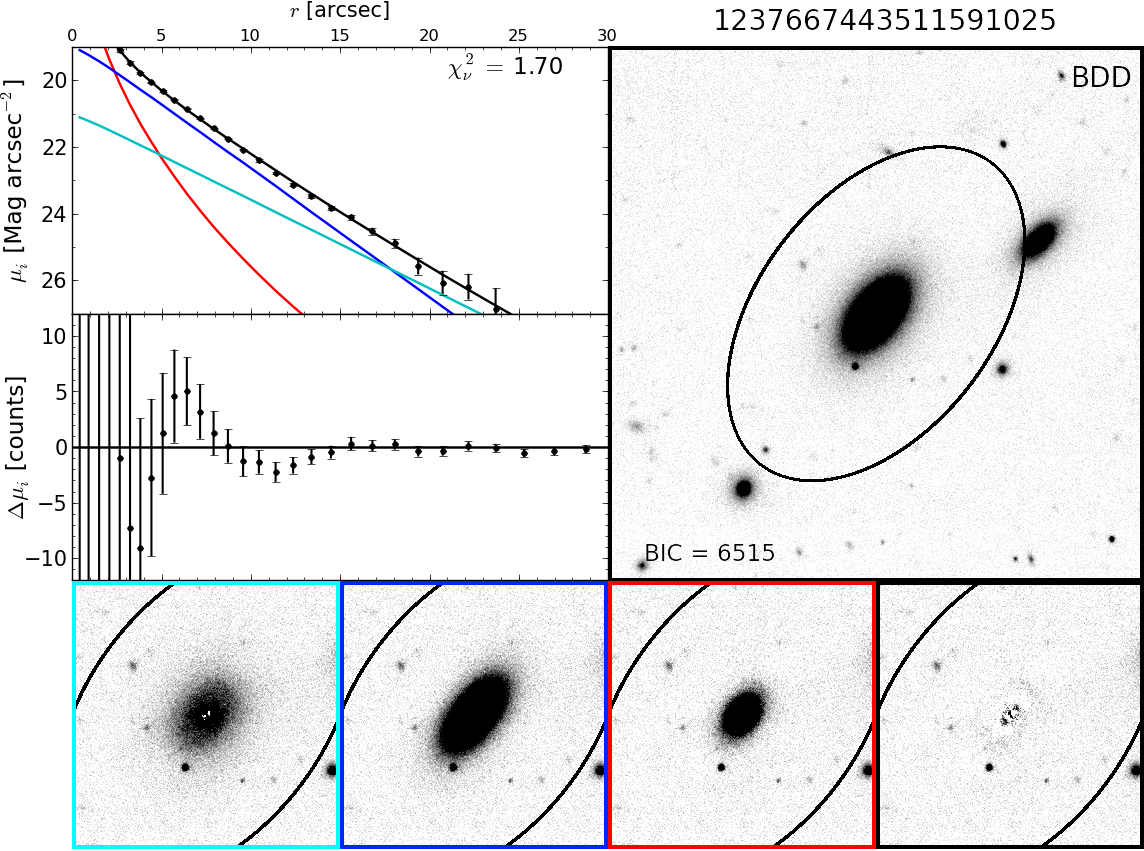}
	\includegraphics[width=.99\linewidth,clip=true]{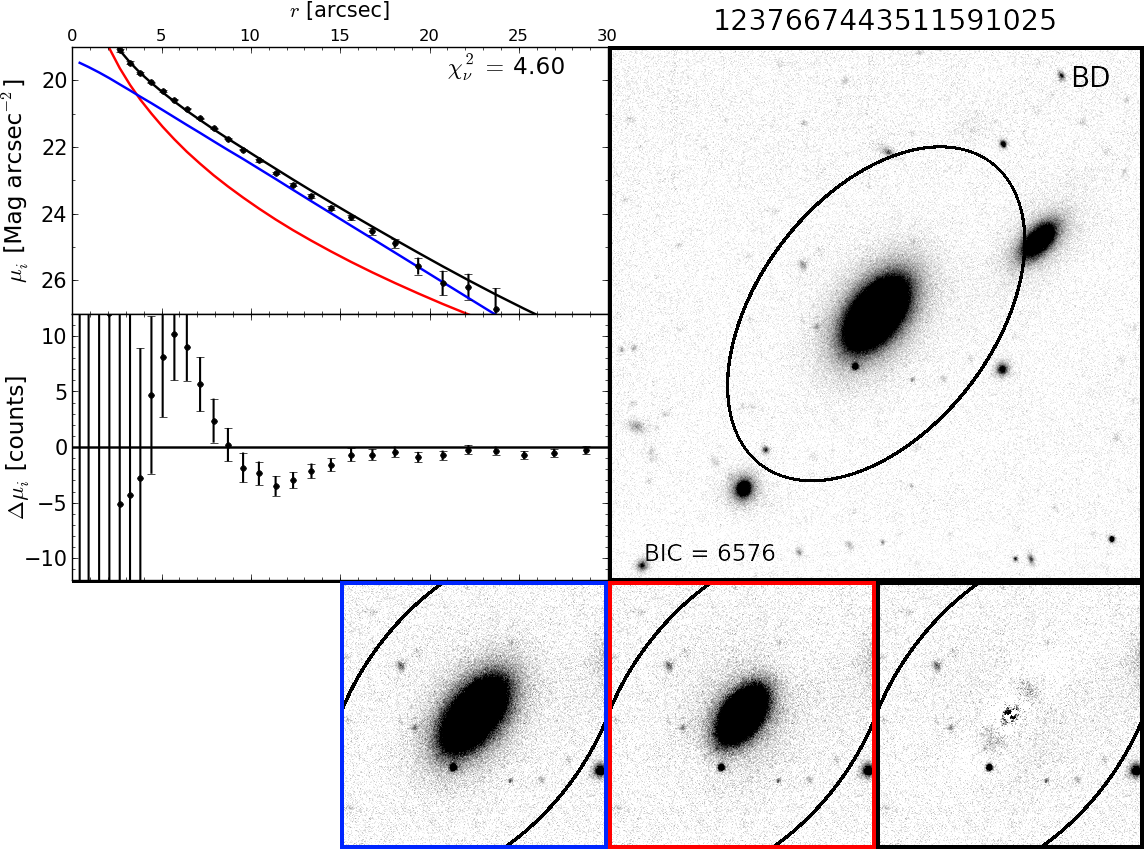}
\end{center}
\caption[Surface brightness profiles, residuals, and galaxy thumbnails for an example BDD galaxy.]{An example galaxy best fit by a {\it BDD} model (SDSS DR8 ObjID 1237667443511591025). As Figure \ref{egB} for a {\it BDD} model ({\bf top:} bulge = red, disc1 = blue, disc2 = cyan), and the corresponding {\it BD} model ({\bf bottom:} bulge = red, disc = blue).}
\label{egBDD}
\end{figure}

\begin{figure}
\begin{center}
	\includegraphics[width=.99\linewidth,clip=true]{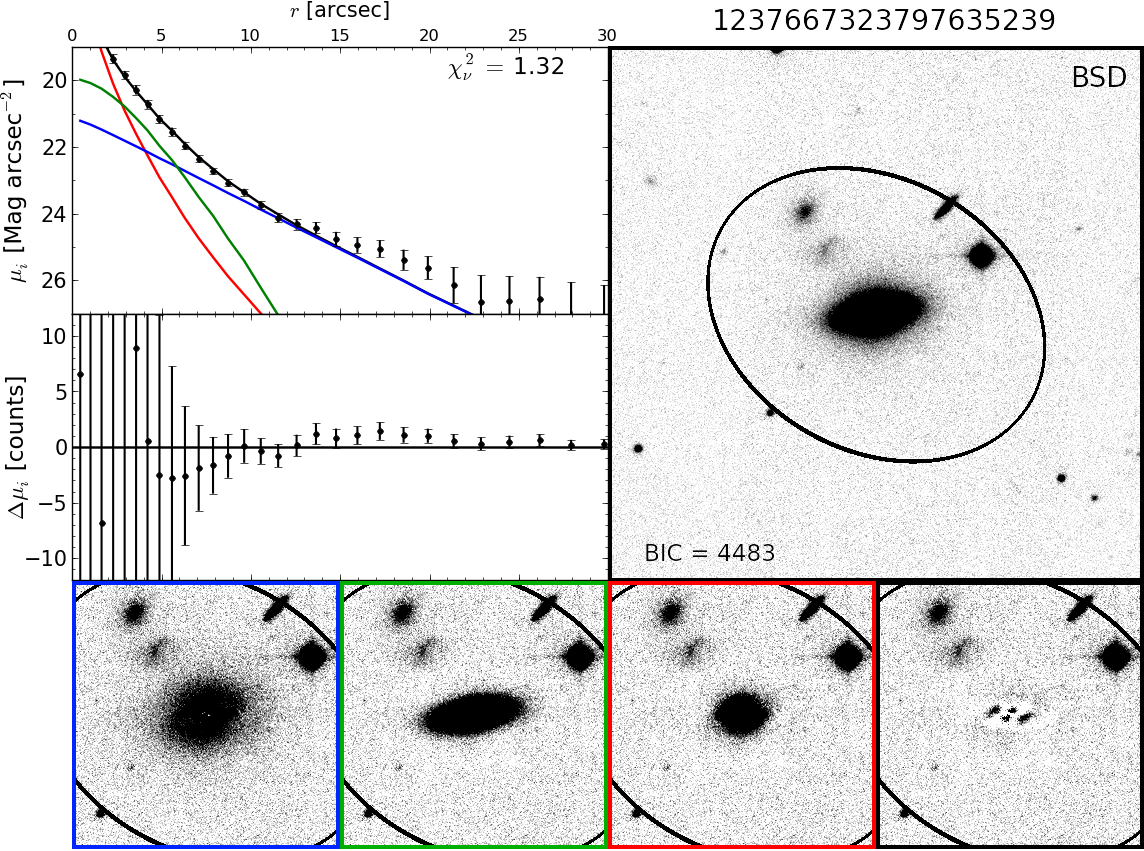}
	\includegraphics[width=.99\linewidth,clip=true]{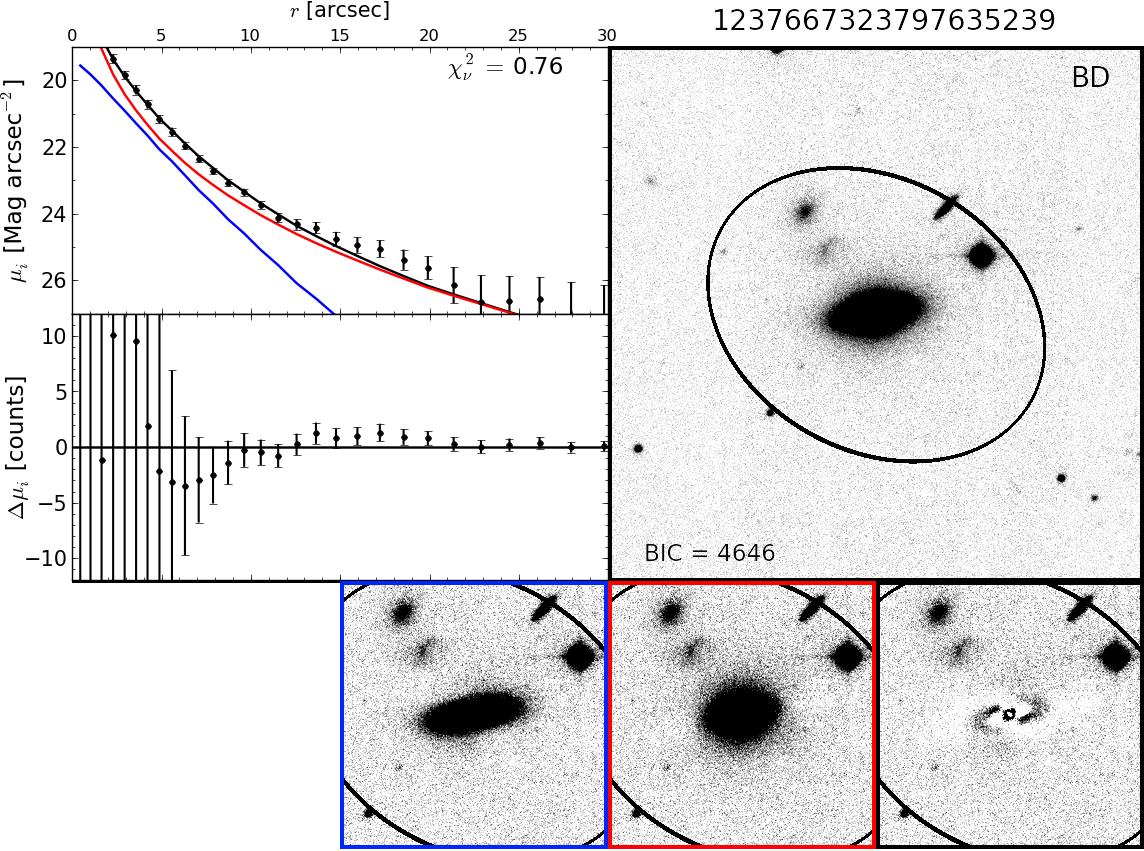}
\end{center}
\caption[Surface brightness profiles, residuals, and galaxy thumbnails for an example BSD galaxy.]{An example galaxy best fit by a {\it BSD} model (SDSS DR8 ObjID 1237667323797635239). As Figure \ref{egB} for a {\it BSD} model ({\bf top:} bulge = red, bar = green, disc = blue), and the corresponding {\it BD} model ({\bf bottom:} bulge = red, disc = blue).}
\label{egBSD}
\end{figure}

\begin{figure}
\begin{center}
	\includegraphics[width=.99\linewidth,clip=true]{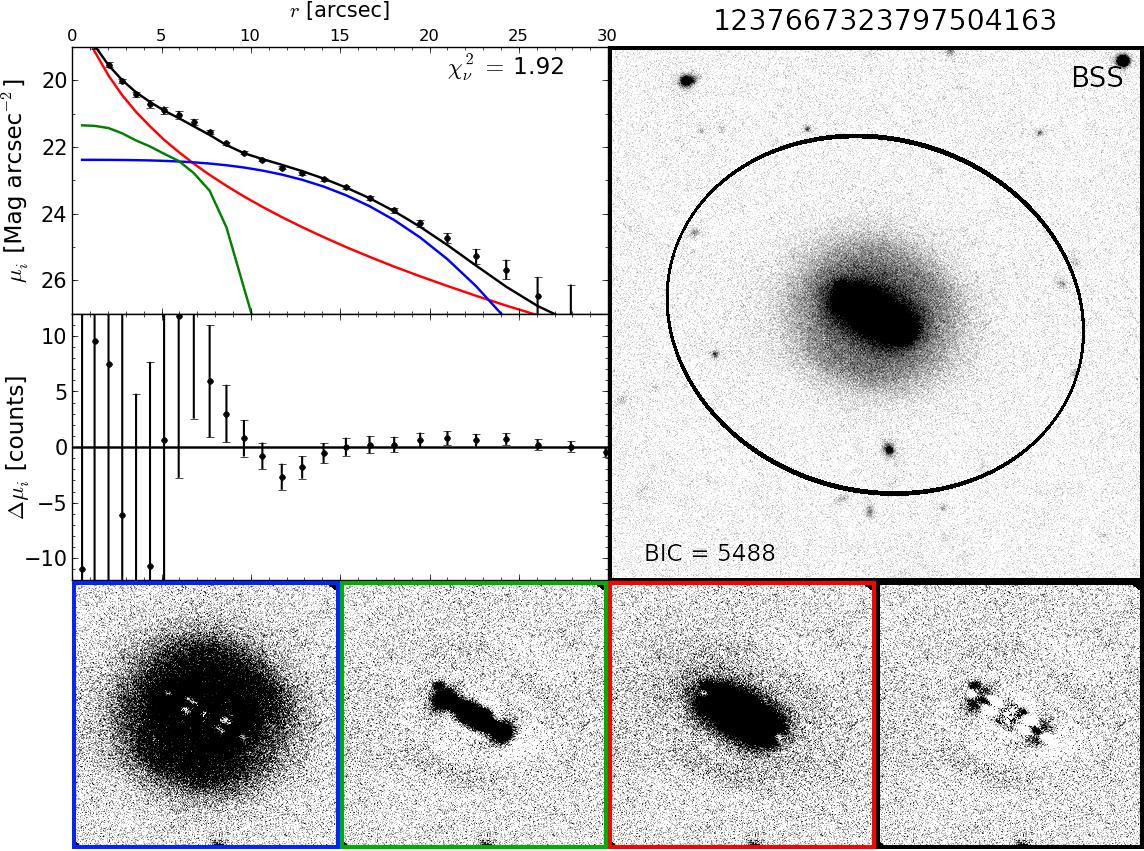}
	\includegraphics[width=.99\linewidth,clip=true]{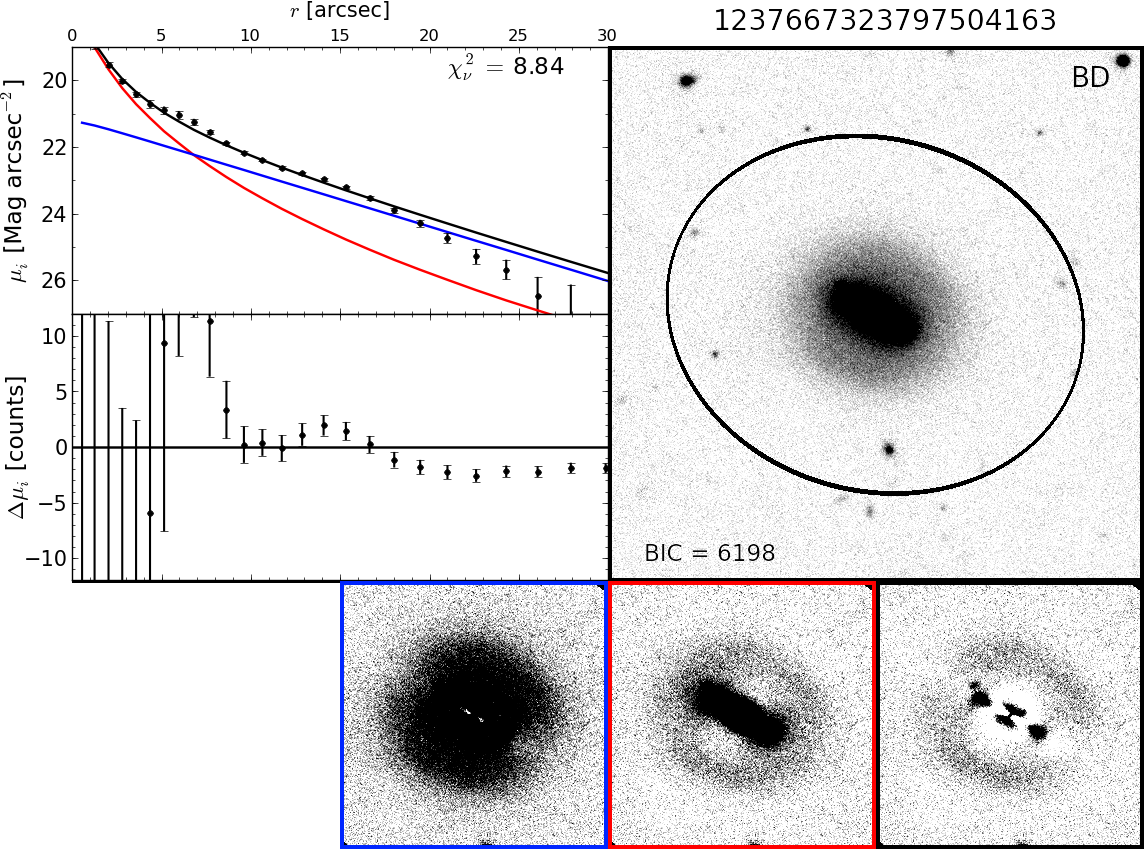}
\end{center}
\caption[Surface brightness profiles, residuals, and galaxy thumbnails for an example BSS galaxy.]{An example galaxy best fit by a {\it BSS} model (SDSS DR8 ObjID 1237667323797504163). As Figure \ref{egB} for a {\it BSS} model ({\bf top:} bulge = red, \Sersic1 = green, \Sersic2 = blue), and the corresponding {\it BD} model ({\bf bottom:} bulge = red, disc = blue).}
\label{egBSS}
\end{figure}

\begin{figure}
\begin{center}
	\includegraphics[width=.99\linewidth,clip=true]{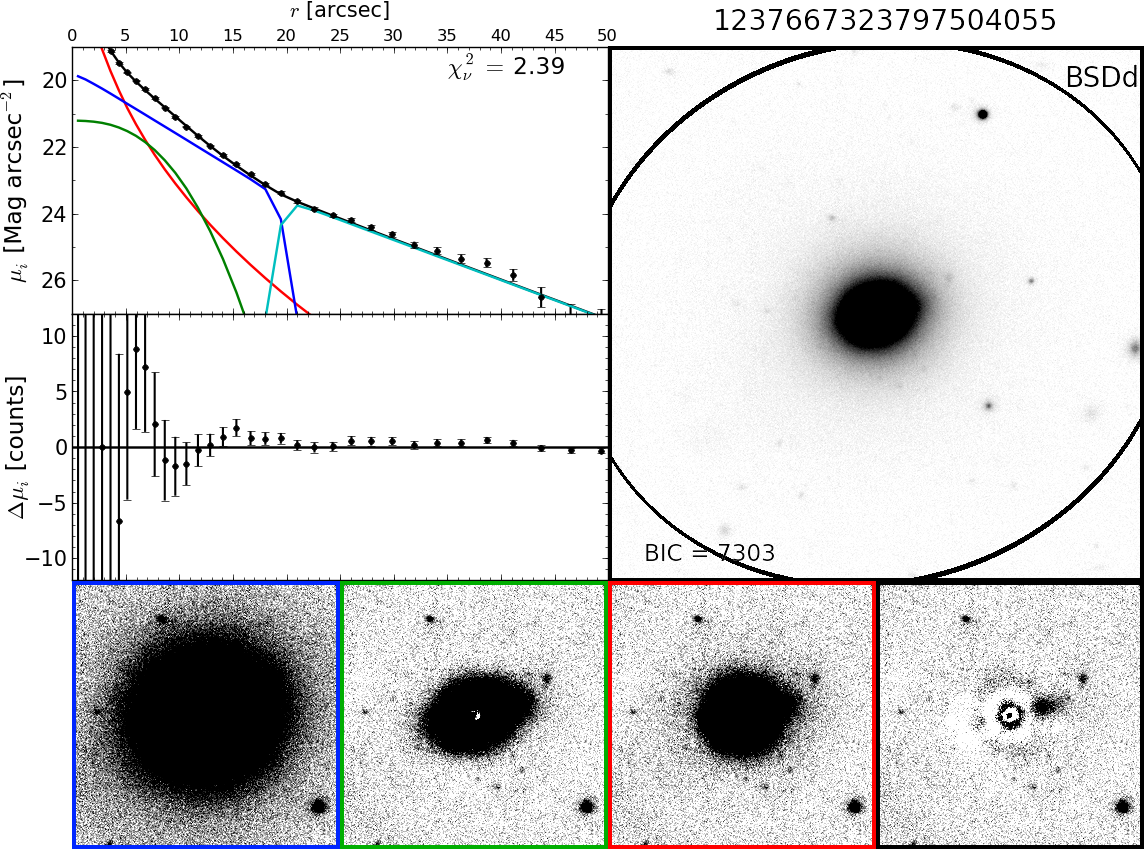}
	\includegraphics[width=.99\linewidth,clip=true]{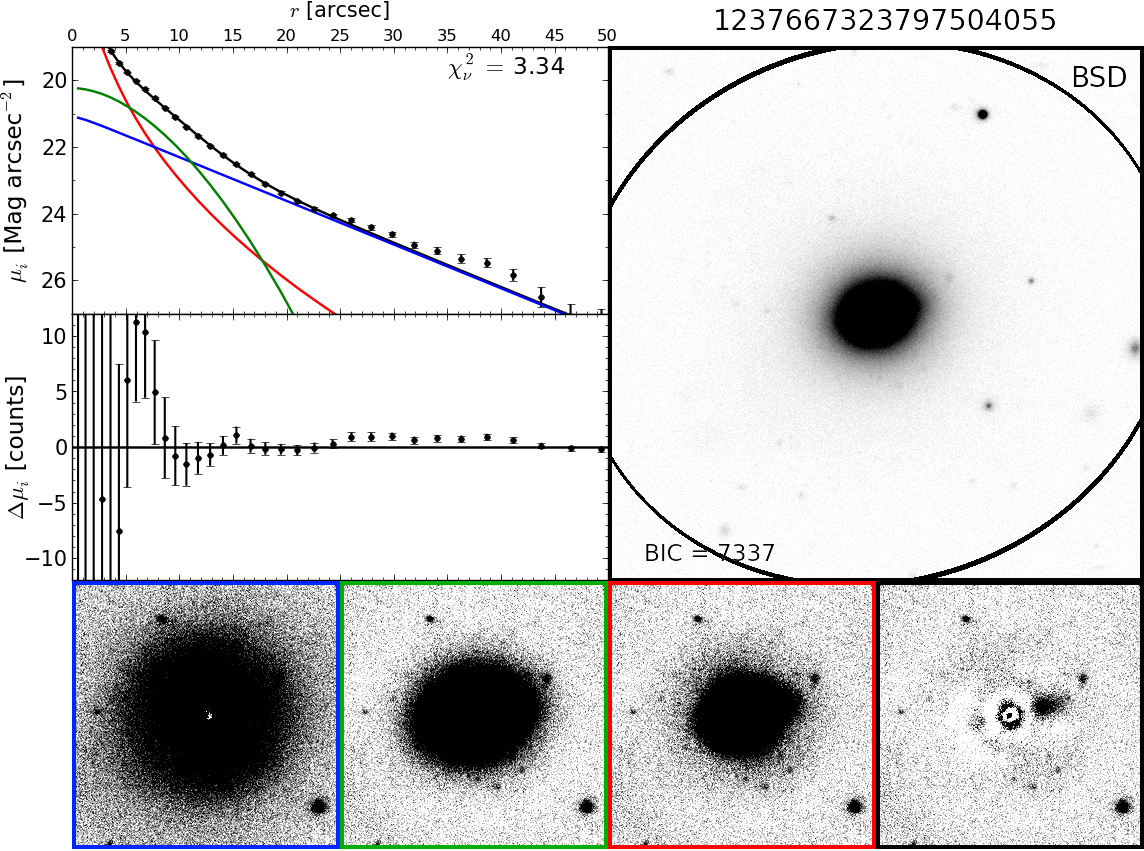}
\end{center}
\caption[Surface brightness profiles, residuals, and galaxy thumbnails for an example BSDd galaxy.]{An example galaxy best fit by a {\it BSDd} model (SDSS DR8 ObjID 1237667323797504055). As Figure \ref{egB} for a {\it BSDd} model ({\bf top:} bulge = red, bar = green, inner/outer disc = blue/cyan), and the corresponding {\it BSD} model ({\bf bottom:} bulge = red, bar = green, disc = blue).}
\label{egBSDd}
\end{figure}

Equivalent example plots for galaxies best fit by all other model types (except {\it CD}, and {\it CDd} due to small sample sizes) are presented in Figures \ref{egBD}-\ref{egBSDd}. Each best fit model (upper panels) is compared to its next simplest equivalent model (in terms of number of model components; lower panels). Hence, best-fit {\it BD} (Figure \ref{egBD}) and {\it BS} (Figure \ref{egBS}) models are compared to (underfit) {\it S} models; best-fit {\it BDd} (Figure \ref{egBDd}), {\it BDD} (Figure \ref{egBDD}), {\it BSD} (Figure \ref{egBSD}), and {\it BSS} (Figure \ref{egBSS}) models are compared to (underfit) {\it BD} models; and the best-fit {\it BSDd} model (Figure \ref{egBSDd}) is compared to an (underfit) {\it BSD} model.

\section{Details of Fitting}\label{filt_app}
\subsection{1D Break Parameterisation}\label{1D}
A simple 1D (outer) profile fitting procedure was used as a preliminary method of disk break detection. This was used primarily to produce realistic input parameter values for the 2D broken disk model fitting (see Section \ref{brk_disk}), but also identifies a sample of candidate broken disk galaxies. 

Galaxy surface brightness profiles (as measured along the major axis in $45^{\circ}$ wedges) were fit with a simple linear or broken linear model (analogous to exponential or broken exponential). Fitting was restricted to the range $3.54\second <r < r_{\rm sky}$ (where $r_{\rm sky}$ is the radius at which the total model surface brightness is equal to 4.94$\times$ the sky uncertainty, following the methodology in \citealp{Erwin2012}) to avoid contamination of the surface brightness profile by the bulge or low level sky background uncertainty. The inner limit ($3.54\second$) comes from the radius at which the bulge contribution, B/T($r$), of an average archetypal galaxy (as determined in preceding chapters) drops below $1\%$. The outer limit is increased relative to the analysis presented in previous chapters to allow the outer regions of galaxy surface brightness profiles to be characterised.

A 1D BIC was used to identify cases where the additional degrees of freedom afforded by the profile break significantly improved the model goodness-of-fit. For such broken galaxies, inner and outer disk scale length values were calculated from the inner and outer slopes of the best-fit broken linear models. The break radius was measured directly from the point at which the linear model switches from the inner to the outer slope.

Following 1D break detection, 215 galaxies (from an initial sample of 631 Coma cluster galaxies) were selected as candidate broken disks. Subsequent analysis stages also include galaxies with no 1D-detected break, however such galaxies must use generic input parameter values for broken disk model fitting. 

\subsection{GALFIT}\label{GALF}
\subsubsection{Initial Processing}
To measure the structural and photometric parameters of galaxy bulges and discs, galaxy decomposition has been carried out using {\footnotesize{\tt GALFIT}} (version 3.0.4), a 2D fitting routine \citep{GALFIT}. Given a user-specified model (of arbitrary complexity), {\footnotesize{\tt GALFIT}} varies parameters based on a non-linear chi-squared minimisation algorithm until no significant reduction in chi-squared ($\chi_{\nu}^2$) is found. The parameter values of this best-fit 2-component model are used to estimate the underlying structure and photometry of the target galaxy. 

For GALFIT's primary data input, $\sim100\second\times 100\second$ thumbnail images were extracted from the MegaCam image frames, centred on each target galaxy. Secondary data products, as derived from the imaging data, were used to improve fitting robustness. These data products are described in detail in Paper I. In brief: The local background sky and the underlying statistical noise map were independently-determined from each galaxy thumbnail. In addition, the image point spread function (psf) was characterised from stars in the MegaCam fields (no further than $5\minute$ from each galaxy), and the zero point of the magnitude scale was calibrated using aperture photometry.

Absolute rest-frame magnitudes were calculated by subtracting the distance modulus ($m-M =35.09$), and applying galactic dust extinction (using \citealp{Schlafly2011}; 0.014 mag in the $i$ band) and $k$-corrections (using \citealp{KCorr,Kcorr2}; typically $\lesssim 0.01$ mag).

\subsubsection{Initial Conditions}
\begin{figure}
\begin{center}
	\includegraphics[width=0.7\linewidth,clip=true]{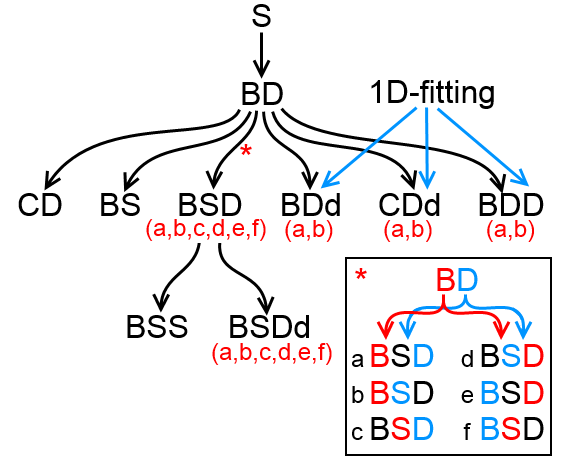}
\end{center}
\caption[Illustration of multi-component fitting model dependencies.]{Graphical illustration of the relation between the models during multi-component decomposition. Black arrows indicate which models take input parameter values from the best fit of a simpler model. Blue arrows indicate models which also take input parameter values from external sources. Models with multiple input variants (differing in their interpretation of progenitor model components) are noted in red. The inset illustrates multiple input generation for {\it BSD} models from the best fit {\it BD} components. 1D fitting is described in Appendix \ref{1D}}
\label{mod_flow}
\end{figure}

For our analysis the initial conditions for the multicomponent fits are based on the best-fit bulge + disc models presented in Paper I.
The iterative build-up of model complexity from the best-fit values of simpler models is the convention recommended for reliable results from GALFIT, and is used to provide a sensible starting point for the shapes (axial ratios), sizes, and intensity of additional model components. Hence, unlike Paper I, the model fitting procedure in this work was not extended (i.e. model parameters are not perturbed and re-fit to more thoroughly investigate the parameter space) as such an approach becomes computationally expensive (and highly sensitive to parameter degeneracies) for 3+ component models. Thus, the results of each input model were the product of one {\footnotesize{\tt GALFIT}} cycle and instead care was taken to generate sensible initial parameter values. In addition to building model complexity iteratively, multiple input models were generated for a single model type if the prior model's components could be interpreted ambiguously. For example, a best-fit {\it BD} model's bulge (or disc) structure can be used as the basis for the bulge, bar, or disc for an input {\it BSD} model. This build-up of model complexity is illustrated in Figure \ref{mod_flow}.
%Unlike Paper I, the model fitting procedure in this work was not extended (i.e. model parameters are not perturbed and re-fit to more thoroughly investigate the parameter space) as such an approach becomes computationally expensive (and highly sensitive to parameter degeneracies) for 3+ component models. Thus, the results of each input model were the product of one {\footnotesize{\tt GALFIT}} cycle. Instead, greater care was taken to generate sensible initial parameter values: in addition to building model complexity iteratively, multiple input models were generated for a single model type if the prior model's components could be interpreted ambiguously. For example, a best-fit {\it BD} model's bulge (or disc) structure can be used as the basis for the bulge, bar, or disc for an input {\it BSD} model. This build-up of model complexity is illustrated in Figure \ref{mod_flow}.

\subsubsection{Parameter Errors}
While GALFIT provides an estimate of the parameter errors these are underestimate by a large factor \citep{Haussler2007}. The formal calculation of the parameter errors is both complex and very computationally expensive, and has not been carried out in this study. In our analysis in Section 4 where necessary we adopt the approach of using the scatter about the observed trends as an upper estimate of the statistical uncertainties in the parameters. For example, in Figure 5a the observed scatter in the bulge \sersic index found at each luminosity bin is $\sim$0.7 and hence if there is no intrinsic scatter this is a reasonable estimate of the \sersic index error. In future work we will analyse mock images of galaxy with similar multi-component structures
found here in order to fully characterise the parameter uncertainties.

\subsubsection{Internal Dust Attenuation}
In this paper we have not considered the possible effects of internal dust attenuation on the observed photometric structures. While this can bias measured structural parameters, particularly at bluer wavebands for spiral galaxies \citep{Driver2007,Mollenhoff2006,Pastrav2013}, over 90\% of our Coma sample are cluster early-type galaxies where the dust content is likely to be small \citep{Kaviraj2012}. The ($B-R$) colours of cluster red-sequence galaxies can be nearly fully accounted for by the observed spectroscopically-determined stellar population trends to within an rms scatter of only $\simeq$0.02 mag \citep{Smith2009}. Such homogeneity in colour is unlikely to occur unless the internal extinction is uniformly small. Furthermore in our analysis highly inclined galaxies, and those with strong dust lanes or strong asymmetries are excluded (see Section \ref{filt_intro}) in order to minimise the possible effects of dust on our conclusions.

\subsection{Surface Brightness Profile Typing}\label{Allen+}
\begin{figure} 
\begin{center}
\includegraphics[width=\linewidth,clip=true]{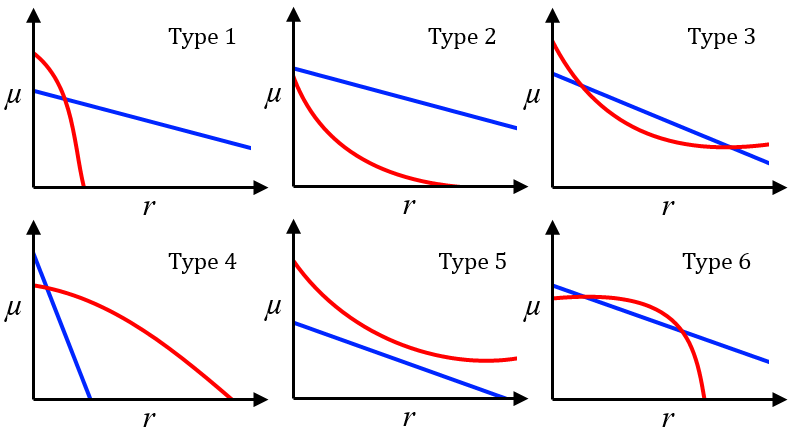}
\end{center}
\caption[Cartoon surface brightness plots for \sersic + disc systems of each Allen type \protect\citep{Allen2006}.]{Cartoon surface brightness plots for \sersic (red) + disc (blue) systems of each Allen type \protect\citep{Allen2006}. Type 1 profiles are termed `archetypal', while all other profiles are described as `atypical'. Profile Types 4 and 6 are inversions of Types 1 and 3 (respectively), and may indicate erroneous fitting results.}
\label{allen}
\end{figure}

\begin{figure}
\begin{center}
	\includegraphics[width=\linewidth,clip=true]{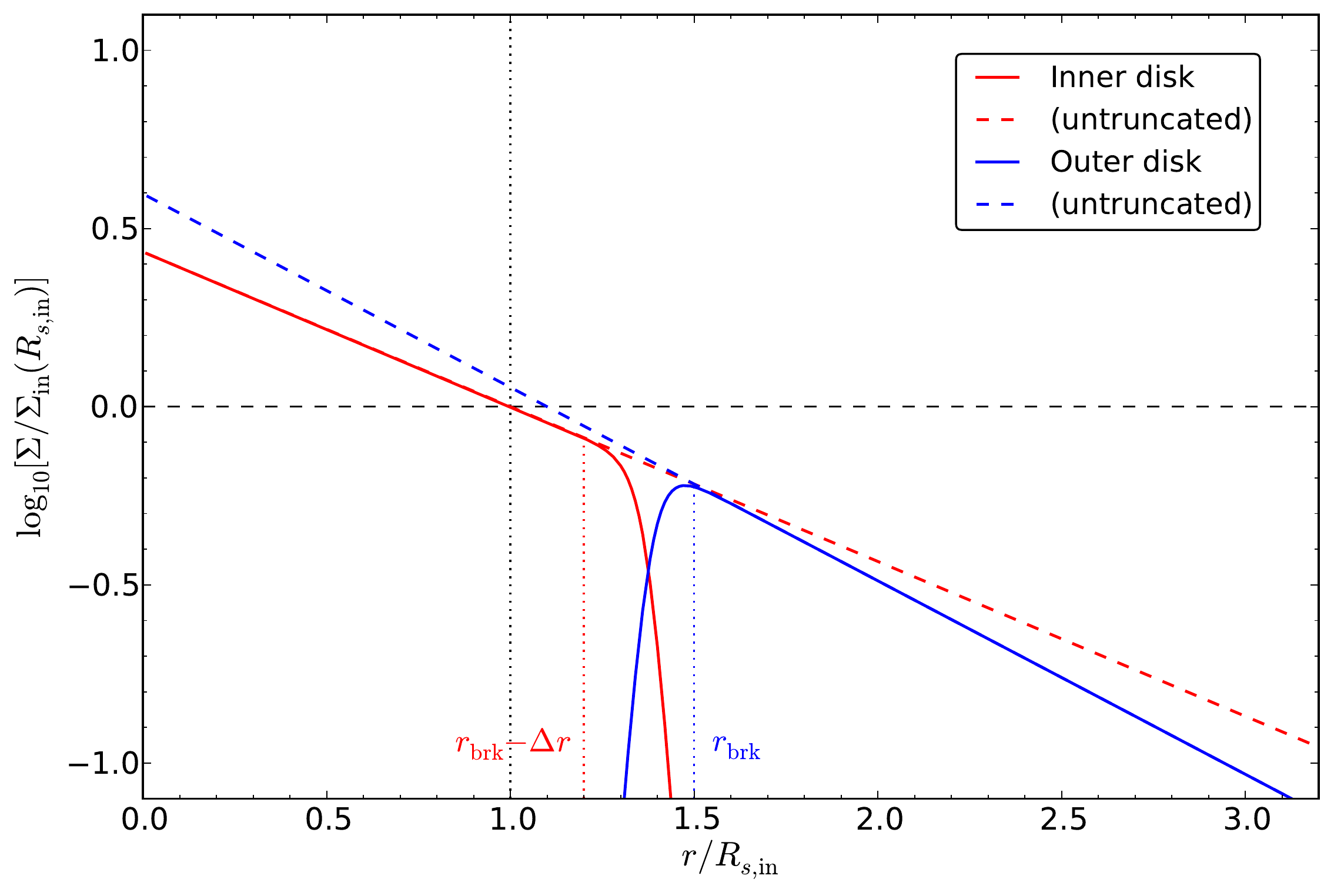}
\end{center}
\caption[Cartoon of the broken disk profile, indicating surface brightnesses of the inner and outer disks.]{Cartoon example of the broken disk profile, indicating surface brightnesses of the inner (red) and outer (blue) disks (and their untruncated forms). $r$ is normalised to the inner disk scale length, $R_{s,\rm{in}}$ (black dotted line), and $\Sigma$ is normalised to the inner disk surface brightness at $R_{s,\rm{in}}$ (black dashed line). The inner ($r_{\rm brk} - \Delta r$) and outer ($r_{\rm brk}$) truncation radii are indicated by red and blue dotted lines. In this example, $R_{s,{\rm out}} = 0.8 R_{s,{\rm in}}$, $r_{\rm brk} = 1.5R_{s,{\rm in}}$, and $\Delta r = 0.3 R_{s,{\rm in}}$.}
\label{brk_toon}
\end{figure}

For a multi-component system, it is convenient to describe the combined model in terms of its component surface brightness profiles. Profile types for \sersic + exponential models where first formalised in \cite{Allen2006} based on which component dominates at $r=0$, and how many times the component profiles intersect (see Figure \ref{allen}). Type 1 profiles correspond to the archetypal central bulge + outer disc structure of spirals and S0s. Type 2/Type 5 profiles represent dominant discs/bulges at all radii, with sub-dominant bulges/discs. Conversely, the centrally-dominant \sersic component in Type 3 profiles re-dominates the model at large radii. These profiles may be non-physical representations of more complex (3+ component) systems. Profile Types 4 and 6 are equivalent to Types 1 and 3 with the roles of the \sersic and exponential components swapped. As such, these inverted profiles may be symptoms of erroneous fitting pathways, rather than true physical structures.

\subsection{2D Broken Disk Model}\label{brk_disk}
Fitting a broken disk structure requires a model profile with distinct inner and outer exponential scale radii, connected via a smooth transition. In {\footnotesize{\tt GALFIT}}, this profile is implemented by linking two exponential disk profiles ($\Sigma_{\rm in}$ and $\Sigma_{\rm out}$) with (hyperbolic) truncation functions at some break radius. This (pixel surface brightness) profile can be expressed as:

\begin{equation}
\Sigma(r) = T_1(r)\Sigma_{\rm in}(r) + T_2(r)\Sigma_{\rm out}(r)
\end{equation}

\noindent
where $T_1$ and $T_2$ are the outer and inner truncation functions available for {\footnotesize{\tt GALFIT}} (see \citealp{GALFIT}). The full functional form of the broken disk profile is:

\begin{multline}
\Sigma(r)=\frac{1}{2}\left(1-{\rm tanh}\left[(2-B)\frac{r}{r_{\rm brk}} + B\right]\right)\Sigma_{0,{\rm in}}{\rm exp}\left(\frac{-r}{R_{s,{\rm in}}}\right) \\ 
+ \frac{1}{2}\left({\rm tanh}\left[(2-B)\frac{r}{r_{\rm brk}}+B\right]+1\right)\Sigma_{0,{\rm out}}{\rm exp}\left(\frac{-r}{R_{s,{\rm out}}}\right)
\label{brk_prof}
\end{multline}

\noindent
where $R_{s,{\rm in}}$ and $R_{s,{\rm out}}$ are the inner and outer disk scale radii, $\Sigma_{0,{\rm in}}$ and $\Sigma_{0,{\rm out}}$ are the (untruncated) central surface brightnesses of the inner and outer disks, and $r_{\rm brk}$ is the break radius. Here, $r_{\rm brk}$ is defined as the radius at which the inner and outer disk surface brightnesses are 1\% and 99\% of their untruncated values respectively. Dimensionless parameter $B$ is defined as $B = 2.65 - 4.98\left(\frac{r_{\rm brk}}{\Delta r}\right)$, where $\Delta r$ is the break softening radius (radial difference within which the truncated flux drops from $99\%$ to $1\%$). An example of the broken disk profile is presented in Figure \ref{brk_toon} for a truncated (Type II) disk with a greatly exaggerated $\Delta r$.

The surface brightness of this model component can be fully described by a single {\footnotesize{\tt GALFIT}} input parameter: surface brightness at the break radius, $\mu(r=r_{\rm brk})$. The value of $\mu(r=r_{\rm brk})$ is constrained to be identical for the inner and outer disk structures, ensuring continuity of the total component profile. Additionally, the axis ratios and position angle parameters of both disks are coupled for structural consistency, and $\Delta r$ is fixed at 0.1 pixel ($0.02 \second$). Hence, the broken disk profile includes only two more free fitting parameters ($R_{s,{\rm out}}$, and $r_{\rm brk}$; $k=6$) than the usual exponential disk model ($k=4$; see Table \ref{multi_modT}).

Fitting using a truncation function with {\footnotesize{\tt GALFIT}} yields a component's surface brightness at $r_{\rm brk}$, rather than the total component magnitude. Integrating Equation \ref{brk_prof} to infinity, however, is non-trivial due to the tanh function. Instead the total broken disk profile luminosity can be approximated using:

\begin{equation}
L_{\rm tot} = \int_0^{r_{\rm brk}} \Sigma_{\rm in}(r) 2 \pi dr + \int_{r_{\rm brk}}^{\infty} \Sigma_{\rm out}(r) 2 \pi dr
\end{equation}

\noindent
which approximates the truncation as a step function at $r_{\rm brk}$. The corresponding total profile magnitude is thus:

\begin{multline}
m_{\rm tot} = m_{\rm zp} - 2.5{\rm log}_{10}[2\pi q] \\
- 2.5{\rm log}_{10}[\Sigma_{0,{\rm in}} R_{s,{\rm in}}^2 \gamma(2,\frac{r_{\rm brk}}{R_{s,{\rm in}}}) \\
+ \Sigma_{0,{\rm out}} R_{s,{\rm out}}^2 \left(1- \gamma(2,\frac{r_{\rm brk}}{R_{s,{\rm out}}})\right)]
\end{multline}

\noindent
where $q$ is the common disk axis ratio, and $\gamma$ is the incomplete gamma function.

\subsection{Results Filter}\label{app_filter}
\begin{figure*}
\begin{center}
	\includegraphics[width=0.7\linewidth,clip=true]{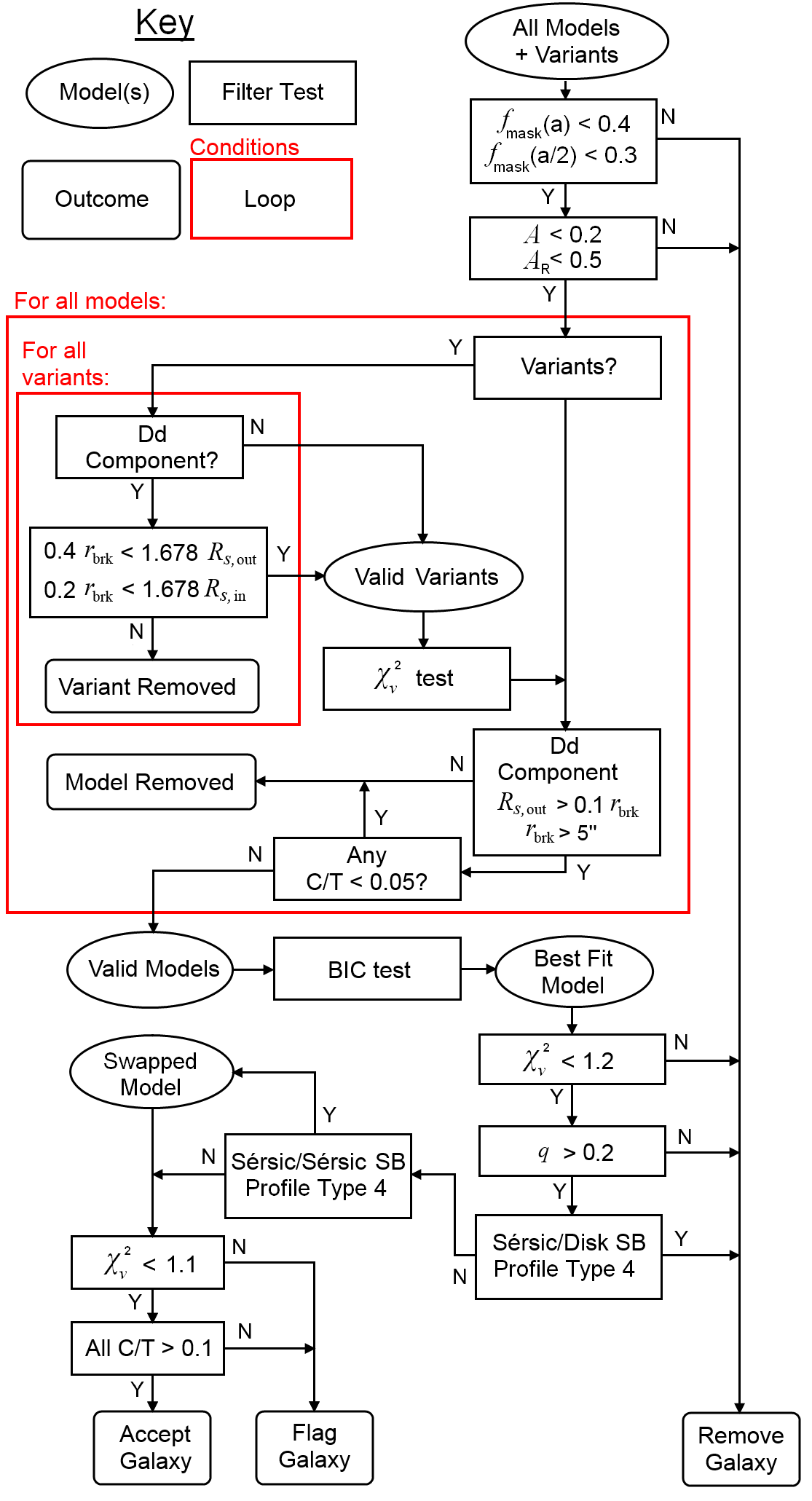}
\end{center}
\caption[Flow chart: Multi-component fitting model selection and filtering.]{Flow chart illustrating multi-component fitting model selection for models {\it S}, {\it BD}, {\it CD}, {\it BS}, {\it BDd}, {\it CDd}, {\it BDD}, {\it BSD}, {\it BSS}, and {\it BSDd}. For profile type definitions, refer to Section \ref{anal} and \citet{Allen2006}.}
\label{flowM}
\end{figure*}

\setcounter{table}{0}
\renewcommand{\thetable}{C\arabic{table}}
\begin{table*}
\begin{center}
\begin{tabular}{|l|l|l|l|l|l|l|l|l|l|}
\hline
\hline
ObjID& RA&Dec.&$z$&$M_{i{\rm{,tot}}}$&$R_{e{\rm ,tot}}$ &$M_{i,1}$&$R_{e,1}$&$n_1$&$q_1$\\
${\rm PA}_1$&$C0_1$&$M_{i,2}$&$R_{e,2}$&$n_2$&$q_2$&${\rm PA}_2$&$M_{i,3}$&$R_{e,3}$&$n_3$\\
$q_3$&${\rm PA}_3$&$R_{e{\rm ,out}}$&$r_{\rm brk}$&$C_1/T$&$C_2/T$&$C_3/T$&Model&Profile&Flag\\
\hline
\hline
1237665427552927881&194.875& 28.7& 0.024& -18.364& 1.1& -16.449& 0.233& 6.96& 0.8\\
118.556& 0.0& -18.16&1.237&1.0&0.826& 121.312& 999.0& 999.0&999.0\\
999.0& 999.0& 999.0&999.0&0.171& 0.829& 0.0& BD& 3& 0\\ 
\hline
1237665427552927902&195.926& 28.906& 0.022& -19.763& 502.0& -17.618& 0.53&1.008&0.955\\
97.153&0.0& -17.089& 4.686&0.128&0.769& 13.057&-19.488& 7.865&1.0\\
0.824& 79.766&2.417&5.914&0.139& 0.085& 0.776& BSDd&312& 1\\ 
\hline
1237665427552993436&195.018& 28.603& 0.023& -20.068& 3.73&-18.354& 1.221& 1.927&0.68\\
167.424& 0.0& -18.758& 2.917&0.374&0.838& 8.609& -19.303& 6.369&0.486\\
0.706& 21.469&999.0&999.0&0.206& 0.299& 0.495& BSS&311& 0\\ 
\hline
1237665427552993478&195.095& 28.574& 0.022& -19.109& 3.3& -18.388& 2.162& 2.124&0.659\\
29.906&0.0& -18.324& 4.354&0.44& 0.858& 111.1& 999.0& 999.0&999.0\\
 999.0& 999.0& 999.0&999.0&0.515& 0.485& 0.0& BS& 3& 0\\ 
\hline
1237665427553124587&195.449& 28.66&0.029& -18.288& 2.143& -18.288& 2.143& 1.649&0.723\\
137.698& 0.0& 999.0& 999.0&999.0&999.0& 999.0& 999.0& 999.0&999.0\\
 999.0& 999.0& 999.0&999.0&1.0& 0.0& 0.0& S&5& 0\\ 
\hline
\hline
\end{tabular}
\end{center}
\caption[Structural $i$ band multi-component model fitting results.]{The structural and photometric parameters of multi-component models fits ($i$ band) for the entire galaxy sample. The column headings are described in Table \ref{resM_key}. This table displays the first 15 data rows only; the complete version will be made available online.}
\label{result_M}
\end{table*}

Results filtering is applied to ensure that only galaxies which can be reliably characterised by (one of) the smooth, symmetric candidate models are considered for analysis. Model selection (i.e. the identification of the most statistically meaningful candidate model) is also a key function of this filter. This process is based on the sample filtering in Paper I, which describes a number of test parameters in greater detail (notably $A$, $A_{\rm res}$, and $f_{\rm mask}$).  

\begin{figure}
\begin{center}
	\includegraphics[width=\linewidth,clip=true]{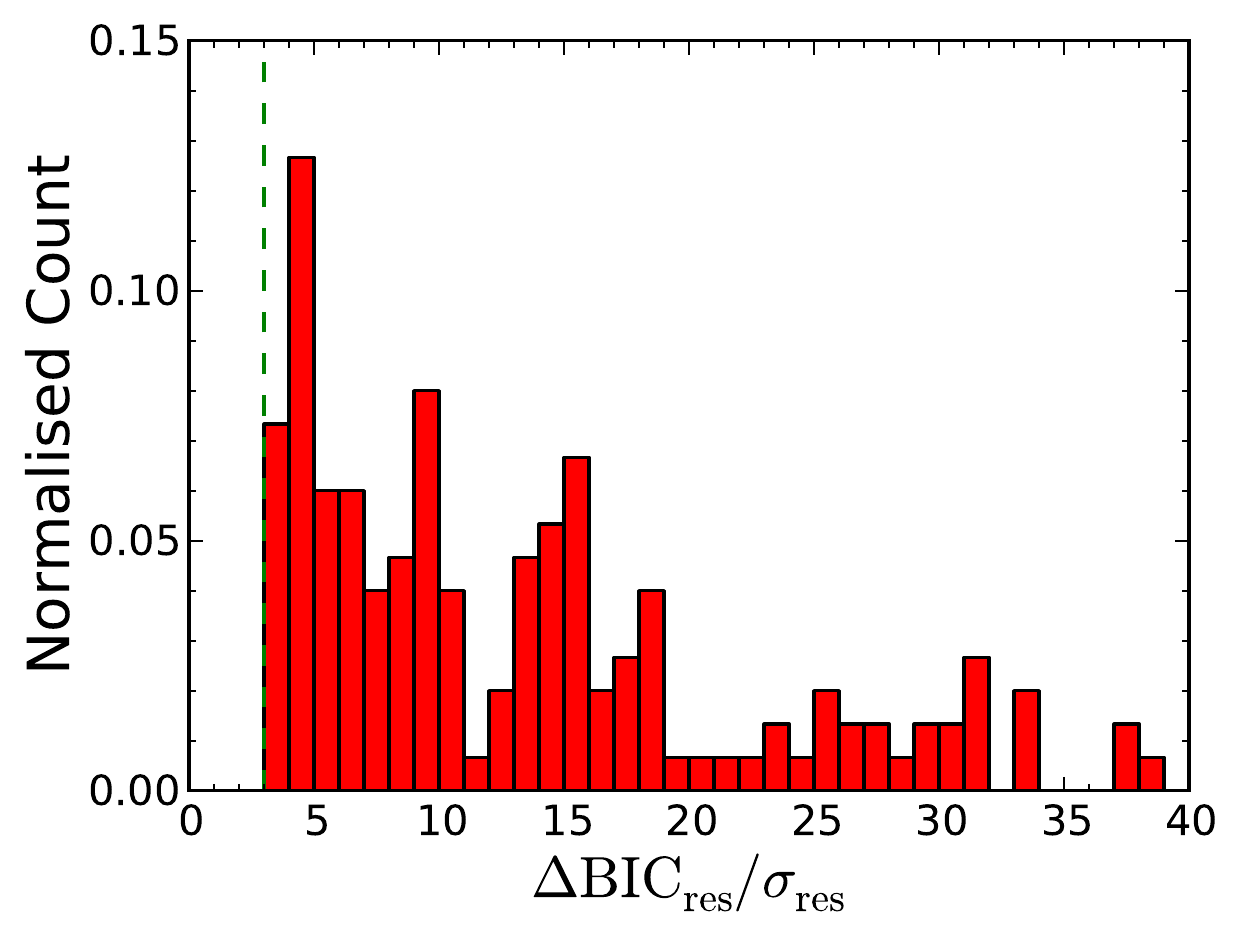}
\end{center}
\caption[Histogram of $\Delta$BIC values relative to the uncertainty on $\Delta$BIC.]{Histogram of $\Delta{\rm BIC}_{\rm res}$ values for well-fit, 2+ component models in the filtered sample ($N=344$). This compares each best-fit model to the next simplest (valid) model, plotted here relative to the uncertainty in $\Delta{\rm BIC}_{\rm res}$, $\sigma_{\rm res}$. A dashed green line is included to indicate the $3\sigma_{\rm res}$ cutoff for model acceptance.}
\label{dBIC}
\end{figure}

The filtering process for multi-component fits is illustrated in Figure \ref{flowM}, and summarised as follows:
\begin{enumerate}
\item Galaxies are excluded if contaminated by nearby sources based on the number of masked galaxy thumbnail pixels within the target ellipse ($f_{\rm{mask}}(a_{\rm{target}},q_{\rm{target}})\geq0.4$) and within the inner quarter of the target ellipse ($f_{\rm{mask}}(a_{\rm{target}}/2,q_{\rm{target}})\geq0.3$). 
\item Asymmetrical galaxies are also removed ($A>0.2$), but the threshold for removing galaxies based on {\it BD}-residual asymmetry was raised to $A_{\rm{res}}>0.5$, as moderate residual asymmetry may simply indicate the presence of unfitted structural components. 
\item For models with multiple variants (e.g. {\it BSD}$_{\rm a - f}$; see Figure \ref{mod_flow}), a single (best fit) model is selected for analysis based on a simple $\chi^2$ test. However, for models with broken discs ({\it BDd}, {\it CDd}, {\it BSDd}) model variants are excluded from consideration if $0.4 r_{\rm brk} > 1.678 R_{s, {\rm out}}$ (i.e. the outer disc contributes less than $8\%$ of its total flux) or $0.2 r_{\rm brk} > 1.678 R_{s, {\rm in}}$ (i.e. less than $0.3\%$ of the inner disc's total flux is truncated). These cuts remove anomalous model structures resulting from the broken disc component fitting to unintended structures.
\item For broken disc ({\it Dd}) models, galaxies are removed if $R_{s,{\rm out}} < 0.1r_{\rm brk}$ as a bug in {\footnotesize{\tt GALFIT}}'s truncation yields an additional (strong) central point source in this regime.
\item Additionally, broken discs with $r_{\rm brk} < 5\second$ are removed, as the inner disc of such systems behave like point sources.
\item A BIC test (see Equation \ref{BICeq}) is applied to select the best-fit model, which introduces the least extra fitting parameters. When comparing any two models, the least complex (lowest $k$) model is preferred unless the BIC value of the higher $k$ model is at least $3\sigma_{\rm res}$ lower. For a range of (valid) candidate models, each model is paired and tested (in increasing order of complexity) with all other models until a best fit is found.
\item Models with (one or more) component-to-total ratios, C/T$<0.05$ are removed from consideration during the BIC test due to high parameter uncertainty. This is similar to the B/T cut for the selection of \Sersic-only models in Paper I, but does not make assumptions regarding the preferred `simpler' model.
\item The $\chi_{\nu}^2$ limit for (BIC-selected) models is lowered to $\chi_{\nu}^2 > 1.2$, while galaxies are now flagged if $1.1 < \chi_{\nu}^2 < 1.2$. This more critical cut in model $\chi_{\nu}^2$ has been calibrated through visual examination of model residuals.
\item Galaxies with disc/outer component axis ratios, $q < 0.2$ are removed, as multi-compo\-nent decomposition cannot be meaningfully applied to edge-on systems.
\item Models with Type 4 \Sersic/disc profiles (i.e. Type 4, $x4x$, $xx4$) are removed due to swapping of the bulge/bar and disc roles of the structural components.
\item Models with Type 4 \Sersic/\sersic profiles (e.g. Type 4, $4xx$) have their components swapped (e.g. bar and bulge swap) to maintain the `inner' role of the bulge component (or `inner'/`middle'/`outer' roles for components 1, 2, and 3 in {\it BSS} models). Galaxy models modified in this way are not removed or flagged.
\item Remaining models with $0.05< {\rm C/T} < 0.1$ are flagged as unreliable.
\end{enumerate}

\subsection{BIC Test Results}
\setcounter{table}{1}
\begin{table}
\begin{center}
\begin{tabular}{ll}
\hline
\hline
Column name & Description  \\
\hline
\hline
ObjID& SDSS DR8 Object ID\\
RA& Object Right Ascension [degrees]\\
Dec.& Object Declination [degrees]\\
$z$&Object SDSS Redshift\\
$M_{i{\rm{,tot}}}$&Total rest-frame magnitude\\
$R_{e{\rm ,tot}}$ &Upper limit total half-light radius [kpc]\\
$M_{i,1}$&Component 1 rest-frame magnitude\\
$R_{e,1}$&Component 1 half-light radius [kpc]\\
$n_1$&Component 1 \sersic index\\
$q_1$&Component 1 axis ratio ($b/a$)\\
${\rm PA}_1$&Component 1 position angle [degrees]\\
$C0_1$&Component 1 boxiness\\
$M_{i,2}$&Component 2 rest-frame magnitude\\
$R_{e,2}$&Component 2 half-light radius [kpc]\\
$n_2$&Component 2 \sersic index\\
$q_2$&Component 2 axis ratio ($b/a$)\\
${\rm PA}_2$&Component 2 position angle [degrees]\\
$M_{i,3}$&Component 3 rest-frame magnitude\\
$R_{e,3}$&Component 3 half-light radius [kpc]\\
$n_3$&Component 3 \sersic index\\
$q_3$&Component 3 axis ratio ($b/a$)\\
${\rm PA}_3$&Component 3 position angle [degrees]\\
$R_{e{\rm ,out}}$&Outer disk half-light radius [kpc]\\
$r_{\rm brk}$&Disk break radius [kpc]\\
$C_1/T$&Component 1 light fraction\\
$C_2/T$&Component 2 light fraction\\
$C_3/T$&Component 3 light fraction\\
Model&Best-fit model\\
Profile&(B/D) \protect\cite{Allen2006} type\\
Flag& Fitting flag\\
\hline
\hline
\end{tabular}
\end{center}
\caption[Column headings of Table \ref{result_M}.]{This table describes the column headings for Table \ref{result_M}, presenting multi-component fitting results in the $i$ band. Best-fit model types are described in Section \ref{anal}, (inner component/outer component) \protect\cite{Allen2006} types are described in Section \ref{anal}, and fitting flags are described in Table \ref{resM_flag}.}
\label{resM_key}
\end{table}

\begin{table}
\begin{center}
\begin{tabular}{llc}
\hline
\hline
Flag code & Description& Condition\\
\hline
\hline
0&Normal fit&N/A\\
1&Bad fit (removed)&See Figure \ref{flowM}\\
2&High chi-squared&$1.1<\chi^2_{\nu} < 1.2$\\
3&Low component fraction&Any $0.05 < {\rm C/T} < 0.10$\\
4&Small break radius&$r_{\rm brk} < 5\second$\\
\hline
\hline
\end{tabular}
\end{center}
\caption[Table of multi-component fitting flag codes.]{This table describes the multi-component fitting flag codes, as used in Table \ref{result_M}.}
\label{resM_flag}
\end{table}

The results of the BIC test used to select the most statistically meaningful model for a given galaxy is illustrated in Figure \ref{dBIC} for all multi-component filtered sample galaxies ($N=344$; i.e. excluding asymmetric galaxies, contaminated images, and bad fits). Here, we plot the difference in ${\rm BIC}_{\rm res}$ between the selected `best fit' and the next simplest (lower $k$) valid model, relative to the uncertainty in that $\Delta{\rm BIC}_{\rm res}$. A green dashed line is included to indicate the 3$\sigma_{\rm res}$ limit, below which a model would not be chosen over a simpler alternative. This plot is comparable with Figure B1 in Paper I, which plots $\Delta{\rm BIC}_{\rm res}$ for \Sersic + disc and \Sersic-only models.

While a number of galaxy models cluster close to the selection limit, only $\sim$20\% of models exhibit an improvement of less than 5$\sigma_{\rm res}$ when compared to a less complex model. The results of the present work are thus insensitive to slight changes to the $\Delta{\rm BIC}_{\rm res}$ selection limit. Therefore, model selection based on a BIC test is robust for comparing multi-component galaxy models.

A more detailed discussion of the BIC test for model selection is available in Paper I and \cite{HeadThesis}. These works provide further details on the formulation of Equation \ref{BICeq}, and include comparison of BIC-selected models with by-eye selection, and F-test model selection.

\section{Fitting Results Catalogue}\label{Cat}
Multi-component $i$ band fitting results for the extended Coma cluster sample ($N=631$, including blue galaxies) are presented in Table \ref{result_M} (column descriptions in Table \ref{resM_key}). The structural parameters of the best-fit model (indicated by `Model') are presented for each galaxy, including values for the total luminosity and combined half-light radius\footnote{This value is an upper bound to the true value based on the assumption that major axes of all model components are aligned on the sky.}. A value of 999.0 indicates a parameter is not present in the relevant best-fit model (e.g. disk break radius in an unbroken {\it BD} galaxy). Fit quality flags (`Flag') are explained in Table \ref{resM_flag}.

%%%%%% End of document %%%%%%
%\bsp
\label{lastpage}
\end{document}